\documentclass[a4paper,12pt,titlepage,numbers=noenddot,bibliography=totoc,twoside,BCOR=6mm]{scrreprt} 

\setkomafont{captionlabel}{\bfseries}   
\setcapindent{0em}

\usepackage{mathptmx}
\usepackage[scaled=.92]{helvet}
\usepackage[T1]{fontenc}
\usepackage[utf8]{inputenc}             
\usepackage{amsmath}
\usepackage{amssymb}
\usepackage[ngerman,british]{babel}     
\usepackage{graphicx}                   
    
\usepackage{url}                        
\usepackage{txfonts}                    

\usepackage{multirow}                   

\usepackage[squaren]{SIunits}           
\usepackage{caption}
\usepackage{subcaption}
\usepackage{cancel}
\usepackage{color}
\usepackage{dsfont}			
\usepackage{makeidx}			

\usepackage{hyperref}
\hypersetup{
    colorlinks,
    citecolor=black,
    filecolor=black,
    linkcolor=black,
    urlcolor=black
}

\usepackage{breqn}
\usepackage{booktabs}
\usepackage{slashed} 
\usepackage{pdfpages}

\usepackage{feynmp}
\makeatletter
\def\endfmffile{
  \fmfcmd{\p@rcent\space the end.^^J
          end.^^J
          endinput;}
  \if@fmfio
    \immediate\closeout\@outfmf
  \fi
  \ifnum\pdfshellescape=\@ne
    \immediate\write18{mpost \thefmffile}
  \fi}
\makeatother

\newcommand{\abgabedatum}{7. September 2013}   
\newcommand{\tn}[1]{\textnormal{#1}}         

\newcommand{\nlo}{nächst\-füh\-ren\-de}
\newcommand{\nnlo}{nächst"=zu"=nächst\-füh\-ren\-de}
\newcommand{\topq}{Top-Quark}
\newcommand{\W}{\ensuremath{\text{W}}-Boson}
\newcommand{\imag}{\ensuremath{\text{i}}}
\newcommand{\ampli}{\ensuremath{\mathcal{T}_{fi}}}

\newcommand{\intmas}{\ensuremath{{\mu^{4-D}\over \text{i}\piup^{{D\over 2}}\text{r}_{\Gamma}}}\ensuremath{\int \mathrm{d}^D l}}

\DeclareMathAlphabet{\mathbbold}{U}{bbold}{m}{n}

\graphicspath{{../images/}{../plots/}}

\begin{document}
	\pagenumbering{roman}

	\newcommand{\HRule}{\rule{\linewidth}{0.5mm}}

\begin{titlepage}
	\centering
	
	\vspace*{5mm}
        \HRule
        \vspace*{6mm}

	\textsf{\textbf{\Large{
	  Produktion einzelner Top-Quarks\\
	  in nächst-zu-nächstführender Ordnung der QCD: \\
	  Der Beitrag der Einschleifenamplituden quadriert \\
	}}}

        \vspace*{6mm}
        \HRule
        \vspace*{10mm}
	
	\textsf{\textbf{\large{ MASTERARBEIT \\ \vspace{8mm}
	zur Erlangung des akademischen Grades \\
	Master of Science (M. Sc.) im Fach Physik\\}}}
	
	\vspace{10mm}	
	\includegraphics[width=6cm]{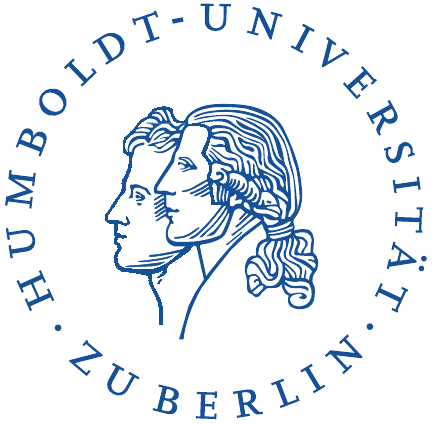}
	\vspace{8mm}
	
	Humboldt-Universität zu Berlin \\
	Mathematisch-Naturwissenschaftliche Fakultät I \\
	Institut für Physik \\
	
	\enlargethispage{2.5cm}
	\vfill

	\raggedright\parbox[t][7em][b]{\textwidth}{
		\begin{tabular}{ll}
			Eingereicht von: & Anne-Sophia Schade \\
			Geboren am:      & 4. Januar 1988 in Berlin \\
			\\
			Gutachter:       & Prof. Dr. Peter Uwer \\
			                 & Prof. Dr. Jan Plefka \\
			\\
			Eingereicht am:  & \abgabedatum
		\end{tabular}
	}
	
\end{titlepage}


\selectlanguage{british}
\clearpage
\begin{abstract}
In particle physics the fundamental
interactions between elementary particles are studied.
The Standard Model describes the weak, electromagnetic
an the strong interactions.
To check theoretical phenomena of the Standard Model,
it is important to precisely measure observables
of the theory.
The precision which can be obtained in experiments
determines the required accuracy of the theoretical 
calculation for the physical observables.

Top quark physics are an appropriate laboratory to
study phenomena of the Standard Model and to test
the limits of this theory. 

To obtain a higher precision for top quark predictions,
the next-to-next-to-leading order (NNLO) in the pertubative expansion of the 
top quark production cross-section has to be calculated.

The next-to-leading order (NLO) results are already available 
and provide a rather good accuracy \cite{Harris,Kidonakis}.
But this precision might be accidental, because at NLO 
not all the possible colour exchanges which occur
at the one-loop level are considered and they could have significant impact 
on the cross-section and its differential distributions.
In order to have precise predictions of the single top quark production
the complete NNLO calculation is necessary.

In this thesis the contributions of the one-loop amplitudes squared 
in NNLO are calculated for one production channel (\textit{t}-channel).
The other possible production channels of the single top quark
are also investigated.

With the presented setup well known results in literature
are reproduced. Intermediate results are calculated with different independent
methods and, if possible, compared with results from literature.

The occurring ultraviolet divergencies are dealt with in conventional
dimensional regularisation (CDR) in the \(\overline{\text{MS}}\) scheme
("`Modified Minimal Substraction Scheme"').
The final result still has infrared divergencies.
An anti-commuting scheme for the calculation with the \(\gamma_5\) matrix in CDR is used.
An alternative scheme for the treatment of \(\gamma_5\) is the
Larin scheme \cite{Larin}, which will be introduced and discussed.
\end{abstract}


\selectlanguage{ngerman}

\begin{abstract}

In der Elementarteilchenphysik beschäftigt man sich mit den grundlegenden
Wechselwirkungen in der Natur. Das Standardmodell, welches diese beschreibt,
umfasst die schwache, elektromagnetische und die starke Wechselwirkung
zwischen den Elementarteilchen. 
Um Phänomene der Theorie zu überprüfen, ist es wichtig genaue Messungen
vorzunehmen. Die Genauigkeit mit der man ein Phänomen messen kann, legt
fest, wie genau die dazugehörigen theoretischen Vorhersagen erfolgen müssen.

Die \topq-Physik ist ein geeignetes Labor, um Phänomene des Standardmodells
zu testen und um möglichweise Hinweise für Grenzen dieser Theorie zu finden.

Genauere theoretische Vorhersagen in der \topq-Physik bedeuten die Berechnung
der \nnlo n Ordnung (NNLO) der Produktion von \topq s.

Die \nlo\,Ordnung (NLO) für den Wirkungsquerschnitt
der Produktion einzelner \topq s ist berechnet und liefert eine 
hohe Genauigkeit \cite{Harris,Kidonakis}.
Die Präzision der NLO Rechnung könnte aber zufällig sein, da nicht 
alle möglichen Farbaustäusche für das Einschleifenniveau
beitragen werden. Mit einem Farbaustausch geht auch immer ein
Impulsaustausch einher, der einen signifikanten Einfluss auf den 
hadronischen Wirkungsquerschnitt
und seine differentiellen Verteilungen haben kann.
Um genauere Vorhersagen für die Produktion einzelner \topq s
machen zu können, ist es notwendig die 
vollständige NNLO Rechnung durchzuführen.

In dieser Arbeit wird für die Produktion
einzelner \topq s in einem Produktionskanal (\textit{t}-Kanal) 
der Beitrag der Einschleifenamplituden
quadriert in NNLO berechnet. 
Die anderen möglichen Produktionskanäle einzelner \topq s
werden ebenfalls berücksichtigt und besprochen.

Während der Rechnung konnten bekannte Ergebnisse reproduziert werden,
Zwischenergebnisse werden mit verschiedenen und unabhängigen 
Methoden ermittelt und wo möglich mit Ergebnissen 
aus der Literatur verglichen. 

Ultraviolette Divergenzen werden im Rahmen der konventionellen
dimensionalen Regularisierung (CDR) im \(\overline{\text{MS}}\)-Schema 
("`Modified Minimal Substraction Scheme"') renormiert.
Das Endergebnis weist noch infrarote Divergenzen auf.
Für die Rechnung wurde ein antikommutierendes 
Vertauschungsschema für die \(\gamma_5\)-Matrix
im Rahmen der dimensionalen Regularisierung benutzt.
Ein alternatives Schema nach Larin \cite{Larin} wird vorgestellt
und diskutiert.
\end{abstract}

        \cleardoublepage
	\tableofcontents
	\listoffigures

	\newpage	
	\pagenumbering{arabic}
	\chapter{Einleitung}

Das Hauptziel der Elementarteilchenphysik ist das Verständnis
der fundamentalen Wechselwirkungen der elementaren
Bestandteile der Materie.
Das Modell zur Beschreibung ist das Standardmodell, es beschreibt die elektromagnetische,
schwache und die starke Wechselwirkunge.
Die Beschreibung ist nicht vollständig.
Es beschreibt nicht die Gravitation für Quantenobjekte
und auch weitere Phänomene existieren (z.B. eine nicht verschwindende Neutrinomasse, das
vorhandene Verhältnis von Materie und Antimaterie),
die mit dem Standardmodell nicht ausreichend erklärt werden können.

Das Standardmodell ist eine Eichtheorie mit einer 
lokalen Eichsymmetrie der Eichgruppe
\(\text{SU}(3)_{\text{{\tiny Farbe}}}\times \text{SU}(2)_{\text{\tiny Links}}
\times \text{U}(1)_{\text{Y}}\)\footnote{\(\text{SU}(2)_{\text{\tiny Links}}\) 
beschreibt die Eichgruppe
der chiralen Symmetrie für linkshändige Fermionen, 
\(Y\) ist die schwache Hyperladung. Nach Ref. \cite{peskin}
ist die schwache Hyperladung \(Y = Q - T_3\), wobei \(Q\) die elektrische Ladung
und \(T_3\) die dritte Komponente des schwachen Isospins sind.}, 
diese bildet die mathematische Grundlage des Standardmodells.
\(\text{SU}(2)_{\text{\tiny Links}}\times \text{U}(1)_{\text{Y}}\) wird
durch einen von Null verschiedenen Vakuumerwartungswert des 
Higgsfeldes zur Gruppe \(\text{U}(1)_{\text{\tiny em}}\)\footnote{Die 
Abkürzung em steht für elektromagnetisch.} spontan gebrochen.
Die zur \(\text{SU}(3)_{\text{{\tiny Farbe}}}\)-gehörige Symmetrie 
ist ungebrochen. Die zugehörige Quantenfeldtheorie nennt man
Quantenchromodynamik (QCD), sie beschreibt die starke Wechselwirkung. 
Die dieser Theorie zugrunde liegenden Feldgleichungen konnten noch 
nicht analytisch gelöst werden, der Zugang zur QCD ist somit im Moment
nur störungstheoretisch möglich\footnote{Die Aussage gilt für 
die \(\text{SU}(2)_{\text{\tiny Links}}\) ebenso. Die durch die 
Störungstheorie bestimmbaren Korrekturen der QCD sind größer 
als die der schwachen Wechselwirkung. Die Berechnung der QCD-Korrekturen
hat also eine größere Priorität.}.
%

Genaue Überprüfungen der Phänomene des Standardmodells und darüber
hinaus sind essentiell, um Theorien falsifizieren
und um das grundlegende Verständnis dieser Wechselwirkungen vertiefen
zu können.
Im Experiment ist man daher bemüht, immer genauere Mess- und
Auswertungsmethoden zu entwickeln.
In der Theorie hingegen bedeutet es, höhere Korrekturen von
möglichen Prozessen zu berechnen, um damit genauere Vorhersagen
für das Experiment liefern zu können.

Ein geeignetes Labor bietet dabei die \topq-Physik.
Als das schwerste entdeckte Quark, mit einer Masse von
rund \(\unit{173}{GeV}\),
entsteht das \topq \, erst bei sehr
hohen Energien. Am LHC ("`\textbf{L}arge \textbf{H}adron 
\textbf{C}ollider"') am CERN
("`\textbf{C}onseil \textbf{E}uropéen 
pour la \textbf{R}echerche \textbf{N}ucléaire"', Europäische 
Organisation für Kernforschung) werden 
Protonen bei einer Schwerpunktenergie von \(\unit{7}{TeV}\) oder \(\unit{8}{TeV}\)
(geplant sind bis zu \(\unit{14}{TeV}\)) zur 
Kollision gebracht. 
Wegen der zu erwartenden hohen Produktionsrate der einzelnen \topq s am LHC 
benötigt man genauere theoretische Vorhersagen, um der hohen Statistik
im Experiment gerecht werden zu können und somit einen Vergleich zwischen
Theorie und Experiment zu ermöglichen.

Da die Lebenszeit des \topq s kürzer als die 
typische Zeitskala der starken Wechselwirkung ist
(als Folge der hohen Masse des \topq s), hadronisiert
das \topq \, nicht, d.h. das \topq \, geht
keine Bindungszustände mit anderen Quarks ein.
Infolgedessen ist das \topq \, das einzige Quark, welches
als ein (quasi) freies Quark angesehen werden kann. 
Aus dieser besonderen Eigenschaft ergeben sich Möglichkeiten zur Untersuchung
des Standardmodells. Zum Beispiel sind detaillierte Untersuchungen
zum Spin der Zerfallsprodukte des \topq s und andere Eigenschaften
möglich, die nicht durch die Hadronisierung eines Quarks verloren gehen
könnten.

Das \topq \,wird überwiegend durch die starke Wechselwirkung in Paaren
erzeugt. Darüber hinaus können einzelne \topq s durch die schwache
Wechselwirkung erzeugt werden.
Die Untersuchung der Produktion einzelner \topq s über die schwache Wechselwirkung
ist interessant, weil sie eine direkte Messung des
Cabibbo-Kobayashi-Maskawa-Matrixelements (CKM-Matrixelements) 
\(\vert V_{\text{tb}}\vert\) erlaubt.
Die CKM-Matrix beschreibt Übergangswahrscheinlichkeiten zwischen 
den drei Quarkfamilien. Falls das CKM-Matrixelement \(V_{\text{tb}}\) 
einen von eins verschiedenen Wert annimmt, kann das ein Hinweis
auf eine weitere Quarkfamilie sein und die Verletzung der Unitarität 
der CKM-Matrix bedeuten.

Die Yukawa-Kopplung, die eine Kopplung des Standardmodell-Higgs-Dubletts
mit den Fermionen darstellt und somit die Massengenerierung der Fermionen
im Standardmodell beschreibt, weist bei der Kopplung mit dem \topq\,eine 
Kopplungskonstante von rund eins auf. 
Die Ursache für diese Kopplung im Vergleich
zu den sehr kleinen Kopplungen zu den anderen Quarks
ist unklar und gibt Raum für die Erforschung des Mechanismus der
elektroschwachen Symmetriebrechung.

Weitere Gründe sich mit dem \topq \, zu beschäftigen, sind
zum Beispiel der Test der Vektor-Axialvektor-Struktur des Vertexes zwischen
\topq , b-Quark und W-Boson,
mögliche anomale Kopplungen in Theorien jenseits des Standardmodells,
Produktion einzelner \topq s als Quelle für polarisierte Quarks
und der Zugang zu den Partonverteilungsfunktionen des b-Quarks.

Wie eingangs erwähnt, ist es für alle diese Phänomene 
notwendig präzise Vorhersagen zu treffen,
um sie mit dem Experiment vergleichen zu können.
Für die Entstehung einzelner \topq s bedeuten präzise Vorhersagen
im Speziellen die Berechnung
der zweiten Ordnung in der Störungstheorie (\nnlo\, Ordnung, 
im Englischen oft abgekürzt mit NNLO, "`next-to-next-to-leading order"').
Der aktuelle Stand der Berechnungen bei der Produktion einzelner \topq s liegt bei
der führenden und der \nlo n Ordnung (siehe Ref. \cite{Harris,Kidonakis}).
Die Unsicherheit der in der \nlo n Ordnung bestimmten Beiträge wird mit einer Variation
der Renormierungs- und Faktorisierungsskala (\(\mu_{\tn{R}}\) und \(\mu_{\tn{F}} \approx m_t \))
(siehe Kapitel \ref{Kap:wq}) im
Bereich \([\tfrac{1}{2}m_t,2m_t]\) abgeschätzt.
Diese Abschätzung der Unsicherheit aus der Variation
der Renormierungs- und Faktorisierungsskala
berücksichtigt nur die in dieser Ordnung auftretenden Strukturen.
Strukturen, die erst in der nächst-höheren Ordnung der Störungstheorie
beitragen werden, können einen größeren Beitrag haben als die Abschätzung
der Unsicherheit der niedrigeren Ordnung liefern kann.
Es ist daher unter Umständen nicht möglich die Größenordnung des Beitrags 
der \nnlo n Ordnung aus der \nlo n Ordnung abzuleiten.
Eine genauere Angabe des Wirkungsquerschnitts gelingt somit nur, indem die Beiträge 
der nächst-höheren Ordnung bestimmt werden.

Im speziellen Fall der Produktion einzelner
\topq s tragen in NNLO Farbstrukturen
bei, die in NLO gerade aufgrund der Art der 
Farbstruktur keinen Beitrag in der Interferenz mit dem Bornprozess leisteten
(siehe Kapitel \ref{einschleifen}). 
Die Berechnung der \nnlo n Ordnung in der Produktion einzelner
\topq s liefert somit einen wichtigen Beitrag zur Genauigkeit des Wirkungsquerschnitts.
Diese Arbeit stellt die Berechnung eines Teils der \nnlo n
Ordnung dar, den Beitrag der Einschleifenamplituden quadriert.

Die Arbeit gliedert sich in folgende Abschnitte:
Zunächst werden in führender Ordnung die Amplitudenquadrate
der Produktion einzelner \topq s in den verschiedenen
Produktionskanälen bestimmt. Darauf aufbauend
werden auch die zugehörigen partonischen und hadronischen Wirkungsquerschnitte
ermittelt und diskutiert.

Im Hauptteil der Arbeit werden dann Prozesse von höherer Ordnung betrachtet.
Nachdem alle Einschleifendiagramme im \textit{t}-Kanal mit ihren Farbstrukturen
eingeführt wurden, werden die Beiträge der Vertexdiagramme bestimmt.
Die bei dieser Rechnung auftretenden Tensorintegrale werden mittels
der Passarino-Veltman-Reduktion \cite{Passarino:1978} zu skalaren Masterintegralen reduziert.
UV-Divergenzen, die bei den Vertexkorrekturen vorhanden sind, werden im Kapitel
zur Renormierung besprochen und entfernt.
In dem Zusammenhang auftretende Besonderheiten, wie zum Beispiel 
die Anwesenheit des Axialvektorstroms (\(\gamma_\mu\gamma_5\))
im Rahmen der dimensionalen Regularisierung werden diskutiert.
In Anschluss werden die virtuellen Beiträge zur \nlo n Ordnung
bestimmt.

Nach der Behandlung der Vertexdiagramme werden die Beiträge der
Boxdiagramme berechnet. Besonderheiten bei der Wahl der Eichung und
das Verhalten der verschiedenen Infrarotdivergenzen wird in diesem
Abschnitt beschrieben.

Im Anschluss daran wird es eine kurze Zusammenfassung des verwendeten
Schemas im Rahmen der konventionellen dimensionalen Regularisierung geben.
Die Beiträge in \nnlo r Ordnung der assoziierten tW-Produktion eines 
einzelnen \topq s werden in dem darauffolgenden Abschnitt behandelt.
%

Im letzten Teil dieser Arbeit wird es noch eine kompakte Zusammenfassung der
erzielten Ergebnisse sowie einen kurzen Ausblick geben.
Im Anhang der Arbeit befinden sich Bezeichnungen der verwendeten Symbole und
der Kinematik sowie zusätzliche Grafiken, Definitionen von auftretenden 
Masterintegralen und die Amplitude der Boxdiagramme.
	\chapter{Grundlagen}

Die Theorie der QCD beschreibt
die Wechselwirkung zwischen Quarks und Gluonen. Die Struktur der Theorie lässt sich
aus einer lokalen Eichinvarianz ableiten. Die zu der Symmetrie zugehörige Eichgruppe
\(\text{SU}(3)_{\text{\tiny Farbe}}\) hat die zugeordnete Quantenzahl 
Farbladung (z.B. rot, grün, blau), sie beschreibt die Invarianz 
unter einer Rotation im Farbraum.
Die QCD ist eine nicht-abelsche Eichtheorie, das bedeutet, dass die Generatoren
der Gruppe nicht miteinander kommutieren. Die lokale Eichinvarianz führt auf
acht masselose Eichfelder, die in der QCD Gluonen genannt werden. Jedes Gluon
trägt eine Farbladung, die sich aus einer Farbe und einer Anti-Farbe zusammensetzt
\cite{peskin}.
Jedes Quark trägt eine Farbe und jedes Antiquark eine Anti-Farbe. 

Es existieren drei Quarkfamilien mit je zwei Quarks. Jedes dieser Quarks
trägt eine Flavour-Quantenzahl. In der Natur werden nur gebundene Quarkzustände
aus mehreren Quarks (Baryonen oder Mesonen)\footnote{Baryonen bestehen 
aus drei Quarks verschiedener Farbe oder Antiquarks verschiedener Anti-Farbe.
Mesonen hingegen bestehen aus einem Quark mit Farbe und einem Antiquark mit zugehöriger
Anti-Farbe.} beobachtet.

In der QCD ist die Flavour-Quantenzahl erhalten. Nur unter Beteiligung der schwachen
Wechselwirkung ist eine Änderung der Flavour-Quantenzahl möglich.
Genau dieser Fall der Flavouränderung tritt bei der 
Produktion einzelner \topq s auf. Ein W-Boson
koppelt an Quarks und ändert die Flavour-Quantenzahl.

Die Kopplungskonstante der QCD ist bei kleinen Werten der Renormierungskonstante 
(kleine Energieskala der betrachteten Wechselwirkung) sehr
groß und bei hohen Werten der Renormierungskonstante (große Energieskala)
sehr klein. Als Konsequenz sind Quarks bei sehr
hohen Energien (bei sehr kleinen Abständen) als nahezu frei zu betrachten 
(asymptotische Freiheit) und bei kleinen Energien (großen Abständen)
nur als gebundene Zustände vorzufinden, was man als „Confinement“
bezeichnet. 
Diese besondere Eigenschaft führt darauf, dass man die Quarks als Bestandteile 
(Partonen) der Hadronen betrachten kann.
Außerdem erlaubt die Energieabhängigkeit der Kopplungskonstanten 
der QCD, die Theorie bei hohen Energien in Ordnungen der Kopplungskonstante 
der QCD zu entwickeln. Ein pertubativer Zugang zur QCD ist somit möglich.

Im Experiment werden hadronische Wirkungsquerschnitte und ihre
differentiellen Verteilungen bestimmt. Teil des hadronischen
Wirkungsquerschnitts ist der partonische Wirkungsquerschnitt
(siehe Gl. \eqref{eq:hadwq}).
Der partonische Wirkungsquerschnitt kann in Ordnungen der 
Kopplungskonstanten der starken Wechselwirkung entwickelt werden:
\begin{align} 
 \hat{\sigmaup} = \hat{\sigmaup}^{\text{{\tiny LO}}} + \alpha_s \cdot \hat{\sigmaup}^{\text{\tiny NLO}} +
                  \alpha_s^2 \cdot \hat{\sigmaup}^{\text{\tiny NNLO}}  + 
                  \alpha_s^3 \cdot \hat{\sigmaup}^{\text{\tiny NNNLO}} +
                  \mathcal{O}(\alpha_s^4)\,.
\label{eq:SumPartWQ}                  
\end{align}

Der partonische Wirkungsquerschnitt ist je betrachteter Ordnung (\(x\)) abhängig vom
Amplitudenquadrat der zugehörigen Prozesse \cite{peskin}:
\begin{align}
 \hat{\sigmaup}^{\text{\tiny{\(\mathcal{O}(\alpha_s^x)\)}}} = 
  \int \mathrm{d}\, \Pi\,\, \vert \ampli^{\text{\tiny{\(\mathcal{O}(\alpha_s^x)\)}}} \vert^2 \,,
\label{eq:PartWQ}  
\end{align}
wobei {\small\( \int \mathrm{d}\, \Pi\)} die Integration über den Phasenraum darstellt. 
Um eine Observable des Experiments theoretisch berechnen zu können, müssen
die Amplitudenquadrate der zugehörigen Prozesse bestimmt werden.
Im nächsten Abschnitt soll es darum gehen das Amplitenquadrat der führenden Ordnung (LO),
also die nullte Ordnung in \(\alpha_s\) {\small\(\left(\mathcal{O}(\alpha_s^0)\right)\)},
der Produktion einzelner \topq s zu bestimmen\footnote{Wie für die Teilchenphysik üblich wird in  
natürlichen Einheiten gerechnet: Die Lichtgeschwindigkeit \(c\) wird zu eins gesetzt,
ebenso das reduzierte Plancksche Wirkungsquantum \,\(\hbar\),
Massen und Energien werden damit in Elektronenvolt eV angegeben.}.
\setlength{\unitlength}{0.70mm}
\begin{figure}[t]
\centering
 \begin{subfigure}[t]{0.4\textwidth}
  \centering
 \begin{fmffile}{schannel}
  \begin{fmfgraph*}(70,35)
  \fmfleftn{i}{2} \fmfrightn{o}{2}
  \fmflabel{$\text{q}(p_1)$}{i1} \fmflabel{$\overline{\text{q}}'(p_2)$}{i2}
  \fmflabel{$\overline{\text{b}}(p_3)$}{o1} \fmflabel{$\text{t}(p_4)$}{o2}
  \fmf{fermion}{i1,v1,i2}
  \fmf{boson,label=$\text{W}(q)$}{v1,v2}
  \fmf{fermion}{o1,v2}
  \fmf{heavy}{v2,o2}
  \fmfdotn{v}{2}
  \end{fmfgraph*}
 \end{fmffile}
 \caption{\textit{s}-Kanal.}
 \label{sch}
\end{subfigure}
  \hspace{1.5cm}
  \begin{subfigure}[t]{0.4\textwidth}
   \centering
  \begin{fmffile}{tchannel}
  \begin{fmfgraph*}(70,35)
  \fmfleftn{i}{2} \fmfrightn{o}{2}
  \fmflabel{$\text{b}(p_2)$}{i1} \fmflabel{$\text{q},\overline{\text{q}}(p_1)$}{i2}
  \fmflabel{$\text{t}(p_4)$}{o1} \fmflabel{$\text{q}',\overline{\text{q}}'(p_3)$}{o2}
  \fmf{fermion}{i1,v1}
  \fmf{heavy}{v1,o1}
  \fmf{boson,tension=0,label=$\text{W}$}{v1,v2}
  \fmf{fermion}{i2,v2,o2}
  \fmfdotn{v}{2}
  \end{fmfgraph*}
 \end{fmffile}
 \caption{\textit{t}-Kanal.}
 \label{tch}
\end{subfigure} 
 \caption[\textit{s}- und \textit{t}-Kanal der Produktion einzelner \topq s]
         {\textit{s}- und \textit{t}-Kanal der Produktion einzelner \topq s.} 
 \label{stkanal}
 \vspace{5mm}
\end{figure}
\setlength{\unitlength}{0.70mm}
\begin{figure}[t!]
\centering
 \begin{subfigure}[t]{0.4\textwidth}
  \centering
 \begin{fmffile}{tW1}
  \begin{fmfgraph*}(70,35)
  \fmfleftn{i}{2} \fmfrightn{o}{2}
  \fmf{fermion}{i2,v1}
  \fmf{heavy}{v2,o1}
  \fmf{fermion,label=$\text{b}(k_1)$}{v1,v2}
  \fmflabel{$\text{b}(p_2)$}{i2}
  \fmflabel{$\text{t}(p_4)$}{o1}
  \fmfdotn{v}{2}
  \fmf{gluon}{i1,v1}
  \fmf{boson}{v2,o2}
  \fmflabel{$\text{g}(p_1)$}{i1}
  \fmflabel{$\text{W}(p_3)$}{o2}
  \end{fmfgraph*}
 \end{fmffile}
  \vspace{5mm}
 \caption{1. Möglichkeit mit b-Propagator.}
 \label{tWB}
\end{subfigure}  
  \hspace{1.5cm}
  \begin{subfigure}[t]{0.4\textwidth}
  \centering
 \begin{fmffile}{tW2}
  \begin{fmfgraph*}(70,35)
  \fmfleftn{i}{2} \fmfrightn{o}{2}
  \fmf{gluon}{i1,v1}
  \fmf{heavy}{v1,o1}
  \fmf{heavy,tension=0,label=$\text{t}(k_2)$}{v2,v1}
  \fmf{fermion}{i2,v2}
  \fmf{boson}{v2,o2}
  \fmflabel{$\text{g}(p_1)$}{i1} \fmflabel{$\text{b}(p_2)$}{i2}
  \fmflabel{$\text{W}(p_3)$}{o2} \fmflabel{$\text{t}(p_4)$}{o1}
  \fmfdotn{v}{2}
  \end{fmfgraph*}
 \end{fmffile}
  \vspace{5mm}
 \caption{2. Möglichkeit mit t-Propagator.}
 \label{tWT}
\end{subfigure} 
 \caption[Zwei mögliche tW-Produktionen]
         {Zwei mögliche tW-Produktionen.} 
 \label{twproduktion}
\end{figure}
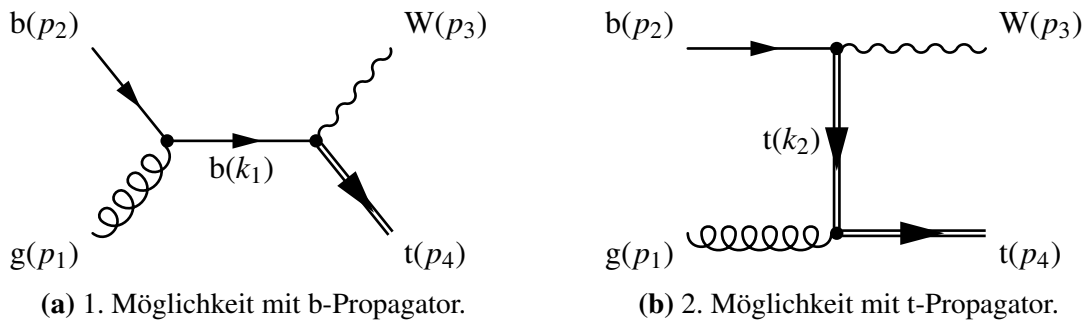
\section{Produktionskanäle einzelner Top-Quarks}

In führender Ordnung kann das einzelne \topq\, durch drei verschiedene 
Produktionskanäle erzeugt werden.
Bei dem sogenannten \textit{s}-Kanal, Abb. \ref{sch}, 
wechselwirken ein Quark und ein Antiquark (mit anderem Quark-Flavour)
über ein \W \, mit einem \topq\, (t) und ein Anti-Bottom-Quark (\(\bar{\text{b}}\)). 

Durch Kreuzen erhält man den \textit{t}-Kanal, Abb. \ref{tch}. Hier wechselwirken
ein Quark (q) (Antiquark \(\bar{\text{q}}\)) und ein Bottom-Quark (b) über ein \W\, miteinander 
und erzeugen so und ein \topq\, und ein Quark (Antiquark) mit anderem Flavour.

In beiden Kanälen ändert sich ein Quark-Flavour, da jeweils das \W\, 
koppelt.
Die Kanäle haben ihre Namen von den unterschiedlichen Impulsüberträgen, die
sich mit Hilfe der Mandelstam-Variablen (siehe \ref{A:mandelstam}) \textit{s} und 
\textit{t} ausdrücken lassen.

Das einzelne \topq\, kann noch über die assoziierte tW-Produktion (Abb. \ref{twproduktion}) entstehen.
Aus einem b-Quark und einem Gluon (g) im Anfangszustand entstehen hier ein
\topq\, und ein \W\, im Endzustand. Hier existieren zwei Möglichkeiten: 
Die erste Möglichkeit besteht darin, dass das propagierende Fermion
ein b-Quark darstellt (Abb. \ref{tWB}). Es ist auch möglich, dass ein
\topq\, der Fermionpropagator ist (Abb. \ref{tWT}).
Eine tW-Produktion mit Cabibbo-unterdrückten\footnote{Cabibbo-unterdrückt
heißt, dass ein Prozess bei dem ein Übergang zwischen den Quarkfamilien auftritt,
unterdrückt wird. Die entsprechenden CKM-Matrixelemente sind wesentlich kleiner als 
die Mischungsfaktoren innerhalb der Quarkfamilien 
(z.B. \(V_{\text{tb}} \sim 0{,}999 > V_{\text{td}} \sim 0{,}008 \)).} 
Beiträgen ist auch möglich.

Das \topq \, entsteht also über diese drei möglichen Kanäle,
abhängig von der Kinematik, haben die Kanäle unterschiedliche Anteile
im gesamten hadronischen Wirkungsquerschnitt (Kapitel \ref{Kap:wq}, Abb. \ref{ratio}). 
Um diese qualitative Aussage quantitativ
zu belegen, ist es notwendig den hadronischen Wirkungsquerschnitt, 
\(\sigmaup_{\text{{\tiny Had}}}\), zu bestimmen, was Thema der folgenden
Abschnitte sein soll. 

\section[Berechnung des Amplitudenquadrats im \textit{s}-Kanal]
        {Berechnung des Amplitudenquadrats im \textit{s}-Kanal}
\label{Kap:ampliquborn}   

\setlength{\unitlength}{0.85mm}
\begin{figure}[h]
\vspace{3mm}
\centering
 \begin{fmffile}{schannelud}
  \begin{fmfgraph*}(70,35)
  \fmfleftn{i}{2} \fmfrightn{o}{2}
  \fmflabel{$\text{u}(p_1)$}{i1} \fmflabel{$\overline{\text{d}}(p_2)$}{i2}
  \fmflabel{$\overline{\text{b}}(p_3)$}{o1} \fmflabel{$\text{t}(p_4)$}{o2}
  \fmf{fermion}{i1,v1,i2}
  \fmf{boson,label=$\text{W}(q)$}{v1,v2}
  \fmf{fermion}{o1,v2}
  \fmf{heavy}{v2,o2}
  \fmfdotn{v}{2}
  \end{fmfgraph*}
 \end{fmffile}
 \vspace{5mm}
 \caption[\textit{s}-Kanal mit $\text{u}$- und $\overline{\text{d}}$-Quark]
         {\textit{s}-Kanal mit $\text{u}$- und $\overline{\text{d}}$-Quark.}
 \label{skanalud}
\end{figure}
Der partonische Wirkungsquerschnitt \(\hat{\sigmaup}\) eines Prozesses 
ist, wie weiter vorn erwähnt, proportional zum Betragsquadrat der 
Streuamplitude \ampli \,(Gl. \eqref{eq:PartWQ}), 
wobei sich \ampli \, aus den Feynmanregeln\footnote{Die Feynmanregeln können
aus der Lagrangedichte des Standardmodells (\cite[Gl. (22-123)]{Nachtmann}) 
abgeleitet werden, zum Beispiel mittels
des Pfadintegralformalismus. In Gleichung \eqref{eq:LagrangeQCD}
ist die Lagrangedichte der QCD angegeben.} 
für das jeweilige Feynmandiagramm ergibt.

Anhand der Produktion des einzelnen \topq s im \textit{s}-Kanal 
mit einem \(\text{u}\)- und \(\overline{\text{d}}\)-Quark wird ersichtlich, 
wie sich das Betragsquadrat einer
Streuamplitude berechnen lässt. In Abb. \ref{skanalud} kann man die den 
Teilchen zugeordneten Impulsen entnehmen, wobei sich der Impuls \(q\) des
\W s, wie folgt, ergibt:

\begin{align*}
q = p_1 + p_2 \rightarrow q^2 = s \,.
\end{align*}
Die verwendeten Feynmanregeln wurden aus Ref. \cite{Nachtmann} entnommen.
Der Ausdruck von \ampli \, ergibt sich durch die Übersetzung des
Feynmangraphen in die Feynmanregeln:
\small
\begin{align*}
\ampli = &\left[\overline{u}_{\text{t}}(p_4)\*\left(-\imag {e\*V_{\text{tb}}\over\sqrt{2}\*\sin \theta_W}\*
    \gamma^\mu\*{\mathbbold{1}-\gamma_5\over2}\right)\*\varv_{\overline{\text{b}}}(p_3)\right]\* \\
    &\left[\imag{-g_{\mu\nu} + {q_\mu\*q_\nu\over m_{\text{W}}^2}\over q^2 -m_{\text{W}}^2 + \imag\*\epsilon}\right]\* \\
    &\left[\overline{\varv}_{\overline{\text{d}}}(p_2)\*\left(-\imag {e\*V_{\text{ud}}^*\over\sqrt{2}\*\sin \theta_W}\*
     \gamma^\nu\*{\mathbbold{1}-\gamma_5\over2}\right)\*u_{\text{u}}(p_1)\right]\,.
\end{align*}
\normalsize
Dieser Ausdruck zusammengefasst ist dann:
\small
\begin{align*}
\ampli = -\imag {e^2\*V_{\text{tb}}\*V_{\text{ud}}^*\over 2\*\sin^2 \theta_W\*(s - m_{\text{W}}^2)}\*
     \left[\overline{u}_{\text{t}}(p_4)\*\gamma^\mu\*{\mathbbold{1}-\gamma_5\over2}\*\varv_{\overline{\text{b}}}(p_3)\right]\*
     \left[-g_{\mu\nu} + {q_\mu\*q_\nu\over m_{\text{W}}^2}\right]\*
     \left[\overline{\varv}_{\overline{\text{d}}}(p_2)\*\gamma^\nu\*{\mathbbold{1}-\gamma_5\over2}\*u_{\text{u}}(p_1)\right]\,.
\end{align*}
\normalsize
Der Anteil \(\tfrac{q_\mu\*q_\nu}{m_{\text{W}}^2}\) des \W-Propagators 
trägt in der Rechnung der führenden Ordnung nichts bei. 

Die hier verwendeten Feynmanregeln sind in der unitären Eichung \cite{Nachtmann} angegeben.
In der unitären Eichung läuft der Eichparameter gegen unendlich \(\xi\rightarrow \infty\).
Als Folge daraus sind einzelne mögliche Diagramme höherer Ordnung nicht mehr offenkundig renormierbar,
in der Summe sind diese Diagramme wieder in einer renormierbaren Theorie eingebettet.
Eine weitere Folge ist, dass aufgrund der Wahl des Eichparameters unphysikalische,  
aufgrund der Eichung zusätzlich auftretende Felder in der unitären Eichung nicht vorhanden sind.
Man nennt die unitäre Eichung aus diesem Grund auch physikalische Eichung \cite{Nachtmann}.

Für endliche Parameter von \(\xi\) führt man den Begriff der \(R_{\xi}\)-Eichung ein.
Das \(R\) steht für "`renormierbar"', für endliche Eichparameter sind alle Diagramme
offenkundig renormierbar \cite{ryder}.
Für spezielle, endliche Eichparameter führt man noch andere Namen für die \(R_{\xi}\)-Eichung ein:
Für \(\xi = 1\) nennt man die Eichung Feynman-Eichung (oder auch Feynman-ähnliche Eichung)
in Analogie zur abelschen Eichtheorie. 
Wurde für \(\xi = 0\) gewählt, nennt man diese Eichung Landau-Eichung.

In dieser Arbeit wurde die Feynman-ähnliche Eichung gewählt.
Für den Propagator des \W s trägt dann nur der \(g_{\mu\nu}\)-Term 
bei \cite{Denner}.  
Die Rechnung wird übersichtlicher und technisch einfacher,
gerade in höheren Ordnungen der Rechnung.
Eine andere Eichung bedeutet andere Feynmanregeln, das heißt es können im Allgemeinen
weitere Beiträge und zusätzliche Wechselwirkungen auftreten.
Im Fall der Feynman-ähnliche Eichung treten zusätzliche Wechselwirkungen mit 
skalaren Teilchen auf.
Diese Teilchen haben eine Kopplung an Quarks proportional
zur Masse der gekoppelten Quarks. 
In dieser Rechnung werden nur die \topq-Masse und die \W-Masse berücksichtigt. Die 
Quarkmassen der anderen beteiligten Quarks werden zu Null gesetzt, ihre Massen sind im Vergleich zur
\topq- und der \W-Masse vernachlässigbar 
(\(m_{\text{t}} \sim \unit{172}{GeV} > m_{\text{u}} \sim \unit{2\cdot 10^{-5}}{GeV}\)).
Aufgrund der verschwindenden Wechselwirkung der zusätzlichen Teilchen mit masselosen 
Quarks, liefert die \topq-Erzeugung mit den skalaren, unphysikalischen Teilchen keinen
Beitrag.

Zur Berechnung des Betragsquadrats der Amplitude wird die Amplitude komplex konjugiert
und mit sich selbst multipliziert. Da die Amplitude lediglich eine komplexe Zahl
darstellt, ist es gleichgültig, ob man das komplex konjugierte oder 
adjungierte\footnote{Adjungieren entspricht komplex konjugieren und 
transponieren.} der Amplitude bildet. Letzteres hat den Vorteil, dass man beim
Bilden des Betragsquadrats nach Summation über die die Spins der Teilchen
die Spur über die Fermionlinien erhält.
Die adjungierte Amplitude \(\ampli^\dagger\) ist: 
\small
\begin{align*}
\ampli^\dagger = +\imag {e^2\*V_{\text{tb}}^*\*V_{\text{ud}}\over 2\*\sin^2 \theta_W\*(s - m_{\text{W}}^2)}\*
            \left[\overline{u}_{\text{u}}(p_1)\*{\mathbbold{1}+\gamma_5\over2}\*\gamma^\sigma\*\varv_{\overline{\text{d}}}(p_2)\right]\*
            \left[-g_{\sigma\rho} \right]\*
            \left[\overline{\varv}_{\overline{\text{b}}}(p_3)\*{\mathbbold{1}+\gamma_5\over2}\*\gamma^\rho\*u_{\text{t}}(p_4)\right]\,.
\end{align*}
\normalsize
Da im Experiment mit unpolarisierten Anfangszuständen gearbeitet wird und
die Polarisation des Endzustandes nicht festzustellen ist, wird über 
alle Spinzustände summiert.
Der Wirkungsquerschnitt, der im Experiment gemessen wird,  
stellt daher nur ein Ergebnis summiert über alle Spinzustände dar. 
In dieser Rechnung gibt es vier mögliche Spinzustände der einlaufenden Teilchen, daher der
Vorfaktor \({1\over 4}\).
Implizit steht also an jedem Spinor ein Spinindex, über den summiert werden muss.

Das spingemittelte Amplitudenquadrat im \textit{s}-Kanal ist unter Berücksichtigung
der eben beschriebenen Aspekte wie folgt:
\small
\begin{align*}
 {1\over4} \sum_{\text{\tiny Spins\small}} \ampli\*\ampli^\dagger = &
 {e^4\*V_{\text{tb}}^2\*V_{\text{ud}}^2\over 4^2\*4^2\sin^4 \theta_W\*
 (s - m_{\text{W}}^2)^2}\times \\
 & \sum_{\text{\tiny Spins\small}}
 \left[\overline{u}_{\text{t}}(p_4)\*\gamma^\mu\*(\mathbbold{1}-\gamma_5)
  \*\varv_{\overline{\text{b}}}(p_3)\right]\* 
 g_{\mu\nu} \* 
 \left[\overline{\varv}_{\overline{\text{d}}}(p_2)\*\gamma^\nu
 \*(\mathbbold{1}-\gamma_5)\*u_{\text{u}}(p_1)\right]\times \\
 & \hphantom{\sum_{\text{\tiny Spins\small}}} \,
 \left[\overline{u}_{\text{u}}(p_1)\*(\mathbbold{1}+\gamma_5)
 \*\gamma^\sigma\*\varv_{\overline{\text{d}}}(p_2)\right]\* 
 g_{\sigma\rho}\* 
 \left[\overline{\varv}_{\overline{\text{b}}}(p_3)\*(\mathbbold{1}+\gamma_5)
 \*\gamma^\rho\*u_{\text{t}}(p_4)\right]\,. 
\end{align*}
\normalsize
Nach der Spinmittelung ergibt sich der Term zu
\small
\begin{align*}
 {1\over4} \sum_{\text{\tiny Spins\small}} \ampli\*\ampli^\dagger = &
    {e^4\*V_{\text{tb}}^2\*V_{\text{ud}}^2\over 256\sin^4 \theta_W\*(s - m_{\text{W}}^2)^2}\* \\
     & \text{Tr}\left[(\slashed{p}_4 + \mathbbold{1}\*m_{\text{t}})\*\gamma^\mu\*(\mathbbold{1}-\gamma_5)\*
        \slashed{p}_3\*(\mathbbold{1}+\gamma_5)\*\gamma^\sigma\right]
     \text{Tr}\left[\slashed{p}_2\*\gamma_\mu\*(\mathbbold{1}-\gamma_5)\*
     \slashed{p}_1\*(\mathbbold{1}+\gamma_5)\*\gamma_\sigma\right]\,,
\end{align*}
\normalsize
nach der Spurbildung zu,
\small
\begin{align*}
 {1\over4} \sum_{\text{\tiny Spins\small}} \ampli\*\ampli^\dagger = 
   {e^4\*V_{\text{tb}}^2\*V_{\text{ud}}^2\over 256\sin^4 \theta_W\*
   (s - m_{\text{W}}^2)^2}\*64\*t\*(t - m_{\text{t}}^2)\,,
\end{align*}
\normalsize
und mit \(e^2 = 4\*\pi\*\alpha\) wird daraus,
\small
\begin{align}
 {1\over4} \sum_{\text{\tiny Spins\small}} \ampli\*\ampli^\dagger = 
   {4\*\pi^2\*\alpha^2\*V_{\text{tb}}^2\*V_{\text{ud}}^2\over \sin^4 \theta_W}\*
   {t\*(t - m_{\text{t}}^2)\over (s - m_{\text{W}}^2)^2}, \text{ mit } t = (p_1 - p_3)^2\,.
\label{eq:borns}
\end{align}
\normalsize
Dieses Ergebnis stimmt mit dem Ergebnis aus Ref. \cite{Kidonakis} überein.
Entsprechend wurden auch die Betragsquadrate des \textit{t}-Kanals 
(mit Quarks und Antiquarks) bestimmt.
Auch hier stimmen die Ergebnisse mit denen aus Ref. \cite{Kidonakis} überein.

Für den \textit{t}-Kanal (\(\text{ub} \rightarrow \text{dt}\)) ergibt sich
(wie man auch leicht durch Kreuzen der kinematischen Variablen 
(\(s \leftrightarrow t\)) erhalten kann):
\begin{align}
{1\over4} \sum_{\text{\tiny Spins\small}} \vert \mathcal{T}_{\text{ub} \rightarrow \text{dt}} \vert ^2 = 
   {4\*\pi^2\*\alpha^2\*\vert V_{\text{tb}}\vert^2\*\vert V_{\text{ud}}\vert ^2\over \sin^4 \theta_W}\*
   {s\*(s- m_{\text{t}}^2) \over (t -m_{\text{W}}^2)^2}\,.
   \label{4DimBorn}
\end{align}
Für den \textit{t}-Kanal mit Antiquarks 
(\(\bar{\text{d}}\text{b} \rightarrow \bar{\text{u}}\text{t}\)) lautet das 
Amplitudenquadrat in führender Ordnung:
\begin{align}
{1\over4} \sum_{\text{\tiny Spins\small}} \vert \mathcal{T}_{\bar{\text{d}}\text{b} \rightarrow \bar{\text{u}}\text{t}} \vert ^2 = 
   {4\*\pi^2\*\alpha^2\*\vert V_{\text{tb}}\vert^2\*\vert V_{\text{ud}}\vert^2\over \sin^4 \theta_W}\*
   {(s +t)^2 -(s + t)\*m_{\text{t}}^2 \over (t -m_{\text{W}}^2)^2}\,.
\label{eq:bornat}   
\end{align}
Wie man im Vergleich von Gl. \eqref{eq:borns} mit Gl. \eqref{4DimBorn} erkennen kann,
kann man den \textit{t}-Kanal aus dem \textit{s}-Kanal erhalten, indem man die 
Mandelstam-Variablen \textit{s} und \textit{t} miteinander vertauscht.
Etwas Vergleichbares gilt für den \textit{t}-Kanal mit und ohne Antiquarks.
Man erhält den \textit{t}-Kanal mit Antiquarks aus dem \textit{t}-Kanal ohne Antiquarks
durch das Vertauschen zweier Impulse (\(p_1 \leftrightarrow -p_3\)).
Man braucht also bei diesen drei möglichen Produktionen nur einen
Kanal berechnen und erhält durch Vertauschen der Kinematik die fehlenden
zwei Kanäle ("`Crossing"' Symmetrie).

Für die Produktion einzelner \topq s muss also noch
die assoziierte tW-Produktion berechnet werden. Dieser Produktionskanal
ist strukturell anders als der \textit{s}- und der \textit{t}-Kanal, da er ein Gluon im
Anfangszustand und ein \W\, im Endzustand aufweist. 

\section[Berechnung der tW-Produktion in führender Ordnung]
        {Berechnung der tW-Produktion in führender Ordnung}
\label{Kap:tWborn}

Bei der Produktion des einzelnen \topq s im tW-Kanal
liegen zwei mögliche Varianten vor
(siehe Abb. \ref{twproduktion}). Daraus folgen zwei
Diagramme (\ampli\(^{(\text{a})}\), \ampli\(^{(\text{b})}\)), 
die miteinander interferieren (vgl. Abb. \ref{tWB} und \ref{tWT}).

Neben dem Quark im Anfangszustand liegt auch ein Gluon vor.
Entsprechend der Feynmanregeln treten
dadurch die Generatoren \(T^a\) der \(\text{SU}(N)_{\text{Farbe}}\)
(\(N = 3\))\footnote{Die Rechnung der farbabhängigen Anteile wird
allgemein für \(\text{SU}(N)\) vorgenommen, um nach ähnlichen Farbstrukturen
zu sortieren. Am Ende wird \(N\) immer auf drei gesetzt.} auf. 
Diesen farbabhängigen Teil kann man vom restlichen Beitrag 
separieren und getrennt berechnen. 
Die Berechnung der Farbe wird hier aus Gründen zur Erhaltung der
Allgemeinheit und zum Erkennen von verschiedenen Farbstrukturen in 
der Abhängigkeit von \(N\) gehalten.
Man erhält dann im Amplitudenquadrat die Spur für den
farbabhängigen Anteil\footnote{Auch bei Wechselwirkungen nur mit Quarks 
(ohne Gluonen) ist eine Farbabhängigkeit vorhanden, hier z.B. im \textit{s}-Kanal.
Diese Abhängigkeit führt im Amplitudenquadrat auf einen konstanten Vorfaktor von eins.}:

\begin{align*}
 \delta_{ab}\* \text{Tr}[T^a\*T^b] &= \delta_{ab}\*{1\over 2}\*\delta^{ab}      
      = {1\over 2}\* (N^2 -1)\,.
\end{align*}
Zusätzlich müssen noch die möglichen Farben für das einlaufende Quark und Gluon
betrachtet werden, das heißt, dass über diese Farben gemittelt wird. 
Da jeweils jede Farbe gleich berechtigt ist, erhält 
man (im Amplitudenquadrat) jeweils den Kehrwert der 
Anzahl der möglichen Farben als zusätzlichen Faktor
(für die Gluonfarbmittelung \({1\over N^2 - 1}\), für die Quarkfarbmittelung \({1\over N}\)).
Insgesamt ergibt sich dann für den farbabhängigen Faktor im Betragsquadrat:
\begin{align}
\text{Farbe} = {1\over N}\*{1\over N^2 - 1}\*\,\delta_{ab}\* \text{Tr}(T^a\*T^b) &= 
{1\over N}\*{1\over N^2 - 1}\*\,{1\over 2}\* (N^2 -1)
       = {1 \over 2\*N}\,.
\label{eq:FarbetW}       
\end{align}

Übersetzt man die Feynmangraphen (Abb. \ref{twproduktion}) wieder 
entsprechend der Feynmanregeln (ganz analog zur Berechnung des 
\textit{s}-Kanals im vorherigen Abschnitt),
ergeben sich die beiden möglichen tW-Produktionsamplituden zu: 
\begin{align*}
 \text{\ampli}^{(\text{a})} &= - {\text{i}\*e\*g\*V_{\text{tb}} \over 2\*\sqrt{2}\*\sin \theta_W} \*
          \overline{u}_{\text{t}}(p_4)\*\gamma^\mu\*(1 - \gamma_5)\*\varepsilon^*_\mu(p_3)\*
          {\slashed{k_1} \over s}\*
          \gamma_\nu\*T^a\*u_{\text{b}}(p_2)\*\varepsilon_{\text{g}}^{a\nu}(p_1),\\
 \text{\ampli}^{(\text{b})} &= - {\text{i}\*e\*g\*V_{\text{tb}} \over 2\*\sqrt{2}\*\sin \theta_W} \*
          \overline{u}_{\text{t}}(p_4)\*\gamma^\nu\*\varepsilon^a_\nu(p_1)\*
          {\slashed{k_2} + m_{\text{t}} \over k_2^2 - m_{\text{t}}^2}\*
          \gamma_\mu\*T^a\*(1 - \gamma_5)\*u_{\text{b}}(p_2)\*\varepsilon_{\text{W}}^{*\mu}(p_3)\,.
\end{align*}
Das Amplitudenquadrat der gesamten tW-Produktion ergibt sich dann formal zu:
\begin{align*}
{1\over 4}\*\sum_{\text{Spins}} \vert \mathcal{T}_{\text{bg} \rightarrow \text{tW}} \vert ^2 &= 
{1\over 4}\*\sum_{\text{Spins}} \vert \text{\ampli}^{(\text{a})} + \text{\ampli}^{(\text{b})} \vert ^2 \\
&= {1\over 4}\*\sum_{\text{Spins}} \left(\text{\ampli}^{(\text{a})}\*
    \text{\ampli}^{(\text{a})\dagger} +
    \text{\ampli}^{(\text{b})}\*\text{\ampli}^{(\text{b})\dagger} +
    \text{\ampli}^{(\text{a})}\*\text{\ampli}^{(\text{b})\dagger} + 
    \text{\ampli}^{(\text{b})}\*\text{\ampli}^{(\text{a})\dagger} \right)\\
&= {1\over 4}\*\sum_{\text{Spins}} \left(\text{\ampli}^{(\text{a})}\*\text{\ampli}^{(\text{a})\dagger}
    + \text{\ampli}^{(\text{b})}\*\text{\ampli}^{(\text{b})\dagger} +
    2\*\text{Re}(\text{\ampli}^{(\text{a})}\*\text{\ampli}^{(\text{b})\dagger})\right) \,.
\end{align*}

Bei der Berechnung des vollen Amplitudenquadrats des tW-Kanals in führender Ordnung
ist zu beachten, dass analog zur Spinmittelung der Fermionen im Anfangszustand
die möglichen Polarisationen des Gluons im Anfangszustand 
und die möglichen Polarisationen des \W s im Endzustand summiert werden müssen.
Die Polarisationssumme für das Gluon ist nach Ref. \cite{WeinzierlPolSumme}: 
\begin{align}
\sum_{s = 1}^2  \varepsilon^{*}_\kappa(p,s)\,\varepsilon_\lambda(p,s)  = 
-g_{\kappa \lambda} + {p_\kappa n_\lambda + n_\kappa p_\lambda \over p\cdot n} - 
n^2 {p_\kappa p_\lambda \over (p\cdot n)^2}\,,
\end{align}
\(n_\lambda\) ist ein beliebiger Vierervektor, die Abhängigkeit von \(n_\lambda\)
verschwindet in eichinvarianten Ausdrücken. Aus diesem Grund werden die Terme, die von
\(n_\lambda\) abhängig sind, nicht weiter berücksichtigt.
Für die Polarisationssumme des \W s ergibt sich nach Ref. \cite{Nachtmann}:

\begin{align}
\sum_{s = 1}^3  \varepsilon^{*}_\kappa(p,s)\,\varepsilon_\lambda(p,s) = 
-g_{\kappa \lambda} + {p_\kappa p_\lambda \over m_{\text{W}}^2}\,.
\end{align}

Das Amplitudenquadrat (\(\text{bg} \rightarrow \text{tW}\)) ist
unter Berücksichtigung der genannten Aspekte:
\begin{align}
{1\over 4}\*\sum_{\text{Spins}} \vert \mathcal{T}_{\text{bg} \rightarrow \text{tW}} \vert ^2 = 
   {4\*\pi^2\*\alpha_s\*\alpha\*\vert V_{\text{tb}}\vert^2\*\over 3\* m_{\text{W}}^2\*\sin^2 \theta_W}\*
      \left({C_1\over u_1^2} - {2\*C_2 \over u_1\*s} +
     {2\*C_3 \over s^2}\right)\,,
\end{align}
wobei die Koeffizienten \(C_1,\,C_2,\,C_3\) und \(t_1,\,u_1,\,u_2\), wie folgt, definiert sind:
\begin{align}
 C_1 &= - u_2\*(s-m_{\text{t}}^2-m_{\text{W}}^2 - 4\*u_2\*m_{\text{t}}^2)\*
           \left(m_{\text{W}}^2+{m_{\text{t}}^2\over 2}\right) - {t_1 \over 2}\*
         (-2\*m_{\text{W}}^4+m_{\text{W}}^2\*m_{\text{t}}^2 + m_{\text{t}}^4) , \nonumber\\
 C_2 &= - t_1\*m_{\text{W}}^2(m_{\text{t}}^2-m_{\text{W}}^2) - {m_{\text{t}}^2\over 2}\*(t_1\*u_2+u_1\*u_2)
          - s\*m_{\text{t}}^2\left(m_{\text{W}}^2 + {m_{\text{t}}^2 \over 2}\right),\nonumber\\
 C_3 &= - {1\over 2}u_1\*s\*\left(m_{\text{W}}^2 + {m_{\text{t}}^2 \over 2}\right),\nonumber\\
 t_1 &= - m_{\text{t}}^2 + m_{\text{W}}^2 - s - u_1,\nonumber\\
 u_1 &= u - m_{\text{t}}^2,\nonumber\\
 u_2 & = u - m_{\text{W}}^2 \,\text{ mit }\, u = (p_1 - p_4)^2 \,\text{ und }\, t = (p_1 - p_3)^2\,.
\label{tWkoeff} 
\end{align}
Dieses Ergebnis für die führende Ordnung des tW-Kanals
stimmt mit dem aus Ref. \cite{Kidonakis} überein.

Die Amplitudenquadrate aller Produktionskanäle des 
einzelnen \topq s sind in führender Ordnung bestimmt.
Es ist nun möglich partonische und hadronische
Wirkungsquerschnitte mit diesen Ergebnissen 
zu berechnen und daraus Vorhersagen für 
physikalische Observablen zu erhalten.

	\section{Wirkungsquerschnitte in führender und \nlo r Ordnung}

\setlength{\unitlength}{0.45mm}
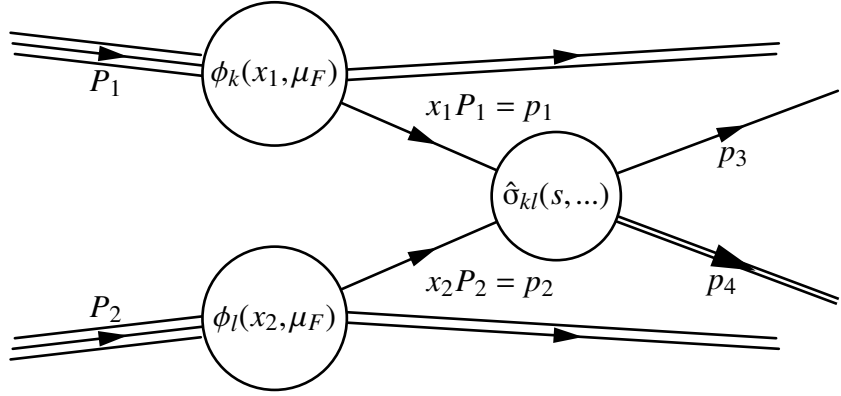
\begin{figure}[h]
	\centering
	\vspace*{5mm}
	\begin{fmffile}{factorisation}
		\begin{fmfgraph*}(280,90)
			\fmfleft{P2,P1}
			\fmfright{P2',p6,p5,p4,p3,P1'}

			\fmf{fermion,label=\(P_1\),tension=3}{P1,g1}
			\fmf{fermion,label=\(P_2\),tension=3}{P2,g2}

			\fmfv{decor.shape=circle,decor.filled=empty,decor.size=19mm,
			      label=\(\phi_{k}(x_{1},,\mu_F)\),l.angle=90, l.dist=0}{g1}
			\fmfv{decor.shape=circle,decor.filled=empty,decor.size=19mm,
			      label=\(\phi_{l}(x_{2},,\mu_F)\),l.angle=90, l.dist=0}{g2}
			\fmfv{decor.shape=circle,decor.filled=empty,decor.size=17mm,
			      label=\(\hat{\sigmaup}_{kl}(s,,...)\),l.angle=90, l.dist=0}{hs}
			      
			\fmf{fermion,label=\(x_{1}P_{1} = p_1\),l.side=left}{g1,hs}
			\fmf{fermion,label=\(x_{2}P_{2} = p_2\),l.side=right}{g2,hs}
			\fmf{fermion,label=\(p_3\)}{hs,p3}
			\fmf{heavy,label=\(p_4\)}{hs,p6}

			\fmf{fermion}{g1,P1'}
			\fmf{fermion}{g2,P2'}

			\fmffreeze

			\renewcommand{\P}[3]{\fmfi{plain}{vpath(__#1,__#2) shifted (thick*(#3))}}

			\P{P1}{g1}{0.4,-2}
			\P{P1}{g1}{-0.4,2}
			\P{P2}{g2}{-0.4,-2}
			\P{P2}{g2}{0.4,2}

			\P{g1}{P1'}{-0.5,-2}
			\P{g2}{P2'}{-0.5,2}
		\end{fmfgraph*}
	\end{fmffile}
	\vspace{5mm}
	\caption[Faktorisierungstheorem]
	        {Das Faktorisierungstheorem veranschaulicht:
	        Die kollidierenden Hadronen tragen die Impulse \(P_1\) und \(P_2\).
	        Je Hadron trägt je ein Parton einen gewissen Anteil \(x_{1,2}\) vom
	        Hadronimpuls \(P_{1,2}\). Die betrachteten Partonen wechselwirken
	        miteinander, was durch den partonischen Wirkungsquerschnitt 
	        \(\hat{\sigmaup}_{kl}\) beschrieben wird.
	        Die PDFs \(\phi_{k,l}\) beschreiben den Partoninhalt der
	        betrachteten Hadronen bei der Energieskala \(\mu_F\).}
	\label{factorisation}
\end{figure}
Am LHC  finden Proton-Proton-Kollisionen
bei hohen Schwerpunktenergien (\(\unit{7}{TeV}\) und \(\unit{8}{TeV}\)) statt. 
In der um das Partonmodell verbesserten QCD sind die 
Bestandteile der Hadronen (am LHC: Protonen) als
quasi-frei zu betrachten. Die Ursache liegt in der Natur der
starken Wechselwirkung: Bei hohen Energien wird die
Kopplungskonstante der starken Wechselwirkung \(\alpha_s\)
sehr klein.
Der Impuls des betrachteten Hadrons teilt sich
auf die Konstituenten des Hadrons auf.
Für Vorhersagen im Standardmodell ist die Information über die
Wahrscheinlichkeitsverteilung der Partonen in einem Hadron wichtig.
Dies wird durch die Partonverteilungsfunktion beschrieben
(im Weiteren PDF genannt, aus dem Englischen für "`Parton Distribution Function"').
Vereinfacht gesagt, stellt die PDF eine Wahrscheinlichkeitsdichte
dar. Diese beschreibt die Wahrscheinlichkeit ein Parton im Hadron mit
gewissem Impulsanteil (\([x,x+dx]\)) des Hadrons zu messen.
Die PDFs sind nicht perturbativ berechenbar und werden daher experimentell bestimmt.
Aus diesem Grund existieren verschiedene PDFs von unterschiedlichen Gruppen.
Unterschiede sind in den verwendeten Daten zur Bestimmung zu suchen sowie in den
unterschiedlichen Annahmen beim Fitten der PDFs an die Daten.
Eine Liste von PDF-Sets kann zum Beispiel Ref. \cite{lhapdf} entnommen werden.

Der hadronische Wirkungsquerschnitt ergibt sich aus der Faltung mit dem
partonischen Wirkungsquerschnitt \(\hat{\sigmaup}\) und
der Partonverteilungsfunktion \(\phi_{(k,l)}(x_j,\mu_F)\) mit
(\(j = 1,2\) und \(k,l = u,d,c,s,b,f\))\footnote{\(k,l\) beinhalten
die verschiedenen Quarkflavours.
\(j\) kennzeichnet das betrachtete Hadron.} \cite{Campbell:2006wx,Collins:1989gx}: 
\begin{align}
  \sigmaup_{\text{Had}} = \sum_{k,l} \int \mathrm{d}x_1 \mathrm{d}x_2 \phi_{k}(x_1,\mu_F) \phi_{l}(x_2,\mu_F) 
   \times \hat{\sigmaup}_{kl}(x_1\*x_2\*s_{\text{Had}},\mu_F,\alpha_s,m_{\text{t}},m_{\text{W}})\,.
\label{eq:hadwq}
\end{align}
Die Größen im hadronischen Wirkungsquerschnitt in der Gl. \eqref{eq:hadwq} 
sind die Massen des \topq s \(m_{\text{t}}\) und des \W s \(m_{\text{W}}\), 
die hadronische Schwerpunktenergie \(s_{\text{Had}}\) (\(s = x_1 x_2 s_{\text{Had}}\)),
die Faktorisierungsskala\index{Faktorisierungsskala} \(\mu_F\) und
\(\alpha_s\), die Kopplungskonstante der starken Wechselwirkung.
Die Variablen \(x_1\) und \(x_2\), über die integriert wird,
sind der Impulsanteil der Partonen (1 und 2) am hadronischen Impuls des Protons.
Eine Veranschaulichung der Gl. \eqref{eq:hadwq} ist in Abb. \ref{factorisation}
dargestellt.

Die Faktorisierungsskala gibt die Energieskala an, bei der
die Partonverteilungsfunktion ausgewertet wird.
Würde die Störungsreihe zu allen Ordnungen berücksichtigt werden, hinge
der physikalische Wirkungsquerschnitt nicht von
der Faktorisierungsskala ab. Alle Rechnungen brechen jedoch an
einem gewählten Punkt der Störungsreihe ab, so dass der bestimmte
Wirkungsquerschnitt von der Wahl der Faktorisierungsskala residual abhängt,
das heißt die Abhängigkeit ist formal von höherer Ordnung in der Störungstheorie.
Die Faktorisierungsskala wird gewöhnlich auf die charakteristische Impulsskala
eines Prozesses gesetzt, zum Beispiel im Fall der einzelnen \topq-Produktion
auf die Masse des \topq s.

\subsubsection{Bestimmung des partonischen Wirkungsquerschnitts}
\label{Kap:wq}

Um den hadronischen Wirkungsquerschnitt zu erhalten, muss zunächst der
partonische Wirkungsquerschnitt \(\hat{\sigmaup}\) (Gl. \eqref{eq:PartWQ})
bestimmt werden. Nach Ref. \cite{peskin} ergibt sich der
partonische Wirkungsquerschnitt für einen Zwei-Teilchen-Endzustand 
für Reaktionen symmetrisch um den Kollisionspunkt zu:

\begin{align} 
 \hat{\sigmaup} = {1\over 2\*s\* 16\*\pi} \int_{-1}^{1} {\mathrm{d}\cos \theta} 
                 \,  \sum_{\text{\tiny{Spins}}} \vert \ampli \vert^2\, ,
                 \label{eq:patwq}
\end{align}
wobei \(s\) die quadrierte Schwerpunktenergie der (partonischen) Reaktion darstellt und
der Winkel \(\theta\) der Winkel zur Strahlachse (Flugrichtung des einlaufenden Partons,
Polarwinkel im Schwerpunktsystem) ist. 
In Gl. \eqref{eq:patwq} ist dann nur noch das spingemittelte
Amplitudenquadrat \(\sum_{\text{\tiny{Spins}}} \vert \ampli \vert^2\) 
des betrachteten Prozesses einzusetzen.
Das spingemittelte Amplitudenquadrat der Streuamplitude für die
verschiedenen Produktionskanäle wurde im vorherigen Abschnitten 
\ref{Kap:ampliquborn} und \ref{Kap:tWborn} bestimmt.
Es bleibt noch die Phasenraumintegration auszuführen.
Anhand des folgenden Beispiels für den \textit{t}-Kanal ist erkennbar, welche
typischen Integrale jeweils pro Produktionskanal gelöst werden müssen.
Die \(\cos \theta\)-Abhängigkeit des Integranden steht implizit in der 
Mandelstam-Variablen \textit{t}.
Für den Wirkungsquerschnitt des \textit{t}-Kanals
ohne Antiquarks ergibt sich:
\begin{align*}
\hat{\sigmaup}_t = {\pi\*\alpha^2\*\vert V_{tb}\vert^2\*\vert V_{ud}\vert^2
                 \over 8\*s\* \sin^4 \theta_W }\* 
                (s-mt^2)
                \int_1^{-1} {\mathrm{d}\cos \theta} 
                 { 1\over \left({m_{\text{t}}^2 - s \over 2}\*
                 (1-\cos \theta) -m_{\text{W}}^2\right)^2 }\, .
\end{align*}
Integriert ergibt sich der partonische Wirkungsquerschnitt des \textit{t}-Kanals zu:
\begin{align}
\hat{\sigmaup}_t =  {\pi\*\alpha^2\*\vert V_{tb}\vert^2\*\vert V_{ud}\vert^2
                 \over 4\*\sin^4 \theta_W} 
                {(s-m_{\text{t}}^2)^2 \over s\,\*m_{\text{W}}^2\*(s+m_{\text{W}}^2-m_{\text{t}}^2)}\,.
\end{align}
Für die anderen Produktionkanäle wurden auch die partonischen
Wirkungsquerschnitte bestimmt.
Für den \textit{t}-Kanal mit Antiquarks ergibt sich:
\begin{align}
\hat{\sigmaup}_{at}  = \, & {\pi\*\alpha^2\*\vert V_{tb}\vert^2\*\vert V_{ud}\vert^2
                 \over \sin^4 \theta_W}\*{s -m_{\text{t}}^2\over 4\*s^2}
                 {1\over a_{at}\,\*m_{\text{W}}^2\*(2\*a_{at}+m_{\text{W}}^2)}\nonumber\\ 
                & \left(a_{at}\*(2\*m_{\text{W}}^2 (a_{at} + m_{\text{W}}^2 + s) +
                    s^2\*-m_{\text{t}}^2\*(m_{\text{W}}^2+s)) \right.\nonumber\\
                & \left. + m_{\text{W}}^2\*(2\*a_{at}+m_{\text{W}}^2)\,\,
                (m_{\text{t}}^2-2\*(m_{\text{W}}^2+s))
                {1\over 2}\*\log(1+2\*\tfrac{a_{at}}{m_{\text{W}}^2})\right)\,,
\end{align}
mit \(a_{at}  = {1\over 2}\*(s-m_{\text{t}}^2)\).\\
Für den \textit{s}-Kanal ist der partonische Wirkungsquerschnitt:
\begin{align}
\hat{\sigmaup}_s  = \, & {\pi\*\alpha^2\*\vert V_{tb}\vert^2\*\vert V_{ud}\vert^2
                 \over 24 \sin^4 \theta_W}\*{s -m_{\text{t}}^2\over s^2}  
                  {(2\*s^2-s\*m_{\text{t}}^2 - m_{\text{t}}^4) \over (s - m_{\text{W}}^2)^2}\, .
\end{align}
Für den tW-Kanal führt man
folgende Variablen ein:
\begin{align*}
a & = E\*\sqrt{s} \,,\\
b & = \sqrt{(E^2+m_{\text{t}}^2)\*s} \,,\\
c & = \sqrt{(E^2+m_{\text{W}}^2)\*s} \,,\\
E & = {s-m_{\text{t}}^2+m_{\text{W}}^2\over 2\*\sqrt{s}}\,.
\end{align*}
Für die Variablen \(t_1\, , u_1\, , u_2\) sowie
\(x\), die Integrationsvariable zwischen \([0,1]\), ergibt sich dann:
\begin{align*}
 t_1 &= a\*(2\*x -1) - b \,,\\
 u_1 &= a\*(1-2\*x) - c + m_{\text{W}}^2 -m_{\text{t}}^2 \,,\\
 u_2 &= a\*(1-2\*x) -c \,.
\end{align*}
Nach der Phasenraumintegration ergeben sich somit die Koeffizienten
\(C_1\), \(C_2\) und \(C_3\) aus Gl. \eqref{tWkoeff} zu den folgenden
Koeffizienten:
\begin{align*}
C'_1 = & \, {(m_{\text{t}}^2 + 2\*m_{\text{W}}^2)\* (m_{\text{t}}^2 - m_{\text{W}}^2)\*(-b-c+2\*m_{\text{t}}^2+s)
             \over 2\*(a+c+m_{\text{t}}^2-m_{\text{W}}^2)\*(a-c-m_{\text{t}}^2+m_{\text{W}}^2)} \\
          & + {(2\*m_{\text{t}}^2+s)\*(m_{\text{t}}^2 +2\*m_{\text{W}}^2)\over 2\*a}\*
                \log { a+c+m_{\text{t}}^2-m_{\text{W}}^2\over c-a+m_{\text{t}}^2-m_{\text{W}}^2} \, ,\\
C'_2 = &  \, {1\over 2\*s}\*\left[ m_{\text{t}}^2\*(b+c+m_{\text{t}}^2)+
            m_{\text{t}}^2\*m_{\text{W}}^2-2\*m_{\text{W}}^4\right.\\
          & \left. - {1\over 2\*a}\*(m_{\text{t}}^2+2\*m_{\text{W}}^2)\*
            \left((m_{\text{t}}^2-m_{\text{W}}^2)\*
               (b+c+m_{\text{t}}^2-m_{\text{W}}^2) -  m_{\text{t}}^2\*s\right) 
           \*\log{a+c+m_{\text{t}}^2-m_{\text{W}}^2 \over c-a+
            m_{\text{t}}^2-m_{\text{W}}^2}\right] \, ,\\
C'_3 = & \, (c+m_{\text{t}}^2-m_{\text{W}}^2)\*{m_{\text{t}}^2+2\*m_{\text{W}}^2\over 2s} \,.
\end{align*}
Der partonische Wirkungsquerschnitt entspricht dann folgendem Ausdruck:
\begin{align}
 \hat{\sigmaup}_{tW} =  {\pi\*\alpha_s\*\alpha\*\vert V_{tb}\vert^2
                         \over 24\* m_{\text{W}}^2\*\sin^2 \theta_W}
                         {1 \over s^2}\* 
             \sqrt{m_{\text{t}}^4 + m_{\text{W}}^4 - 2\*m_{\text{t}}^2\*m_{\text{W}}^2 
                  - 2\*m_{\text{t}}^2\*s - 2\*m_{\text{W}}^2\*s  + s^2}\,
              \*(C'_1 -2C'_2 +C'_3) \,.
\end{align}
Nach der Bestimmung der partonischen Wirkungsquerschnitte ist es
nun möglich hadronische Wirkungsquerschnitte 
nach Gl. \eqref{eq:hadwq} zu berechnen.
Im Weiteren wurde zusätzlich die Software \textsc{HatHor} \cite{hathor} zur 
Berechnung der hadronischen Wirkungsquerschnitte
verwendet.

\subsubsection{Hadronischer Wirkungsquerschnitt}

\begin{figure}[t!]
\centering
\includegraphics[trim=0.8cm 0.5cm 0.5cm 0.1cm,width=0.8\textwidth,angle=0]{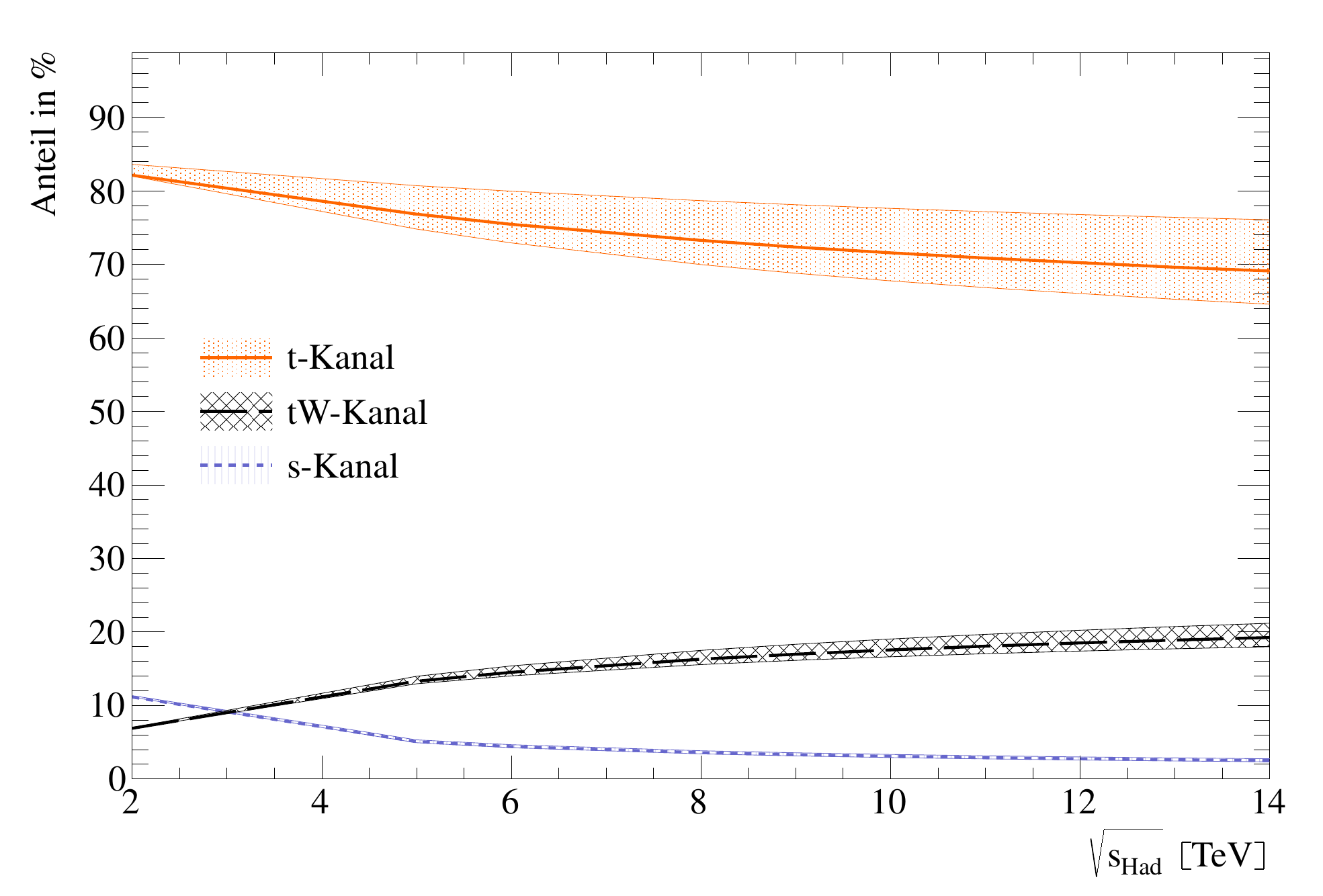}
\vspace{2mm}
\caption[Anteile der verschiedenen Produktionskanäle am gesamten hadronischen Wirkungsquerschnitt]
{Anteile der verschiedenen Produktionskanäle am gesamten hadronischen Wirkungsquerschnitt
 abhängig von der Schwerpunktenergie bei einer angenommenen 
 \topq-Masse von $m_{\text{t}} = \unit{172{,}5}{GeV}$ für einen pp-Beschleuniger (LHC).
 Als PDF-Set wurde MRST2004nnlo genutzt (\(\mu_F = \mu_R = m_{\text{t}}\)).
 Zusätzlich sind die Standardabweichungen der Werte (als Fehlerband) eingezeichnet.}
\label{ratio}
\end{figure}

\begin{figure}[t!]
\centering
\includegraphics[trim=0.8cm 0.5cm 0.5cm 0.1cm,width=0.8\textwidth,angle=0]{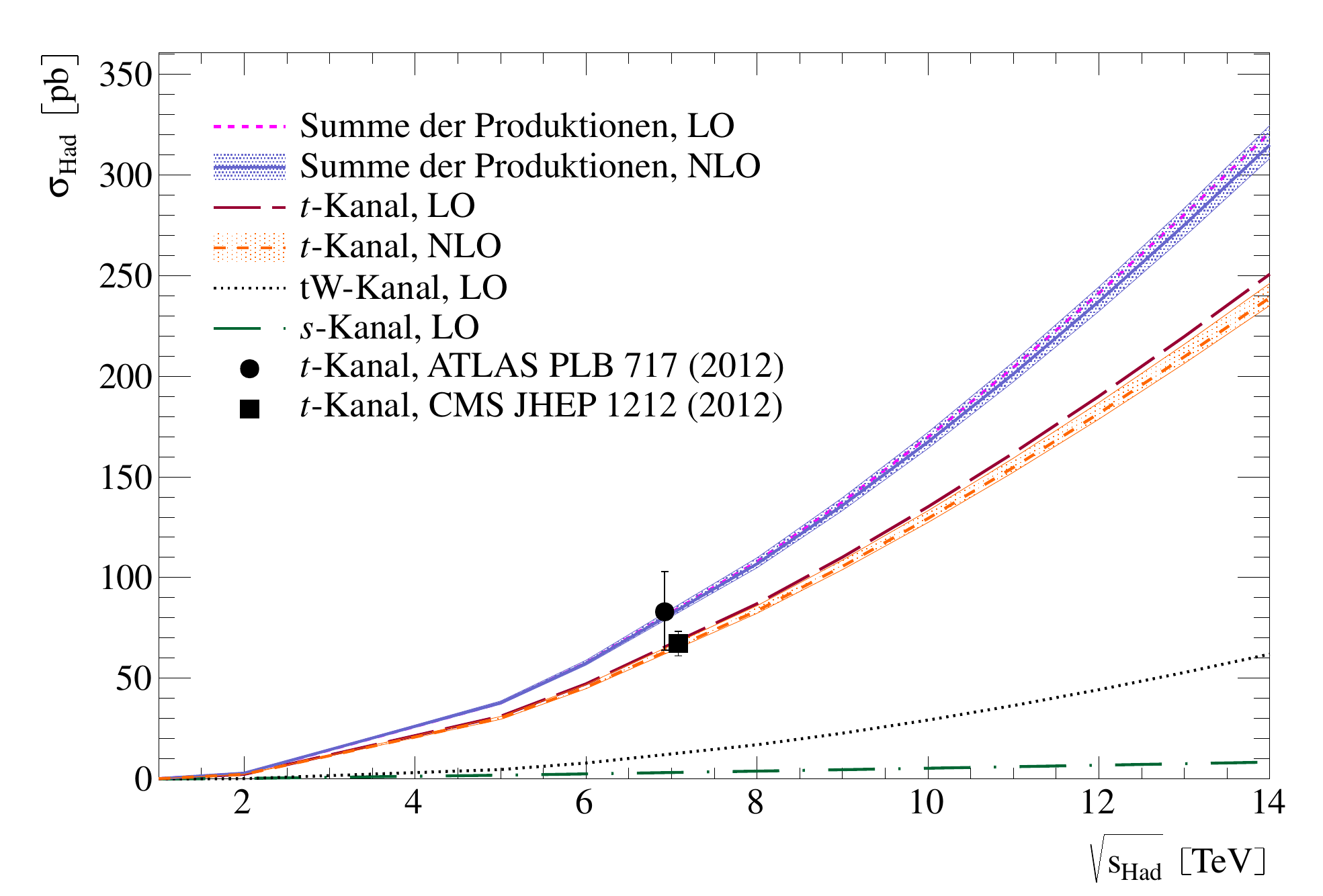}
\vspace{2mm}
\caption[Hadronischer Wirkungsquerschnitt in Abhängigkeit von der Schwerpunktenergie]
        {Hadronischer Wirkungsquerschnitt in Abhängigkeit von der Schwerpunktenergie
         bei einer angenommenen \topq-Masse von $m_{\text{t}} = \unit{172{,}5}{GeV}$ für 
         pp-Beschleuniger (LHC).
         Als PDF-Set wurde MRST2004nnlo genutzt ($\mu_F = \mu_R = m_{\text{t}}$).
         Zusätzlich sind teilweise die Standardabweichungen der Werte 
         (als Fehlerband) eingezeichnet.}
\label{wqcms}
\end{figure}

\nocite{Ellis}
\begin{figure}[t!]
\centering
\includegraphics[trim=1cm 0.5cm 0.8cm 1cm,width=0.8\textwidth,angle=0]{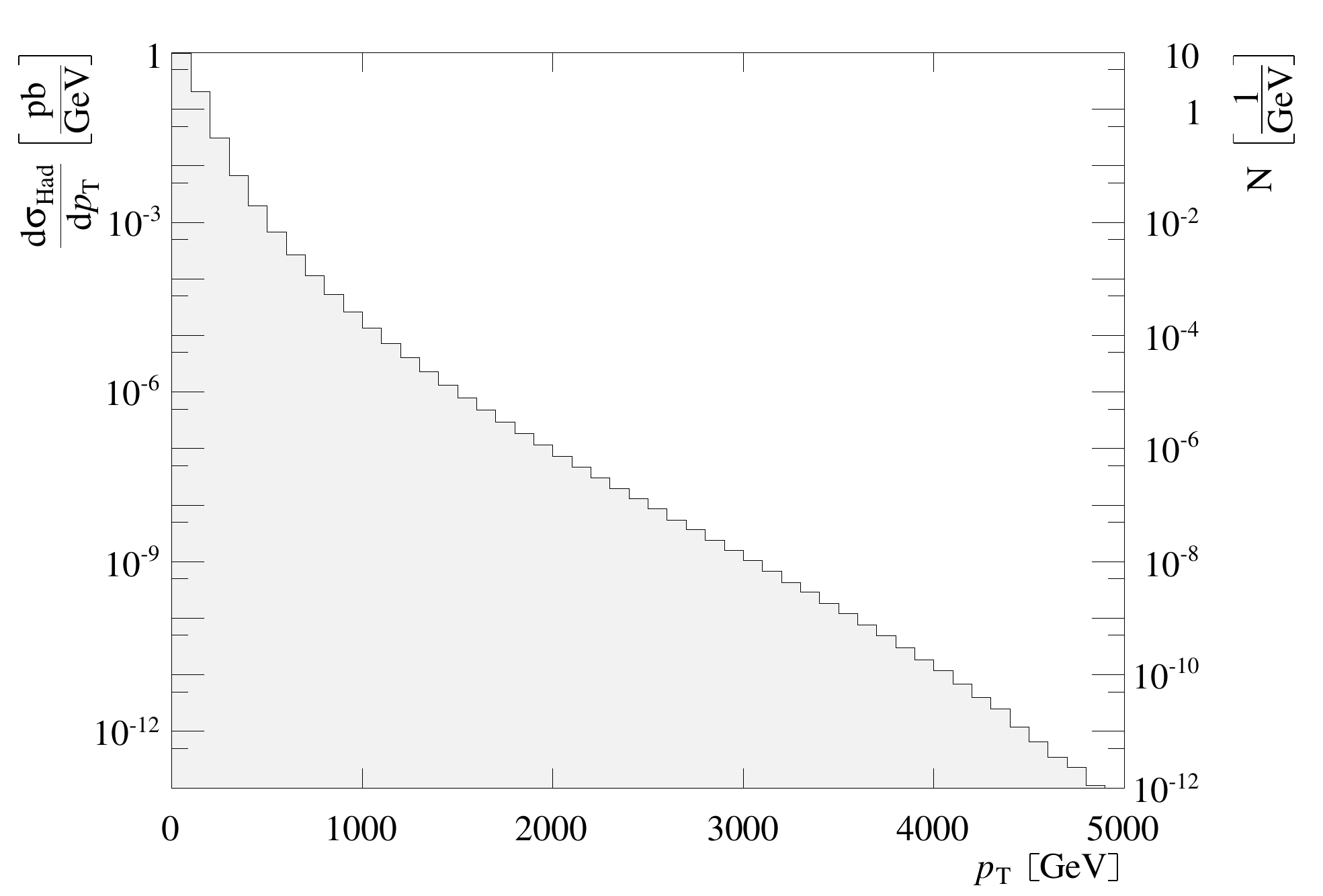}
\vspace{2mm}
\caption[Transversaler Impuls des \topq s im \textit{t}-Kanal (LHC)]
        {Transversaler Impuls des \topq s im \textit{t}-Kanal bei einer
         Schwerpunktenergie von \(\sqrt{s_{\text{\tiny Had}}} = \unit{14}{TeV}\)
         und \(m_{\text{t}} = \unit{172{,}5}{GeV}\)
         für pp-Kollisionen (LHC).
         Auf der linken Ordinate ist der 
         differentielle Wirkungsquerschnitt, auf der rechten die erwartete Anzahl
         von Ereignissen bei einer angenommenen integrierten Luminosität von 
         \(\unit{100}{fb}^{-1}\) 
         dargestellt. Als PDF-Set wurde MRST2004nnlo genutzt (\(\mu_F = \mu_R = m_{\text{t}}\)).}
\label{ptToptch}
\end{figure}
\begin{figure}[t]
\centering
\includegraphics[trim=1cm 1cm 0.8cm 1cm,width=0.8\textwidth,angle=0]{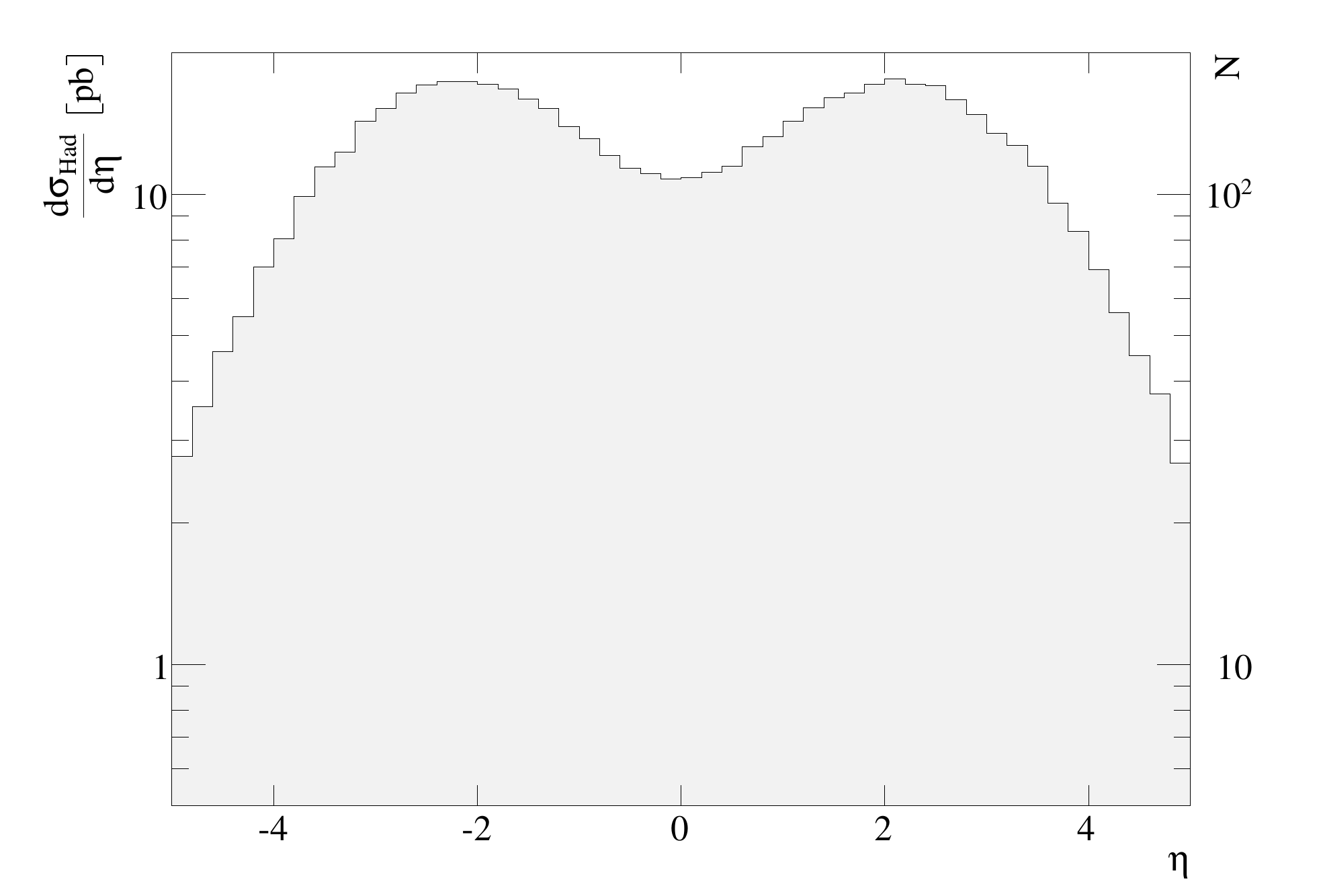} 
\vspace{5mm}
\caption[Pseudorapidität des \topq s im \textit{t}-Kanal (LHC)]
        {Pseudorapidität des \topq s im \textit{t}-Kanal bei einer
         Schwerpunktenergie von \(\sqrt{s_{\text{\tiny Had}}} = \unit{14}{TeV}\)
         und \(m_{\text{t}} = \unit{172{,}5}{GeV}\)
         für pp-Kollisionen (LHC).
         Auf der linken Ordinate ist der 
         differentielle Wirkungsquerschnitt, auf der rechten die erwartete Anzahl
         von Ereignissen bei einer angenommenen integrierten Luminosität von 
         \(\unit{100}{fb}^{-1}\)
         dargestellt. Als PDF-Set wurde MRST2004nnlo genutzt (\(\mu_F = \mu_R = m_{\text{t}}\)).}
\label{rapToptch}
\end{figure}
In den Abb. \ref{ptToptch} und \ref{rapToptch}\footnote{Zur Erzeugung 
der Abb. \ref{ptToptch}, \ref{rapToptch}, \ref{ratio} und \ref{wqcms} wurde 
{\tt ROOT} \cite{ROOT} verwendet.} sind die Verteilungen des transversalen Impulses 
und der Pseudorapidität des \topq s im \textit{t}-Kanal für
pp-Kollisionen dargestellt.
In beiden Abbildungen ist auf der rechten Ordinate die Anzahl der zu erwarteten
Ereignisse bei einer angenommenen integrierten Luminosität von \(\unit{100}{fb}^{-1}\)
aufgetragen (bei \(\sqrt{s_{\text{\tiny Had}}} = \unit{14}{TeV}\)). 
Aktuell hat das ATLAS-Experiment bei einer
Schwerpunktenergie von \(\sqrt{s_{\text{\tiny Had}}} = \unit{8}{TeV}\)
eine integrierte Luminosität  von \(\unit{21{,}70}{fb}^{-1}\) \cite{lumiAtlas}
aufgezeichnet. 
Bei \(\sqrt{s_{\text{\tiny Had}}} = \unit{7}{TeV}\)  liegt die 
integrierte Luminosität bei \(\unit{5{,}25}{fb}^{-1}\) \cite{lumiAtlas}. 

Der Transversalimpuls \(\textit{p}_{\text{T}}\) entspricht dem Betrag der zwei 
Impulskomponenten eines Dreierimpulses, die
senkrecht zur Strahlachse (und somit zur Flugrichtung der 
kollidierenden Hadronen, hier: \(z\)-Richtung) sind:
\[p_{\text{T}} = \sqrt{p_x^2 + p_y^2}.\]
Das erzeugte \topq \, weist wahrscheinlicher kleine Werte für den 
Transversalimpuls auf, hohe Energien sind unwahrscheinlich.
Bei einem Transversalimpuls von rund \(\unit{200}{GeV}\) erwartet man 
nach Abb. \ref{ptToptch} nur noch ein Ereignis bei einer angenommenen
integrierten Luminosität von \(\unit{100}{fb}^{-1}\) 
(bei \(\sqrt{s_{\text{\tiny Had}}} = \unit{14}{TeV}\)). Damit liegen
ungefähr \(97 \%\) der zu erwartenden Ereignisse bei einem 
Transversalimpuls von \(p_{\text{T}} \leq \unit{200}{GeV}\).
\topq s mit höheren Transversalimpulsen sind unter den angegebenen 
experimentellen Rahmenbedingungen nicht messbar.

Die Pseudorapidität, die auch eine Observable im Experiment darstellt,
ist wie folgt definiert (Ref. \cite{Nachtmann}): 
\begin{align}
 \eta  = {1\over 2}\*\log \left({\vert \overrightarrow{p} \vert + p_{\text{l}}\over 
                                  \vert \overrightarrow{p} \vert - p_{\text{l}}}\right)
       = -\log \, \tan {\vartheta \over 2}, \quad
      \text{ mit } \vartheta = \arccos \tfrac{p_{\text{l}}}{\vert \overrightarrow{p}\vert}\,,
\end{align}
dabei entspricht \(\overrightarrow{p}\) dem Dreierimpuls des betrachteten Teilchens 
und \(p_{\text{l}}\)
der longitudinalen Impulskomponente, also der Komponente, die parallel zur Strahlachse
verläuft (\(z\)-Richtung).
Im Wesentlichen gibt diese Größe den Winkel der Flugrichtung des Teilchens
zur Strahlachse an. Für Werte von \(\eta\), die gegen Null gehen, weist das Teilchen
einen Winkel von \({\piup\over 2}\) zur Strahlachse auf. 
Hat \(\eta\) einen hohen Wert, fliegt das Teilchen parallel zur Strahlachse.
Bei kleineren Schwerpunktenergien ist zu erwarten, dass die beiden erkennbaren
"`Peaks"' bei kleineren Beträgen von \(\vert \eta \vert\) zu finden sind. Am LHC bewegen sich
zwei Protonen mit sehr hoher Energie in entgegengesetzter \(z\)-Richtung aufeinander zu.
Aufgrund der hohen kinetischen Energie, ist die Bewegung relativistisch zu beschreiben.
Die Lorentztransformation ("`Boost"') vom Ruhesystem des Teilchen 
in das Laborsystem findet entlang der \(z\)-Richtung statt.
Eine Winkelverteilung in Vorwärts- bzw. Rückwärtsrichtung in Bezug auf die Strahlachse
ist zu erwarten. Je kleiner die Schwerpunktenergie ist, desto kleiner ist der Boost
in \(z\)-Richtung und die Peaks sind bei kleineren \(\vert \eta \vert\)-Werten zu finden.

Aufgrund dessen, dass ein symmetrischer Anfangszustand am LHC vorliegt (Proton-Proton-Streuung),
erwartet man auch für die entstandenen \topq s eine symmetrische Streuung um den Kollisionspunkt.
Abweichungen der Symmetrie in Abb. \ref{rapToptch} sind mit statistischen Schwankungen zu erklären.
Am Tevatron ist die Verteilung (Abb. \ref{TevatronPseudorap}) ähnlich zu der Abb. \ref{rapToptch}:
Man erkennt zwei Peaks, die aufgrund der niedrigeren Schwerpunktenergie 
\(\sqrt{s_{\text{\tiny Had}}} = \unit{1{,}96}{TeV}\) bei kleineren Werten von \(\vert \eta \vert\)
zu finden sind. Jedoch ist die Verteilung asymmetrisch, positive Werte von \(\eta\) treten
häufiger auf. Diese Asymmetrie lässt sich mit dem am Tevatron vorliegenden 
asymmetrischen Anfangszustand erklären, hier wurden Protonen und Antiprotonen zur Kollision
gebracht. Die PDFs für Protonen aus Quarks und Antiquarks sind unterschiedlich: 
In einem Proton aus Quarks liegen Antiquarks nur als Seequarks vor (und umgekehrt für
Protonen aus Antiquarks). Diese  Seequarks weisen einen quantitativ kleineren Verlauf der PDFs
als Valenzquarks auf. Sie sind also mit einer geringeren Wahrscheinlichkeit in dem
betrachteten Hadron zu finden.

In Abb. \ref{ratio} ist der Anteil der verschiedenen Produktionskanäle
am gesamten hadronischen Wirkungsquerschnitt in Abhängigkeit der Schwerpunktenergie
dargestellt.
Der \textit{t}-Kanal hat dabei den größten Anteil, bei höheren Energien wird dieser
Anteil geringer zu Gunsten des tW-Kanals, so dass bei hohen Energien diese
beiden Produktionskanäle relevant sind. Der \textit{s}-Kanal ist im Vergleich 
dazu bei hohen Energien zu vernachlässigen.
Qualitativ verläuft der Wirkungsquerschnitt für einen p\(\bar{\text{p}}\)-Beschleuniger (Tevatron) wie
für einen pp-Beschleuniger (LHC), wie man in den Abbildungen im Anhang \ref{TevWQ} sehen kann.
Quantitativ liegen die Wirkungsquerschnitte für die hohen Schwerpunktenergien, wie sie
etwa bei LHC vorliegen, weitaus über denen, die man bei niedrigeren Schwerpunktenergien
wie beim Tevatron findet. 

Ein quantitativer Unterschied ist die Beobachtung, dass die
Wirkungsquerschnitte für den \textit{s}-Kanal am LHC  unter der tW-Produktion liegen.
Für die Werte am Tevatron ist es genau umgekehrt.
Die Ursache hierfür ist in den Unterschieden der beiden Beschleuniger zu finden. Am Tevatron
werden Proton- und Antiprotonpaare zur Kollision gebracht. Wie für den \textit{s}-Kanal
erforderlich liegt ein Antiquark im Anfangszustand als Valenzquark vor.
Am LHC hingegen werden Protonenpaare zur Kollision gebracht. Das für den \textit{s}-Kanal
benötigte Antiquark liegt nun als Seequark vor. See- und Valenzquarks haben
unterschiedliche PDFs, generell ist die Wahrscheinlichkeit für ein Valenzquark
höher als die für ein Seequark.
Die sich daraus ergebende Wahrscheinlichkeit ein Anti-Seequark im Proton vorzufinden, ist 
wesentlich kleiner als ein Anti-Valenzquark im Antiproton vorzufinden (vgl. Diskussion weiter vorn).
Deswegen ist es plausibel für den \textit{s}-Kanal an einem pp-Beschleuniger 
einen geringeren Wirkungsquerschnitt zu finden.
Die Partonverteilungsfunktionen sind hier ursächlich für die Unterschiede zwischen
den beiden Beschleunigern. Der partonische Wirkungsquerschnitt \(\hat{\sigmaup}\) ist unabhängig 
vom Beschleunigertyp (pp- oder p\(\bar{\text{p}}\)-Beschleuniger).

In der Abb. \ref{wqcms} wird der hadronische Wirkungsquerschnitt
in Abhängigkeit von der Schwerpunktenergie dargestellt. Die Werte 
für den hadronischen Wirkungsquerschnitt wurden mit \textsc{HatHor} \cite{hathor} bestimmt.
Abb. \ref{wqcms} beinhaltet den hadronischen Wirkungsquerschnitt der
drei verschiedenen Produktionskanäle in führender Ordnung (LO) 
(für den \textit{t}-Kanal auch in \nlo n Ordnung (NLO)),
die Summe der verschiedenen Produktionkanäle (in LO und NLO) sowie
zwei aktuelle Messwerte für den \textit{t}-Kanal von den 
Experimenten ATLAS \cite{atlasWert} und CMS \cite{cmsWert}.
Man sieht in der Abb. \ref{wqcms}, dass der hadronische Wirkungsquerschnitt mit
der Schwerpunktenergie zunimmt. Dieses Verhalten erwartet man auch. 
Die Erzeugung eines \topq s ist erst ab einer Energie möglich, die der \topq-Masse
entspricht. Steigt die Schwerpunktenergie darüber hinaus, wird die Erzeugung
eines \topq s immer wahrscheinlicher, da einerseits ausreichend Energie vorhanden ist
und die Möglichkeiten einer Erzeugung des 
\topq s durch die Vergrößerung des Phasenraums zunehmen.

Andererseits weisen die partonischen Wirkungsquerschnitte 
in der Produktion einzelner \topq s (außer im \textit{t}-Kanal)
ein \(\tfrac{1}{s}\)-Verhalten
für die partonische Schwerpunktenergie \(s = x_1 x_2 s_{\text{\tiny Had}}\) auf.
Bis zu einem Maximalwert \(s_{\text{\tiny max}}\) steigt der partonische 
Wirkungsquerschnitt mit der partonischen Schwerpuntenergie an und fällt dann ab.
Die PDFs für die verschiedenen Partonen werden bei kleinen Anteilen \(x_i\) an
der hadronischen Energie groß: Kleine Impulsanteile der Partonen am gesamten 
hadronischen Impuls sind wahrscheinlicher als ein großer Anteil.
Der Maximalwert \(s_{\text{\tiny max}}\) ist fest und je größer nun
die hadronische Schwerpunktenergie wird, desto kleiner werden die zugehörigen
Impulsanteile der Partonen. Damit werden die Werte der PDFs für die Partonen größer.
Man sagt auch, dass sich der Partonfluss durch eine höhere 
hadronische Schwerpuntenergie erhöht.
\begin{table}[t]
	\centering
	\begin{tabular}{cccc}
	\toprule
		\(m_{\text{t}}\) [GeV] & \textit{t}-Kanal [pb] & \textit{s}-Kanal [pb] & tW-Kanal [pb] \\
		\midrule
		170 & \(1{,}134\) & \(0{,}352\)  &  \(0{,}081\)\\ 
		172 & \(1{,}093\) & \(0{,}334\)  &  \(0{,}076\)\\
		175 & \(1{,}037\) & \(0{,}309\)  &  \(0{,}071\)\\
	\bottomrule
	\end{tabular}
	\caption
	{Hadronische Wirkungsquerschnitte der Produktion einzelner \topq s
	in führender Ordnung für verschiedene \topq-Massen in den drei
	Produktionskanälen für p\(\bar{\text{p}}\)-Kollsisionen (Tevatron) bei einer 
	Schwerpunktenergie von \(\sqrt{s_{\text{\tiny Had}}} = \unit{1{,}96}{TeV}\).
	Als PDF-Set wurde MRST2004nnlo genutzt (\(\mu_F = \mu_R = m_{\text{t}}\)).}
	\label{Tab:Kidonakis}
\end{table}

Die theoretisch berechneten Vorhersagen der hadronischen Wirkungsquerschnitte
in Abb. \ref{wqcms} stimmen mit den beiden Messwerten von ATLAS und CMS (\cite{atlasWert,cmsWert})
im Rahmen ihrer Messfehler überein.
Die berechneten Wirkungsquerschnitte
für p\(\bar{\text{p}}\)-Kollisionen
mit dem PDF-Set "`MRST2004nnlo"'
stimmen mit den theoretischen Vorhersagen aus Ref. \cite{Kidonakis} 
auf zwei Nachkommastellen überein (siehe Tabelle \ref{Tab:Kidonakis}).
Die hier berechneten Wirkungsquerschnitte konnten wesentlich genauer
\(\left(\text{Standardabweichung} \sim \mathcal{O}(10^{-13})\right)\)
bestimmt werden, für den
Vergleich mit Ref. \cite{Kidonakis} wurden die Ergebnisse auf 
die dritte Nachkommastelle gerundet.

In der Tabelle \ref{Tab:WqCms} sind ausgewählte hadronische Wirkungsquerschnitte
für die verschiedenen Produktionkanäle angegeben, die im Rahmen dieser
Arbeit berechnet wurden.
Der statistische Fehler (Standardabweichung) der angegeben Werte bezieht 
sich auf die letzte angegebene Stelle.
Die zur Berechnung verwendeten Konstanten befinden sich im Anhang \ref{A:numerik}.

\begin{table}[b!]
	\centering
	\begin{tabular}{cccc}
	\toprule
		\(\sqrt{s_{\text{\tiny Had}}}\) [TeV] & \textit{t}-Kanal [pb] 
		                           & \textit{s}-Kanal [pb] & tW-Kanal [pb] \\
	\midrule
		\(1{,}0\)  & \(0{,}100\)    & \(0{,}021\)   &  \(0{,}002\)\\
		\(1{,}96\) & \(1{,}583\)    & \(0{,}197\)   &  \(0{,}076\)\\
		\(7{,}0\)  & \(44{,}137\)   & \(1{,}914\)   &  \(6{,}062\)\\
		\(8{,}0\)  & \(56{,}240\)   & \(2{,}437\)   &  \(8{,}523\)\\
		\(13{,}0\) & \(137{,}822\)  & \(4{,}218\)   &  \(26{,}341\)\\
		\(14{,}0\) & \(158{,}398\)  & \(5{,}009\)   &  \(31{,}102\)\\
	\bottomrule
	\end{tabular}
	\caption
	{Hadronische Wirkungsquerschnitte der Produktion einzelner \topq s
	in führender Ordnung für verschiedene Schwerpunktenergien \(\sqrt{s_{\text{\tiny Had}}}\) 
	in den drei Produktionskanälen für pp-Kollsisionen bei einer 
	\topq-Masse von \(m_{\text{t}} = \unit{172{,}5}{GeV}\),
	PDF-Set: MRST2004nnlo (\(\mu_F = \mu_R = m_{\text{t}}\)).}        
	\label{Tab:WqCms} 
\end{table}

Für den gesamten hadronischen Wirkungsquerschnitt schließt die NLO-Korrektur
die führende Ordnung mit ein (Abb. \ref{wqcms}).
Der \textit{t}-Kanal hat von den drei möglichen 
Produktionskanälen den größten Anteil (siehe Abb. \ref{ratio}).
Die Korrektur in \nlo r Ordnung für den \textit{t}-Kanal 
schließt im Rahmen ihrer approximierten Unsicherheit die führende Ordnung 
nicht mit ein (Abb. \ref{wqcms}).
Der Unterschied zwischen der führenden und \nlo n Ordnung sind
rund \(4\%\) der führenden Ordnung. 
Das Fehlerband der \nlo n Ordnung, welches aus der Variation
der Faktorisierungs- und Renormierungsskala bestimmt wurde, hat eine Breite
von rund \(7\%\) der \nlo n Ordnung.
Die Abhängigkeit der Variation der Renormierungsskala in \nlo r
Ordnung ist formal eine Abhängigkeit von höherer Ordnung (residuale Abhängigkeit), 
gibt somit also einen Hinweis auf den Beitrag der \nnlo n Ordnung.
Die Unsicherheit beider Ordnungen (LO, NLO) liegt in der gleichen Größenordnung
und daher ist es sinnvoll die \nnlo \, Ordnung zu bestimmen, um
eine exaktere Angabe des hadronischen Wirkungsquerschnitts zu erhalten.

Einen Grund für einen möglichen höheren als erwarteten Beitrag in NNLO
können die zusätzlichen Farbstrukturen liefern, die in \nlo r Ordnung
aufgrund der Interferenz mit dem Bornprozess keinen Beitrag liefern
(siehe Kapitel \ref{kap:nlo}).
In die Fehlerabschätzung von NLO fließen 
nicht die Farbstrukturen ein, die erst bei NNLO aufgrund ihrer
Interferenz miteinander Beiträge liefern.
Ein Farbaustausch ist auch immer mit einem Impulsaustausch verknüpft.
Dieser Impulsfluss kann einen größeren Einfluss auf den Wirkungsquerschnitt
und differentielle Verteilungen 
haben als die Fehlerabschätzung der NLO-Rechnung approximieren kann.

Der Beitrag der \nnlo n Ordnung beinhaltet alle Beiträge der Ordnung 
\(\mathcal{O}(\alpha_s^2)\).
Dazu gehören zunächst die Zweischleifendiagramme, die mit dem Bornprozess
interferieren (Abb. \ref{2loops}). Ein weiterer Beitrag zu den virtuellen Korrekturen stellen
die Einschleifenamplituden quadriert dar (Abb. \ref{1loopsq}).
Neben den virtuellen Korrekturen existieren noch reelle Korrekturen.
Zum einen gibt es die Einschleifenamplituden mit einer reellen Abstrahlung,
die man mit der einer reellen Korrektur interferiert (Abb. \ref{1loop2real}), 
zum anderen hat man \(2 \rightarrow 4\)-Prozesse, 
die man wieder mit sich selbst interferiert (Abb. \ref{2real}),
um auf die Ordnung \(\mathcal{O}(\alpha_s^2)\) zu kommen.
In Abb. \ref{SchemaNNLO} sind alle Beiträge zu NNLO 
schematisch dargestellt. 
In der Produktion einzelner \topq s treten in der Ordnung \(\mathcal{O}(\alpha_s^2)\)
des Amplitudenquadrats folgende Beiträge auf:

\begin{align}
 \vert \ampli^{\text{\tiny NLO}}\vert^2 = 
 \mathcal{T}_{\text{\tiny 2 Loops}}       \times \mathcal{T}_{\text{\tiny Born}}^{\dagger}   \,+\, 
 \mathcal{T}_{\text{\tiny 1 Loop}}        \times \mathcal{T}_{\text{\tiny 1 Loop}}^{\dagger} \,+\, 
 \mathcal{T}_{\text{\tiny 1 Loop 1 Real}} \times \mathcal{T}_{\text{\tiny 1 Real}}^{\dagger} \,+\,
 \mathcal{T}_{\text{\tiny 2 Real}}        \times \mathcal{T}_{\text{\tiny 2 Real}}^{\dagger}.                                    
\label{eq:NNLOampli}                                      
\end{align}
Diese Beiträge sind Teil des partonischen Wirkungsquerschnitts der entsprechenden
Ordnung \(\hat{\sigmaup}^{\text{\tiny NNLO}}\) in Gl. \eqref{eq:SumPartWQ}.

Das Ziel dieser Arbeit ist die Berechnung des
\(\vert \ampli^{\text{\tiny NLO}^2}\vert^2 =
\left(\vert \mathcal{T}_{\text{\tiny 1 Loop}}\vert^2 = 
\mathcal{T}_{\text{\tiny 1 Loop}} \times 
\mathcal{T}_{\text{\tiny 1 Loop}}^{\dagger}\right)\)-Beitrags 
in Gl. \eqref{eq:NNLOampli} und Abb. \ref{1loopsq}.
\setlength{\unitlength}{0.70mm}
\begin{figure}[t]
  \vspace{12mm}
\centering
\begin{subfigure}[h]{0.9\textwidth}
   \centering
 \begin{fmffile}{2loops}
  \begin{fmfgraph*}(60,35)
  \fmfleft{i1,id3,i2} 
  \fmfright{o1,od3,o2}
  \fmf{phantom}{i1,v3,i2}
  \fmf{phantom}{o1,v3,o2}
  \fmf{vanilla}{i1,v1,i2}
  \fmf{phantom,tension=3.7}{v1,v2}
  \fmf{vanilla}{o1,v2,o2}
  \fmfv{decor.shape=circle,decor.filled=empty,decor.size=20mm,
       l.angle=90, l.dist=0}{v3} 
  \fmfv{decor.shape=circle,decor.filled=empty,decor.size=7mm,
     l.angle=90, l.dist=0}{v1}
  \fmfv{decor.shape=circle,decor.filled=empty,decor.size=7mm,
     l.angle=90, l.dist=0}{v2}
 \fmfv{lab=$\mathcal{T}_{\text{\tiny 2 Loops}}\times \mathcal{T}_{\text{\tiny Born}}^{\dagger} =$,lab.dist=0.000009mm}{id3}
 \fmfv{lab=$\times$,lab.dist=1.5mm}{od3}     
  \end{fmfgraph*}
 \end{fmffile}
 \begin{fmffile}{born}
 \begin{fmfgraph*}(60,35)
 \fmfleftn{i}{2}
 \fmfright{o1,o4,o5,o2,o3}
 \fmf{vanilla}{i1,v1,i2}
 \fmfv{decor.shape=circle,decor.filled=shaded,decor.size=17mm,
       l.angle=90, l.dist=0}{v1}
 \fmf{phantom}{o1,v1,o3}
 \fmffreeze
 \fmf{phantom}{v1,v6}
 \fmf{phantom,tension=1.2}{v6,o3}
 \fmffreeze
 \fmf{vanilla,tension=0}{o1,v1}
 \fmf{vanilla}{v1,v6}
 \fmf{vanilla}{v6,o3}
 \end{fmfgraph*}
 \end{fmffile}
 \vspace{5mm}
 \caption{Beitrag der Zweischleifendiagramme interferiert mit dem Bornprozess.}
 \label{2loops}
 \end{subfigure}
\begin{subfigure}[h]{0.9\textwidth}
 \centering
 \vspace{5mm}
  \begin{fmffile}{1loop}
  \begin{fmfgraph*}(60,35)
 \fmfleft{i1,id3,i2}
 \fmfright{o1,od3,o2}
 \fmf{vanilla}{i1,v1,i2}
 \fmf{vanilla}{o1,v1,o2}
 \fmfv{decor.shape=circle,decor.filled=empty,decor.size=17mm,
       l.angle=90, l.dist=0}{v1}
 \fmf{phantom}{i1,v2,i2}
 \fmf{phantom}{o1,v2,o2}
 \fmfv{decor.shape=circle,decor.filled=empty,decor.size=7mm,
       l.angle=90, l.dist=0}{v2} 
 \fmfv{lab=$ \mathcal{T}_{\text{\tiny 1 Loop}} \times \mathcal{T}_{\text{\tiny 1 Loop}}^{\dagger} =$,lab.dist=0.000009mm}{id3}
 \fmfv{lab=$\times$,lab.dist=1.5mm}{od3}         
 \end{fmfgraph*}
 \end{fmffile}
\begin{fmffile}{1loop2}
 \begin{fmfgraph*}(60,35)
 \fmfleft{i1,i2}
 \fmfright{o1,o2}
 \fmf{vanilla}{i1,v1,i2}
 \fmf{vanilla}{o1,v1,o2}
 \fmfv{decor.shape=circle,decor.filled=empty,decor.size=17mm,
       l.angle=90, l.dist=0}{v1}
 \fmf{phantom}{i1,v2,i2}
 \fmf{phantom}{o1,v2,o2}
 \fmfv{decor.shape=circle,decor.filled=empty,decor.size=7mm,
       l.angle=90, l.dist=0}{v2}      
 \end{fmfgraph*}
 \end{fmffile}
 \vspace{5mm}
 \caption{Beitrag durch Einschleifendiagramme interferiert mit sich selbst.}
\label{1loopsq}
\end{subfigure} 
\begin{subfigure}[h]{0.9\textwidth}
\centering
\vspace{5mm}
 \begin{fmffile}{1loopreal}
 \begin{fmfgraph*}(60,35)
 \fmfleft{i1,id3,i2}
 \fmfright{o1,o4,o5,o2,o3}
 \fmf{vanilla}{i1,v1,i2}
 \fmfv{decor.shape=circle,decor.filled=empty,decor.size=17mm,
       l.angle=90, l.dist=0}{v1}
 \fmf{phantom}{o1,v1,o3}
 \fmf{phantom}{i1,v2,i2}
 \fmf{phantom}{o1,v2,o3}
 \fmfv{decor.shape=circle,decor.filled=empty,decor.size=7mm,
       l.angle=90, l.dist=0}{v2} 
 \fmffreeze
 \fmf{phantom}{v1,v6}
 \fmf{phantom,tension=1.2}{v6,o3}
 \fmffreeze
 \fmf{vanilla,tension=0}{o1,v1}
 \fmf{vanilla}{v1,v6}
 \fmf{vanilla}{v6,o3}
 \fmf{plain}{v6,o2}
 \fmfdot{v6}
 \fmfv{lab=$\mathcal{T}_{\text{\tiny 1 Loop 1 Real}} \times \mathcal{T}_{\text{\tiny 1 Real}}^{\dagger} =$,lab.dist=0.000009mm}{id3}
 \fmfv{lab=$\times$,lab.dist=1.5mm}{o5}  
 \end{fmfgraph*}
 \end{fmffile}
 \reflectbox{ 
 \begin{fmffile}{1real}
 \begin{fmfgraph*}(60,35)
 \fmfleftn{i}{2}
 \fmfright{o1,o4,o5,o2,o3}
 \fmf{vanilla}{i1,v1,i2}
 \fmfv{decor.shape=circle,decor.filled=shaded,decor.size=17mm,
       l.angle=90, l.dist=0}{v1}
 \fmf{phantom}{o1,v1,o3}
 \fmffreeze
 \fmf{phantom}{v1,v6}
 \fmf{phantom,tension=1.2}{v6,o3}
 \fmffreeze
 \fmf{vanilla,tension=0}{o1,v1}
 \fmf{vanilla}{v1,v6}
 \fmf{vanilla}{v6,o3}
 \fmf{plain}{v6,o2}
 \fmfdot{v6}
 \end{fmfgraph*}
 \end{fmffile}}
\vspace{5mm}
\caption{Beitrag durch Einschleifendiagramme mit einer reeller Korrektur 
         interferiert mit einer reellen Korrektur.}
\label{1loop2real}
\end{subfigure}
\hspace{2.5cm}
 \begin{subfigure}[h]{0.9\textwidth}
 \centering
 \vspace{5mm}
 \begin{fmffile}{2real}
 \begin{fmfgraph*}(60,35)
 \fmfleft{i1,id3,i2}
 \fmfright{o1,d1,o4,o5,o2,d2,o3}
 \fmf{vanilla}{i1,v1,i2}
 \fmfv{decor.shape=circle,decor.filled=shaded,decor.size=17mm,
       l.angle=90, l.dist=0}{v1}
 \fmf{phantom}{o1,v1,o3}
 \fmffreeze
 \fmf{phantom}{v1,v6}
 \fmf{phantom,tension=1.2}{v6,o3}
 \fmf{phantom}{v1,v7}
 \fmf{phantom,tension=1.2}{v7,o1}
 \fmffreeze
 \fmf{vanilla}{v1,v7}
 \fmf{vanilla}{v7,o1}
 \fmf{plain}{v7,o4}
 \fmf{vanilla}{v1,v6}
 \fmf{vanilla}{v6,o3}
 \fmf{plain}{v6,o2}
 \fmfdot{v6,v7}
 \fmfv{lab=$\mathcal{T}_{\text{\tiny 2 Real}} \times \mathcal{T}_{\text{\tiny 2 Real}}^{\dagger} =$,lab.dist=0.000009mm}{id3}
 \fmfv{lab=$\times$,lab.dist=1.5mm}{o5}
 \end{fmfgraph*}
 \end{fmffile}
 \reflectbox{
 \begin{fmffile}{2real2}
 \begin{fmfgraph*}(60,35)
 \fmfleftn{i}{2}
 \fmfright{o1,d1,o4,o5,o2,d2,o3}
 \fmf{vanilla}{i1,v1,i2}
 \fmfv{decor.shape=circle,decor.filled=shaded,decor.size=17mm,
       l.angle=90, l.dist=0}{v1}
 \fmf{phantom}{o1,v1,o3}
 \fmffreeze
 \fmf{phantom}{v1,v6}
 \fmf{phantom,tension=1.2}{v6,o3}
 \fmf{phantom}{v1,v7}
 \fmf{phantom,tension=1.2}{v7,o1}
 \fmffreeze
 \fmf{vanilla}{v1,v7}
 \fmf{vanilla}{v7,o1}
 \fmf{plain}{v7,o4}
 \fmf{vanilla}{v1,v6}
 \fmf{vanilla}{v6,o3}
 \fmf{plain}{v6,o2}
 \fmfdot{v6,v7}
 \end{fmfgraph*}
 \end{fmffile}}
  \vspace{5mm}
  \caption{Beitrag durch zwei reelle Korrekturen interferiert mit sich selbst.}
\label{2real}
\end{subfigure}
\vspace{5mm}
\caption[Schematische Darstellung der Beiträge der Produktion einzelner 
         \topq s in \nnlo r Ordnung]
         {Schematische Darstellung der Beiträge der Produktion einzelner 
         \topq s in \nnlo r Ordnung.}
\label{SchemaNNLO}
\end{figure}

	\chapter{Berechnung von Prozessen höherer Ordnung}

Die führende Ordnung liefert in der Produktion einzelner \topq s
den größten Beitrag zum Wirkungsquerschnitt in der Störungsreihe.
Um den Wirkungsquerschnitts genauer angeben zu können,
müssen auch immer Prozesse höherer Ordnung
mitberücksichtigt, also berechnet werden.
%

Im Bereich der Hochenergiephysik ist die
Kopplungskonstante der starken Wechselwirkung klein, das heißt
eine hohe Potenz der Kopplungskonstanten bedeutet auch ein sehr kleiner
Faktor für den Wirkungsquerschnitt, woraus folgt, dass Prozesse höherer Ordnung
in der QCD weniger wahrscheinlich sind. 
Mit Ansteigen der zusätzlichen Wechselwirkungen werden sie
immer unwahrscheinlicher und liefern somit einen immer kleiner werdenden Beitrag
zum Wirkungsquerschnitt,
sofern die Beiträge aus den Betragsquadraten der Spinorstrukturen die
zusätzlichen Potenzen der Kopplungskonstante nicht kompensieren können.
Die Berechnung möglichst vieler Ordnungen ist eine Erhöhung der
Genauigkeit des berechneten Wirkungsquerschnitts.
Die höheren Ordnungen sind Störungen der führenden Ordnung und
da die Kopplungskonstante in der QCD klein ist, ist eine Entwicklung in der
Kopplungskonstanten eine Störungsreihe.

Die Berechnung der führenden und \nlo n Ordnung 
sind elementar für die Angabe eines Wirkungsquerschnitts. 
Die zusätzliche Berechnung der \nnlo n Ordnung stellt
eine Erhöhung der Genauigkeit dar und ist somit gleichzeitig auch
ein Kontrollterm für die ersten beiden Ordnungen.

Man unterscheidet bei den Prozessen höherer Ordnung zwischen virtuellen
und reellen Korrekturen zur führenden Ordnung.

Eine virtuelle Korrektur bezeichnet das Auftreten eines zusätzlichen propagierenden
Teilchens, welches also nicht als äußeres Teilchen die Signatur ändert.
Man nennt solche Korrekturen auch Schleifen.
Die niedrigste Ordnung von virtuellen Korrekturen sind demnach Einschleifenkorrekturen.

Bei reellen Korrekturen treten zusätzliche Teilchen im Endzustand auf,
sie ändern somit streng genommen die Signatur des Prozesses.
Ein Detektor, der zum Nachweis eines Prozesses und seiner Signatur verwendet wird,
liefert nur eine gewisse Auflösung.
Liegen zum Beispiel im Endzustand zwei Teilchen sehr dicht beieinander 
(sind sie also kollinear), kann der Detektor sie nur als ein Teilchen auflösen.
Die gemessene Signatur ist somit nicht zu unterscheiden von einem Prozess, der
tatsächlich nur eines der Teilchen im Endzustand aufweist. 
Solche Ereignisse fließen in den gemessenen Wirkungsquerschnitt mit ein und müssen
demnach auch in der Berechnung des Wirkungsquerschnitts mit berücksichtigt werden.

Wie in der Einleitung erwähnt, werden in dieser Arbeit die Einschleifenamplituden
quadriert bestimmt. Sie sind den virtuellen Korrekturen zuzuordnen.

\section[Einschleifendiagramme im \textit{t}-Kanal]
        {Einschleifendiagramme im \textit{t}-Kanal}
\label{einschleifen}
        
\begin{figure}[t]
  \vspace{12mm}
\centering
  \begin{subfigure}[h]{0.4\textwidth}
   \centering
  \begin{fmffile}{tchanneloneloop1}
  \begin{fmfgraph*}(70,35)
  \fmfleftn{i}{2} \fmfrightn{o}{2}
  \fmfv{lab=$\text{b}(p_2)$,label.angle=180}{i1}
  \fmfv{lab=$\text{u}(p_1)$,label.angle=180}{i2}
  \fmfv{lab=$\text{d}(p_3)$,label.angle=0}{o2}
  \fmfv{lab=$\text{t}(p_4)$,label.angle=0}{o1}
  \fmf{fermion}{i1,v1}
  \fmf{heavy}{v1,o1}
  \fmf{boson,tension=0,label=$\text{W}(q)$}{v1,v3}
  \fmf{fermion}{i2,v2}
  \fmf{fermion}{v2,v3}
  \fmf{fermion}{v3,v4}
  \fmf{fermion}{v4,o2}
  \fmf{gluon,left=0.5,tension=0,label=$k$}{v2,v4} 
  \fmfdotn{v}{4}
 \end{fmfgraph*}
 \end{fmffile}
  \vspace{12mm}
 \caption{Vertexkorrektur, leichte Seite.}
 \label{vtx1}
 \end{subfigure}
 \hspace{10mm}
 \begin{subfigure}[h]{0.4\textwidth}
  \centering
  \begin{fmffile}{tchanneloneloop2}
  \begin{fmfgraph*}(70,35)
  \fmfleftn{i}{2} \fmfrightn{o}{2}
  \fmfv{lab=$\text{b}(p_2)$,label.angle=180}{i1}
  \fmfv{lab=$\text{u}(p_1)$,label.angle=180}{i2}
  \fmfv{lab=$\text{d}(p_3)$,label.angle=0}{o2}
  \fmfv{lab=$\text{t}(p_4)$,label.angle=0}{o1}
  \fmf{fermion}{i2,v3}
  \fmf{fermion}{v3,o2}
  \fmf{boson,tension=0,label=$\text{W}(q)$}{v1,v3}
  \fmf{fermion}{i1,v2,v1}
  \fmf{heavy}{v1,v4,o1}
  \fmf{gluon,left=-0.5,tension=0,label=$k$,lab.dist=0.08w}{v2,v4} 
  \fmfdotn{v}{4}
 \end{fmfgraph*}
 \end{fmffile}
 \vspace{12mm}
 \caption{Vertexkorrektur, schwere Seite.}
 \label{vtx2}
\end{subfigure}
 \caption[Vertexkorrekturen der Produktion einzelner \topq s im \textit{t}-Kanal]
          {Vertexkorrekturen im \textit{t}-Kanal.} 
 \label{tkanaloneloop}
\end{figure}
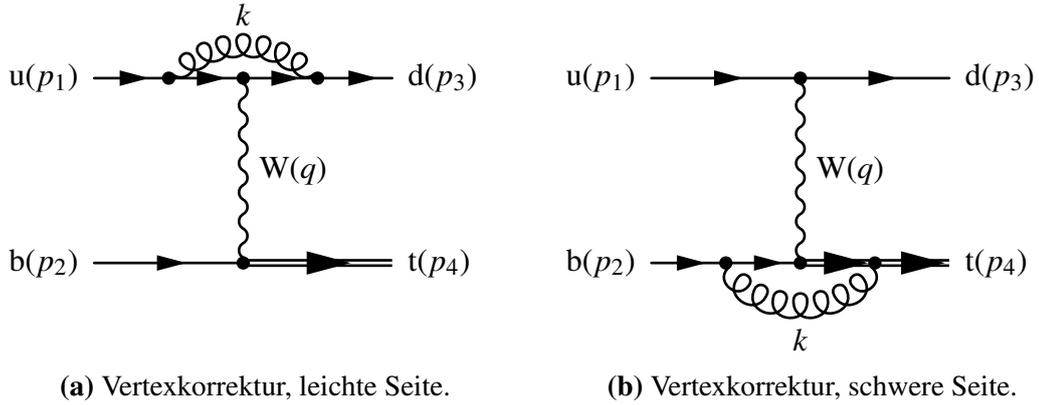

Zur Berechnung der Einschleifenamplituden quadriert ("`\(\text{NLO}^2\)"') des \textit{t}-Kanals
tragen alle Diagramme bei, die im Gegensatz zum Bornprozess einen
zusätzlichen Gluonpropagator aufweisen.  In den Abb. \ref{tkanaloneloop} und \ref{4boxes}
sind alle auftretenden Einschleifendiagramme dargestellt.
\setlength{\unitlength}{0.70mm}
\setlength{\unitlength}{0.70mm}
\begin{figure}
\centering
  \vspace{3mm}
 \begin{subfigure}[b]{0.4\textwidth}
  \centering
  \begin{fmffile}{tchannel3}
  \begin{fmfgraph*}(70,35)
  \fmfleftn{i}{2} \fmfrightn{o}{2}
  \fmfv{lab=$\text{b}(p_2)$,label.angle=180}{i1}
  \fmfv{lab=$\text{u}(p_1)$,label.angle=180}{i2}
  \fmfv{lab=$\text{d}(p_3)$,label.angle=0}{o2}
  \fmfv{lab=$\text{t}(p_4)$,label.angle=0}{o1}
  \fmf{fermion,straight}{i1,v1}
  \fmf{vanilla}{v1,v3}
  \fmf{heavy,straight}{v3,o1}
  \fmf{boson,straight,tension=0,label=$\text{W}$,label.side=right}{v3,v4}
  \fmf{vanilla}{v2,v4}
  \fmf{fermion,straight}{i2,v2}
  \fmf{fermion,straight}{v4,o2}
  \fmf{gluon,straight,tension=0,label=$\text{g}$,label.side=left}{v1,v2} 
  \fmfdotn{v}{4}
 \end{fmfgraph*}
 \end{fmffile}
 \caption{1. Box.}
  \vspace{3mm}
 \end{subfigure}
  \hspace{10mm}
 \begin{subfigure}[b]{0.4\textwidth}
  \centering
  \begin{fmffile}{tchannel4}
  \begin{fmfgraph*}(70,35)
  \fmfleftn{i}{2} \fmfrightn{o}{2}
  \fmfv{lab=$\text{b}(p_2)$,label.angle=180}{i1}
  \fmfv{lab=$\text{u}(p_1)$,label.angle=180}{i2}
  \fmfv{lab=$\text{d}(p_3)$,label.angle=0}{o2}
  \fmfv{lab=$\text{t}(p_4)$,label.angle=0}{o1}
  \fmf{fermion,straight}{i1,v1}
  \fmf{double}{v1,v3}
  \fmf{heavy,straight}{v3,o1}
  \fmf{boson,tension=0,label=$\text{W}$,label.side=left}{v1,v2}
  \fmf{fermion,straight}{i2,v2}
  \fmf{vanilla}{v2,v4}
  \fmf{fermion,straight}{v4,o2}
  \fmf{gluon,straight,tension=0,label=$\text{g}$,label.dist=0.4cm,label.side=right}{v3,v4} 
  \fmfdotn{v}{4}
 \end{fmfgraph*}
 \end{fmffile}
 \caption{2. Box.}
  \vspace{3mm}
 \end{subfigure}
 \begin{subfigure}[b]{0.4\textwidth}
  \centering
  \begin{fmffile}{tchannel5}
  \begin{fmfgraph*}(70,35)
  \fmfleftn{i}{2} \fmfrightn{o}{2}
  \fmfv{lab=$\text{b}(p_2)$,label.angle=180}{i1}
  \fmfv{lab=$\text{u}(p_1)$,label.angle=180}{i2}
  \fmfv{lab=$\text{d}(p_3)$,label.angle=0}{o2}
  \fmfv{lab=$\text{t}(p_4)$,label.angle=0}{o1}
  \fmf{fermion,straight}{i1,v1}
  \fmf{vanilla,tension=0.5}{v1,v3}
  \fmf{heavy}{v3,o1}
  \fmf{boson,tension=0}{v3,v2}
  \fmf{vanilla,tension=0.5}{v2,v4}
  \fmf{fermion,straight}{i2,v2}
  \fmf{fermion,straight}{v4,o2}
  \fmf{gluon,tension=0}{v1,v4} 
  \fmfdotn{v}{4}
 \end{fmfgraph*}
 \end{fmffile}
 \caption{3. Box.}
 \end{subfigure}
  \hspace{10mm}
 \begin{subfigure}[b]{0.4\textwidth}
  \centering
   \vspace{3mm}
  \begin{fmffile}{tchannel6}
  \begin{fmfgraph*}(70,35)
  \fmfleftn{i}{2} \fmfrightn{o}{2}
  \fmfv{lab=$\text{b}(p_2)$,label.angle=180}{i1}
  \fmfv{lab=$\text{u}(p_1)$,label.angle=180}{i2}
  \fmfv{lab=$\text{d}(p_3)$,label.angle=0}{o2}
  \fmfv{lab=$\text{t}(p_4)$,label.angle=0}{o1} 
  \fmf{fermion,straight}{i1,v1}
  \fmf{double,tension=0.5}{v1,v3}
  \fmf{heavy,straight}{v3,o1}
  \fmf{boson,tension=0}{v1,v4}
  \fmf{fermion,straight}{i2,v2}
  \fmf{vanilla,tension=0.5}{v2,v4}
  \fmf{fermion,straight}{v4,o2}
  \fmf{gluon,tension=0}{v3,v2} 
  \fmfdotn{v}{4}
 \end{fmfgraph*}
 \end{fmffile}
 \caption{4. Box.}
 \end{subfigure}
 \caption[Boxdiagramme des \textit{t}-Kanals] 
         {Boxdiagramme des \textit{t}-Kanals.} 
 \label{4boxes}
\end{figure}

Im Vergleich zur Bornrechnung muss beachtet werden, dass
ein zusätzlicher Schleifenimpuls \(k\) des Gluons auftritt, der 
alle möglichen Werte annehmen kann.
Das heißt unter Einbeziehung von Schleifendiagrammen muss der Schleifenimpuls
über alle möglichen Impulse integriert werden.

Für die Berechnung des Beitrags von \(\text{NLO}^2\) müssen a priori 
die Interferenzen der Einschleifendiagramme (Abb. \ref{tkanaloneloop} und \ref{4boxes})
untereinander berücksichtigt werden.
Wie bei der Berechnung des tW-Kanals (Kapitel \ref{Kap:tWborn}) treten hier
farbabhängige Terme durch die Generatoren der \(\text{SU}(3)_{\text{\tiny Farbe}}\) auf.
Dieser farbabhängige Anteil kann wieder getrennt vom restlichen Beitrag bestimmt werden.
Es ist sinnvoll die Berechnung in verschiedene Farbstrukturen zu strukturieren.

\begin{figure}
\vspace{10mm}
\centering
\hspace{2.5cm}
 \begin{subfigure}[t]{0.4\textwidth}
  \centering
  \begin{fmffile}{tchannel1loop}
  \begin{fmfgraph*}(70,35)
  \fmfleft{i1,i3,i2}
  \fmfright{o1,o3,o2}
  \fmfv{lab=$\text{b}(p_2)$,label.angle=180}{i1}
  \fmfv{lab=$\text{u}(p_1)$,label.angle=180}{i2}
  \fmfv{lab=$\text{d}(p_3)$,lab.dist=0.09w,label.angle=0}{o2}
  \fmfv{lab=$\text{t}(p_4)$,lab.dist=0.09w,label.angle=0}{o1}
  \fmfv{lab=$T^a$,lab.dist=0.04w,label.angle=90}{v2}
  \fmfv{lab=$T^a$,lab.dist=0.04w,label.angle=67}{v4}
  \fmf{fermion,foreground=red}{i1,v1}
  \fmf{heavy,foreground=red}{v1,o1}
  \fmfv{foreground=red}{v1}
  \fmf{boson,tension=0}{v1,v3}
  \fmf{fermion}{i2,v2,v3,v4,o2}
  \fmf{gluon,left=0.5,tension=0.}{v2,v4} 
  \fmfdotn{v}{4}
  \fmfv{lab=$\ampli^{\text{\tiny 1. Vertex}}\times \ampli^{\text{{\tiny 1. Box}}\dagger} =$,lab.dist=6mm}{i3}
  \fmfv{lab=$\times$}{o3}
 \end{fmfgraph*}
 \end{fmffile}
 \vspace{5mm}
 \caption{1. Vertex}
\end{subfigure}
\hspace{-9mm}
\begin{subfigure}[t]{0.4\textwidth}
  \centering
  \begin{fmffile}{intvtxbox}
  \begin{fmfgraph*}(70,35)
  \fmfleft{i1,i3,i2} \fmfrightn{o}{2}
  \fmfv{lab=$\text{b}(p_2)$,label.angle=0}{o1}
  \fmfv{lab=$\text{u}(p_1)$,label.angle=0}{o2}
  \fmfv{lab={\color{red}$T^b$},lab.dist=0.035w,label.angle=-80}{v3}
  \fmfv{lab=$T^b$,lab.dist=0.04w,label.angle=61}{v4}
  \fmf{heavy,straight,foreground=red}{i1,v1}
  \fmf{vanilla,foreground=red}{v1,v3}
  \fmfv{foreground=red}{v1,v3}
  \fmf{fermion,straight,foreground=red}{v3,o1}
  \fmf{boson,straight,tension=0,label.side=left}{v1,v2}
  \fmf{vanilla}{v2,v4}
  \fmf{fermion,straight}{i2,v2}
  \fmf{fermion,straight}{v4,o2}
  \fmf{gluon,straight,tension=0,label.side=right}{v3,v4} 
  \fmfdotn{v}{4}
 \end{fmfgraph*}
 \end{fmffile}
 \vspace{5mm}
 \caption{1. Box}
\end{subfigure}
 \caption[Interferenz zwischen einer Vertexkorrektur und einer Box]
         {Interferenz zwischen einer Vertexkorrektur und einer Box 
          verschwindet aufgrund der Farbstruktur der unteren (rot
          gekennzeichneten) Fermionlinie.} 
 \label{interfvtxbox}
\end{figure}
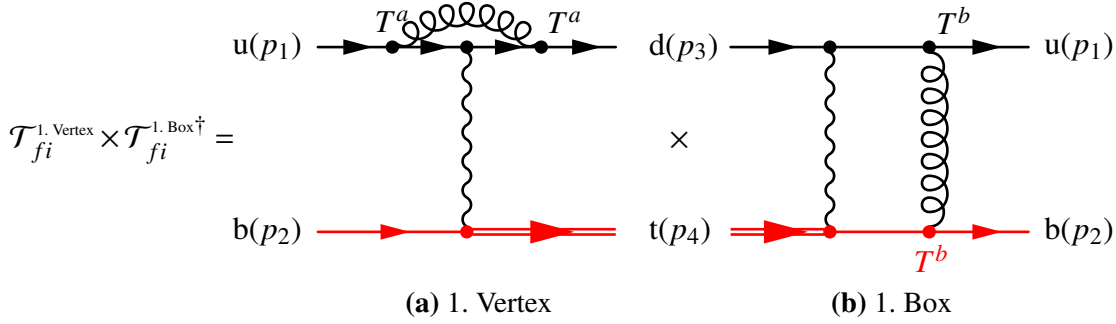

In Abb. \ref{interfvtxbox} ist die Interferenz zwischen der Vertexkorrektur auf der leichten 
Seite und dem 1. Boxdiagramm dargestellt.
\setlength{\unitlength}{0.70mm}
Wie man in Abb. \ref{interfvtxbox}
erkennen kann, wird beim Amplitudenquadrat die Spur über die
untere Spinorlinie nur über einen Generator \(T^b\)
der \(\text{SU}(3)_{\text{\tiny Farbe}}\)-Gruppe gebildet:
\[|\mathcal{T}_{\tn{fi}}|^2 \sim \tn{Tr}(T^b) \cdot \tn{Tr}(T^a T^a T^b) = 0.\]
Die Generatoren sind spurlos und somit verschwindet der Beitrag 
der Interferenz. Für die Interferenzen mit den anderen Boxdiagrammen
ergibt sich ein analoges Bild.

Die vollständige Farbstruktur einer Vertexkorrektur berechnet sich wie folgt,
wobei beachtet werden muss (analog zum tW-Kanal, Gl. \eqref{eq:FarbetW}), 
dass über die Farbe der einlaufenden Quarks gemittelt werden muss. 
Auf Amplitudenniveau ist das ein Faktor von \(\tfrac{1}{\sqrt{N}}\)
pro Quark (also insgesamt \(\tfrac{1}{N}\)):
\begin{align}
{1\over N}\*\delta_{mn}\*\delta_{ab}\*T^a_{ij}\*T^b_{jl} & = 
   {1\over N}\*{1\over 2}\*\left(\delta_{il}\*\delta_{jj} 
   - {1\over N} \delta_{ij}\*\delta_{jl}\right)\*\delta_{mn} \nonumber\\
   & = {1\over N}\*{1\over 2}\*\left(\delta_{il}\*N - {1\over N}\delta_{il}\right) \nonumber\\
   & = {1\over N}\*{1\over 2}\*{N^2-1\over N}\*\delta_{il}\*\delta_{mn} = {1\over N}\*C_F\*C_1 \,.
\label{C1}   
\end{align}
Die volle Farbstruktur einer Box ergibt sich analog:
\begin{align}
 {1\over N}\*T^a_{ij}\*T^b_{kl}\*\delta_{ab} & = 
   {1\over N}\*{1\over 2}\*\left(\delta_{il}\*\delta_{jk} 
   - {1\over N}\*\delta_{ij}\*\delta_{kl}\right) \nonumber \\
   & = {1\over N}\*\left({1\over 2}\*C_2 - {1\over N}\*C_1\right) \,.
\label{C2}   
\end{align}
Wobei die im Amplitudenquadrat auftretenden Produkte von \(C_1\) und \(C_2\),
die folgenden sind:
\begin{align*}
 C_1^2  = C_2^2 & = N^2 \,,\\
 C_1\cdot C_2 & = N \,,
\end{align*}
für die Interferenz der beiden Farbstrukturen folgt:
\begin{align}
 {1\over N^2}\*C_F\*C_1 \cdot \left({1\over 2}\*C_2 - {1\over 2\*N}\*C_1\right)  
  = {1\over N^2}\*\left({1\over 2}\*N - {1\over 2\*N}\* N^2 \right)= 0 \,.
\label{eq:ColourNull}  
\end{align}
Die Interferenz der Farbstrukturen von Vertex- und Boxdiagrammen verschwindet
(nach Gl. \eqref{eq:ColourNull}),
wie erwartet nach der Abschätzung aus Abb. \ref{interfvtxbox}.
Aufgrund dieser Tatsache bietet es sich an, die Berechnung des
Amplitudenquadrats der Einschleifendiagramme getrennt nach Vertex- und Boxdiagrammen
zu organisieren. 
Formal ergibt sich die Amplitude der virtuellen Korrekturen \(\ampli^{\text{\tiny NLO}^2}\) 
wie folgt:
\begin{align}
 \ampli^{\text{\tiny NLO}^2} = & 
 {1\over N}\*C_F\*C_1\*\ampli^{\text{\tiny Vertizes}} + 
               {1\over N}\*\left(\tfrac{1}{2}C_2 - \tfrac{1}{N} C_2\right)\*
               \ampli^{\text{\tiny Boxen}} \nonumber\\
 = & \tilde{C_1}\cdot \ampli^{\text{\tiny Vertizes}} +
     \tilde{C_2}\cdot \ampli^{\text{\tiny Boxen}} \,.
\label{eq:formaleAmpli}
\end{align}
Aufgrund der unterschiedlichen Farbstruktur der Vertex- und Boxdiagramme
besteht das Amplitudenquadrat aus zwei Summanden:
\begin{align}
\vert \ampli^{\text{\tiny NLO}^2} \vert^2 
 &= \tilde{C_1}^2\cdot \vert \ampli^{\text{\tiny Vertizes}}\vert^2 +
    \tilde{C_2}^2\cdot \vert \ampli^{\text{\tiny Boxen}}\vert^2 , \nonumber \\
\text{wobei }\, \tilde{C_1}^2 & = {1\over N^2}\*C_F^2\cdot C_1^2 = \left({N^2 -1\over 2\*N}\right)^2
                                            \,(= {16\over 9} \,\text{ für }\, N = 3) , \nonumber \\
                \tilde{C_2}^2 & = {1\over N^2}\*\left({1\over 2}\*C_2 
                                  - {1\over N}\*C_1\right)^2 = {1 \over 4\*N^2}C_2^2 
                                  - {1\over N^3}C_1\*C_2 + {1\over N^4}C_1^2 = {1\over 4} \,. 
\label{eq:amplisqu}                                            
\end{align}
Die Amplitudenquadrate \(\vert \ampli^{\text{\tiny Vertizes}}\vert^2\) und
\(\vert \ampli^{\text{\tiny Boxen}} \vert^2\) beinhalten nur die Interferenzen
der Vertex- und Boxdiagramme untereinander.
Im nächsten Abschnitt soll es zunächst um die Vertexkorrekturen gehen.

	\section{Vertexkorrekturen}
\label{Kap:Vtxkorrekturen}

In der Abb. \ref{vtx1} ist der Feynmangraph zu einem 
Einschleifendiagramm im \textit{t}-Kanal der Produktion eines \topq s zu sehen.
Zwischen den Quarks u und d wird ein Gluon ausgetauscht.
Der Schleifenimpuls, der durch das Gluon fließt, wird mit \(k\)
bezeichnet.
Wie schon in führender Ordnung schreibt man entsprechend die Feynmanregeln
entlang der Fermionlinien auf.
Nach den Feynmanregeln aus Ref. \cite{Nachtmann} ergibt sich für den Graphen 
in Abb. \ref{tkanaloneloop} folgende Amplitude\footnote{Bei den Propagatoren 
in Gl. \eqref{eq:erste1Loop} steht implizit an den Quarkmassen eine
Einheitsmatrix \(\mathds{1}\), die im Weiteren immer weggelassen wird.}:
\begin{align}
\tilde{C_1}\,\ampli^{\text{\tiny 1. Vertex}} = & \*\left[\overline{u}_{\text{t}}(p_4)\*
       \left(-\imag {e\*V_{\text{tb}} \over \sqrt{2}\*\sin \theta_W}\*
       \gamma^\mu\*{1-\gamma_5 \over 2}\right)\*u_{\text{b}}(p_2)\right]\times \nonumber \\
     & \imag\*{-g^{\mu\nu} + {q^\mu\*q^\nu\over m_{\text{W}}^2} \over 
        q^2 -m_{\text{W}}^2 + \imag\*\epsilon}\times \nonumber \\
     & \int {\mathrm{d}^4 k \over (2\*\pi)^4} 
        \left[\overline{u}_{\text{d}}(p_3)\*(-\imag\*g_s\*\gamma_\sigma)\*
        T^a\*{\imag \over (\slashed{k}+\slashed{p}_3)-m_{\text{d}}+\imag\*\epsilon}
        \times \right.\nonumber \\
     & \left. \left(-\imag\*{e\*V_{\text{ud}}^* \over \sqrt{2}\*\sin \theta_W}\* 
        \gamma^\nu\*{1-\gamma_5\over2}\right)\times \right. \nonumber \\
     & \left.{\imag \over (\slashed{k}+\slashed{p}_1)-m_{\text{u}}+\imag\*\epsilon}\*
       (-\imag\*g_s\*\gamma_\rho)
        \*T^b\*u_{\text{u}}(p_1)\right]\times \nonumber \\
     & \left[\imag\*\delta^{ab}\*\left({-g_{\sigma\rho} \over k^2 +\imag\*\epsilon} +
       {(1-\xi)\*(k)_\sigma\*(k)_\rho \over(k^2+\imag\*\epsilon)^2}\right)\right] \,.
\label{eq:erste1Loop}       
\end{align}
Wie weiter vorn (Kapitel \ref{Kap:ampliquborn}) beschrieben, 
kann für den \W-Propagator eine Eichung (Feynman-ähnliche Eichung) gewählt
werden, in der nur der \(g_{\mu\nu}\)-Term beiträgt. In
der Berechnung der Vertizes liefert der zusätzliche Term des 
\W-Propagators keinen Beitrag, auch wenn man ihn explizit 
mitberücksichtigen würde.
%
Mit der Annahme, dass lediglich das \W\, und das \topq\, 
nicht vernachlässigbare Massen besitzen, vereinfacht sich die Amplitude,
indem die Masse der leichten Quarks \(m_{\text{u}}\) und \(m_{\text{d}}\)
in den Propagatoren zu Null gesetzt werden:
\begin{align}
\ampli^{\text{\tiny 1. Vertex}}  = & {\alpha\*\alpha_s\*V_{\text{tb}}\*V_{\text{ud}}^*
          \over 8\*\sin^2 \theta_W\*(t-m_{\text{W}}^2)}\, \times \nonumber \\
     & \left[\overline{u}_{\text{t}}(p_4)\*\gamma^\mu
        \*(1-\gamma_5)\*u_{\text{b}}(p_2)\right]\, \times \nonumber \\
     & \left[\overline{u}_{\text{d}}(p_3)\*\gamma_\sigma\*\gamma_\alpha\*\gamma_\mu
       \*(1-\gamma_5)\*\gamma^\sigma\*\gamma_\beta\*u_{\text{u}}(p_1)\right] \, \times \nonumber \\
     & \int {\mathrm{d}^4 k \over (2\*\pi)^4}\*
       \left( {(k+p_1+p_2-p_4)^\alpha\*(k+p_1)^\beta \over 
        k^2\cdot(k+p_1+p_2-p_4)^2\cdot(k+p_1)^2}\right) \,. 
     \label{looptchannel}
\end{align}
In der letzten Zeile von Gl. \eqref{looptchannel} ist das zu 
lösende Schleifenintegral zu finden. In vier Dimensionen
ist das Integral divergent im Grenzwert für \(k \rightarrow 0\) (Infrarotdivergenz,
kurz IR-Divergenz) 
und im Hochenergielimit mit \(k \rightarrow \infty\) (Ultraviolettdivergenz, 
kurz UV-Divergenz).
Aus diesem Grund betrachtet man
das Integral in \(d\) Dimensionen und hat das Integral damit konventionell dimensional 
regularisiert ("`conventional dimensional regularisation"' oder kurz CDR\footnote{Andere
mögliche Regularisierungsmethoden sind die Pauli-Villars-Regularisierung und
die "`Cut-off"'-Methode.}). 
Es stellt nun ein endliches Integral für bestimmte \(d \neq 4\) 
(je nach betrachteter Art der Divergenz) dar und kann somit in Summanden
zerlegt werden. Die Behandlung der Divergenzen wird im Abschnitt \ref{Kap:Renorm}
beschrieben.

Die in Gl. \eqref{looptchannel} aufgestellte Amplitude weist ein Tensorintegral
im Schleifenimpuls \(k\) auf. Dieses Tensorintegral möchte man auf skalare
Integrale zurückführen. Im folgenden Abschnitt wird eine Methode zur 
Reduktion von Tensorintegralen vorgestellt.

\subsection{Passarino-Veltman-Reduktion}
\label{pv}

In der Gl. \eqref{looptchannel} ist das betrachtete Schleifenintegral 
ein Tensorintegral.
Das bedeutet im Zähler des Schleifenintegrals tritt eine 
vom Schleifenimpuls abhängige Tensorstruktur (Vektorstruktur) auf.
Man möchte das Tensorintegral in skalare Integrale überführen. 
Folgende Einschleifen-Tensorintegrale und skalaren Integrale 
können allgemein auftreten \cite{Passarino:1978}:
\begin{align}
 A = & {1 \over \imag\*\pi^2}\int \mathrm{d}^d k 
       {1\over (k^2 -m_1^2 + \imag\*\epsilon)}, \nonumber\\
 B_{0;\mu;\mu\nu} = & {1 \over \imag\*\pi^2}\int 
                       \mathrm{d}^d k {[1;k_\mu;k_\mu\*k_\nu] \over 
                       (k^2 -m_1^2 + \imag\*\epsilon)\*
                       \left((k+p_1)^2 -m_2^2 + \imag\*\epsilon\right)}, \nonumber\\ 
 C_{0;\mu;\mu\nu} = & {1 \over \imag\*\pi^2}\int 
                      \mathrm{d}^D k {[1;k_\mu;k_\mu\*k_\nu]
                      \over 
                      (k^2 -m_1^2 + \imag\*\epsilon)\*
                      \left((k+p_1)^2 -m_2^2 + \imag\*\epsilon\right)\*
                      \left((k+p_1+p_2)^2 -m_3^2 + \imag\*\epsilon\right)}, \nonumber \\
 D_{0;\mu;\mu\nu;\mu\nu\rho} = & {1 \over \imag\*\pi^2}\int \mathrm{d}^D k 
                               {[1;k_\mu;k_\mu\*k_\nu;k_\mu\*k_\nu\*k_\rho]
                               \over (k^2 -m_1^2 + \imag\*\epsilon)\*
                               \left((k+p_1)^2 -m_2^2 + \imag\*\epsilon\right)} \times \nonumber\\
                               & {1\over \left((k+p_1+p_2)^2 -m_3^2 + \imag\*\epsilon\right)\*
                               \left((k+p_1+p_2+p_3)^2 -m_4^2 + \imag\*\epsilon\right)} \,.                      
\end{align}
Die Verallgemeinerung für Tensorintegrale höherer Ordnung 
(mit mehreren Propagatoren) lässt sich aus dem oben
erkennbaren Schema ableiten.
Für \(B_0\) folgt z.B. nach dieser Notation:
\[B_{0} = {1 \over \imag\*\pi^2}\int 
        \mathrm{d}^d k {1 \over 
        (k^2 -m_1^2 + \imag\*\epsilon)\*
        \left((k+p_1)^2 -m_2^2 + \imag\*\epsilon\right)} \,,\]
für \(B_{\mu\nu}\) hingegen:
\[B_{\mu\nu} =  {1 \over \imag\*\pi^2}\int 
                       \mathrm{d}^d k {k_\mu\*k_\nu \over 
                       (k^2 -m_1^2 + \imag\*\epsilon)\*
                       \left((k+p_1)^2 -m_2^2 + \imag\*\epsilon\right)} \,.\]
Für die anderen Koeffizienten gilt dieselbe Notation,
die Indizes geben die Tensorstufe des zu reduzierenden Integrals an.

Die Reduktionsmethode nach Passarino und Veltman \cite{Passarino:1978} beruht auf 
der Lorentzsymmetrie, woraus folgt, dass die Lorentzstruktur des
Ergebnisses nur von den externen Impulsen \(p_i\) (für \(i = 1,2,3,4\)) und dem
metrischen Tensor \(g_{\mu\nu}\) abhängen kann.
Man kann daher die Tensorintegrale \(B_{\mu;\mu\nu}, C_{\mu;\mu\nu;\mu\nu\rho},...\)
in Linearkombinationen von Produkten skalarer Koeffizienten, externer Impulse und
dem metrischen Tensor umschreiben:
\begin{align}
 B^\mu & = p_1^\mu \* B_1 ,& \nonumber\\
 B^{\mu\nu} &  =  p_1^\mu\*p_1^\nu\*B_{21} + g^{\mu\nu}\*B_{22} \,,& \nonumber\\
 C^\mu & =  p_1^\mu\*C_{11} + p_2^\mu\*C_{12} \,,& \nonumber\\
 C^{\mu\nu} & =  p_1^\mu\*p_1^\nu\*C_{21} + p_2^\mu\*p_2^\nu\*C_{22} + (p_1^\mu\*p_2^\nu + 
                                                     p_2^\mu\*p_1^\nu)\*C_{23} + g^{\mu\nu}\*C_{24} \,,&\nonumber\\
 C^{\mu\nu\rho} & =  p_1^\mu\*p_1^\nu\*p_1^\rho\*C_{31} + p_2^\mu\*p_2^\nu\*p_2^\rho\*C_{32} +
                  \{p_1\*p_1\*p_2\}^{\mu\nu\rho}\*C_{33} + & \nonumber\\
                & \hspace{5mm} \{p_1\*p_2\*p_2\}^{\mu\nu\rho}\*C_{34} + 
                   \{g\*p_1\}^{\mu\nu\rho}\*C_{35} + \{g\*p_2\}^{\mu\nu\rho}\*C_{36}\,, &  \nonumber \\
 D^\mu & =  p_1^\mu\*D_{11} + p_2^\mu\*D_{12} + p_3^\mu\*D_{13}\,, & \nonumber \\
 D^{\mu\nu} & =  p_1^\mu\*p_1^\nu\*D_{21} + p_2^\mu\*p_2^\nu\*D_{22} + p_3^\mu\*p_3^\nu\*D_{23}
                 + \{p_1\*p_2\}^{\mu\nu}\*D_{24}  + &  \nonumber \\
                 &\hspace{5mm}\{p_1\*p_3\}^{\mu\nu}\*D_{25} + 
                 \{p_2\*p_3\}^{\mu\nu}\*D_{26}  + g^{\mu\nu}\*D_{27}\,, & \nonumber \\
 D^{\mu\nu\rho} & =  p_1^\mu\*p_1^\nu\*p_1^\rho\*D_{31} + p_2^\mu\*p_2^\nu\*p_2^\rho\*D_{32} +
                  \{p_1\*p_1\*p_2\}^{\mu\nu\rho}\*D_{33} 
                  +  & \nonumber \\
                & \hspace{5mm} \{p_1\*p_2\*p_2\}^{\mu\nu\rho}\*D_{34} + 
                  \{p_1\*p_1\*p_3\}^{\mu\nu\rho}\*D_{35} + 
                  \{p_2\*p_2\*p_1\}^{\mu\nu\rho}\*D_{36}
                  +  & \nonumber \\
                & \hspace{5mm}\{p_3\*p_3\*p_1\}^{\mu\nu\rho}\*D_{37}
                \{g\*p_1\}^{\mu\nu\rho}\*D_{311} + \{g\*p_2\}^{\mu\nu\rho}\*D_{312} +                        
                  \{g\*p_3\}^{\mu\nu\rho}\*D_{313}\,, &  \nonumber \\
 \text{wobei} \hspace{3mm} \{p_i\*p_i\*p_j\}^{\mu\nu\rho} & =
                       p_i^\mu\*p_i^\nu\*p_j^\rho + p_j^\mu\*p_i^\nu\*p_i^\rho + 
                       p_i^\mu\*p_j^\nu\*p_i^\rho\,, & \nonumber \\
 \{g\*p_i\} & = g^{\mu\nu}\*p_i^\rho + g^{\nu\rho}\*p_i^\mu + g^{\rho\mu}\*p_i^\nu \,, 
                \text{  für  } \{i,j\} = \{1,2,3\} \,. &    
\label{PVtensors}                
\end{align}
Die auf der rechten Seite in Gl. \eqref{PVtensors} auftretenden Koeffizienten 
lassen sich auf Funktionen von \(A\), \(B_0\), 
\(C_0\) und \(D_0\) zurückführen bzw. reduzieren, 
indem man beide Seiten mit den externen Impulsen und dem metrischen Tensor
kontrahiert.
Auf der linken Seite entstehen dadurch Skalarprodukte zwischen 
dem Schleifenimpuls und den externen Impulsen,
diese kann man in die Form der inversen Propagatoren umschreiben. Dadurch kürzen sich
im Nenner Terme, die vom Schleifenimpuls abhängen.
Nun bleibt noch ein Gleichungssystem zu lösen, welches die Koeffizienten 
als Unbekannte aufweist. Hierfür muss die Matrix, die man auf der rechten Seite der Gleichungen 
erhält, invertierbar sein.
Die Koeffizienten sind dann Funktionen der skalaren Masterintegrale \(A\), \(B_0\), 
\(C_0\) und \(D_0\), ihre Lösung wird weiter hinten diskutiert.

An folgendem kleinen Beispiel lässt sich das Beschriebene leichter nachvollziehen.
Es tritt zum Beispiel folgendes Tensorintegral auf:
\begin{align*}
B^\mu &= \int{d^d k\over (2\pi)^d} 
         {k^\mu \over (k^2 - m_1^2) ((k-p_1)^2 - m_2^2)} = p_1^\mu\*B_1\,.
\end{align*}
Nach der Umschreibung in die Passarino-Veltman-Koeffizienten kontrahiert man
mit den äußeren Impulsen:
\begin{align}
  p_1^2 B_1 & =  \int {d^d k\over (2\pi)^d} {p_1\cdot k\over 
            (k^2 - m_1^2) ((k-p_1)^2 - m_2^2)}\,.
\label{eq:PVKontraktion}            
\end{align}
Ausgenutzt wird nun, dass das Skalarprodukt zwischen äußerem Impuls
und dem Schleifenimpuls als Summe von inversen Propagatoren ausgedrückt werden kann:
\begin{align}
 p_1\*k = -{1\over 2}\left\{((k-p_1)^2-m_2^2)-(k^2-m_1^2)+(m_2^2-m_1^2-p_1^2)\right\}\,.
 \label{SKPkundp}
\end{align}
Man erhält dann für Gl. \eqref{eq:PVKontraktion}:
\begin{align}
 p_1^2 B_1
 & = -{1\over 2}\left( \int {d^d k\over (2\pi)^d} {     
 ((k-p)^2-m_2^2)-(k^2-m_1^2)+(m_2^2-m_1^2-p_1^2) \over 
             (k^2 - m_1^2) ((k-p_1)^2 - m_2^2)}\right) \nonumber\\
 & = -{1\over 2} \left\{\int {d^d k\over (2\pi)^d} {     
 1\over k^2 - m_1^2} - \int {d^d k\over (2\pi)^d} {     
 1\over (k-p_1)^2 - m_2^2} \right. + \nonumber\\
 & \hspace{13mm}  + \left. \int {d^d k\over (2\pi)^d} 
           {(m_2^2-m_1^2-p_1^2) \over (k^2 - m_1^2) ((k-p_1)^2 - m_2^2)}\right\}\nonumber\\
 & = -{1\over 2}\left\{ A(m_1) -A(m_2) + B_0(1,2)\cdot(m_2^2 -m_1^2 -p_1^2)\right\}\,.            
\end{align}
Unter der Annahme, dass alle Teilchen masselos sind, würden hier die beiden
Masterintegrale \(A(m_1)\) und \(A(m_2)\) verschwinden und das betrachtete
Schleifenintegral würde lediglich vom \(B_0(1,2)\)-Integral abhängen.

Für Tensoren zweiter Stufe (bei Zwei-Punkt-Funktionen) erhält man
folgendes Gleichungssystem:
\begin{align*}
   \left(\begin{matrix}
    p^2 & 1\\ p^2 & d 
   \end{matrix}\right)
   \left(\begin{matrix}
    B_{21} \\ B_{22}
   \end{matrix}\right) =
   \left(\begin{matrix}
    -{1\over 2}(m_2^2 -m_1^2-p^2)B_0 - A_0(m_1) - A_0(m_2) \\
     m_1^2 B_0 - A_0(m_2)
   \end{matrix}\right)
   \,.
\end{align*}
Die Lösung des Gleichungssystems erfordert, dass die zu invertierende
Matrix auf der linken Seite keine verschwindende Determinante aufweist.
Als Lösung des Gleichungssystems erhält man wieder Masterintegrale.

Möchte man eine höhere Anzahl von Schleifen berechnen, 
funktioniert diese Umformung nicht mehr, da sich Skalarprodukte
der Schleifenimpulse und der äußeren Impulse nicht in inverse Propagatoren
umschreiben lassen.
Für Tensorintegrale mit mehr als einem Schleifenimpuls
muss eine andere Reduktion als die nach Passarino und Veltman verwendet werden.
Eine allgemeine Reduktionsmethode für mehr als eine Schleife wurde von
Tarasov \cite{Tarasov} entwickelt. Eine weitere
Methode zur Tensorreduktion ist die Projektionsmethode wie in den Ref. \cite{Glover:2004si},
\cite{Anastasiou:2000kg} und \cite{Anastasiou:2000ue} beschrieben.
Neben der Reduktion nach Passarino und Veltman wurde auch die Methode
nach Davydychev \cite{Davydychev:1991va} für Einschleifendiagramme
getestet, bei der Anwendung der verschiedenen Methoden konnte dasselbe
Ergebnis reproduziert werden.

Der für diese Arbeit benutzte Algorithmus basiert auf der Originalarbeit
von Passarino und Veltman \cite{Passarino:1978}.
In der Rechnung wurde eine Implementierung in {\tt Maple} verwendet,
die für beliebige Dimensionen \(d\) gültig ist \cite{puwer}.
\nocite{weinzierl}

Das Schleifenintegral aus Abb. \ref{tkanaloneloop} lässt sich 
unter Verwendung des Algorithmus äquivalent
zu dem eben besprochenen Beispiel umschreiben:
\begin{align}
\ampli^{\text{\tiny 1. Vertex}}  =
        & \, {\alpha\*\alpha_s\*V_{\text{tb}}\*V_{\text{ud}}^*
          \over 8\*\sin^2 \theta_W\*(t-m_{\text{W}}^2)}\, \times \nonumber\\
        & \overline{u}_{\text{t}}(p_4)\*(1+\gamma_5)\*\gamma^\nu\*u_{\text{b}}(p_2)
        \overline{u}_{\text{d}}(p_3)\*(1+\gamma_5)\*\gamma^\nu\*u_{\text{u}}(p_1)
        \, \times \nonumber\\
        & \left[- C_0(1,2,3) \* 2\*t
        - C_{11}\*2\*t
        + C_{22}\* t\*(d-2)
        - C_{23}\* t\*(d-2)
        + C_{24} \* (d-2)^2  \right]\,.
\label{eq:Vtx1PV}        
\end{align}
Man sieht in Gl. \eqref{eq:Vtx1PV}, dass in dem betrachteten Diagramm  nur eine 
Spinstruktur (\(S_3\)) auftritt.
Als Spinstruktur bezeichnet man die Spinoren und die Produkte der Gamma-Matrizen zwischen
den Spinoren in der Amplitude, in Gl. \eqref{eq:Vtx1PV} also folgenden Term:
\[\overline{u}_{\text{t}}(p_4)\*(1+\gamma_5)\*\gamma^\nu\*u_{\text{b}}(p_2)
  \overline{u}_{\text{d}}(p_3)\*(1+\gamma_5)\*\gamma^\nu\*u_{\text{u}}(p_1) = S_3\,.\]
Stellt man die Amplitude mit den Feynmanregeln auf, ist nicht leicht
ersichtlich, dass nur eine Spinstruktur für dieses Diagramm auftritt.
Es erfordert Umformungen mittels der
Antikommutatorrelation der Gamma-Matrizen und der Dirac-Gleichung.
Man versucht eine Reihenfolge und Ordnung der Gamma-Matrizen
festzulegen.
Dabei wird für die \(\gamma_5\)-Matrix ein antikommutierendes Schema
(auch naives Schema genannt)
in \(d\) Dimensionen (CDR) verwendet. Für den hier betrachteten Prozess
ist die Annahme richtig. Im Kapitel \ref{Kap:Larin} wird die 
Besonderheit der \(\gamma_5\)-Matrix in \(d\) Dimensionen
genauer betrachtet.

Für den vollständigen \textit{t}-Kanal treten
nach dieser Umsortierung und Vereinfachung im naiven Schema
sieben verschiedene Spinstrukturen auf.
Diese Anzahl von verschiedenen Spinstrukturen erwartet man für die 
sechs unterschiedlichen Diagramme im \textit{t}-Kanal naiv auch.
Anhand der Anzahl der Gamma-Matrizen und der auftretenden Massen in
der Schleife, erwartet man zunächst sechs Spinstrukturen. Aufgrund
der Tensorstruktur der Schleife hat man noch einen Term, der 
proportional zum metrischen Tensor ist. Dieser Term erzeugt zusätzlich
noch die Bornspinstruktur, womit man insgesamt bei sieben verschiedenen
Spinstrukturen ist.
Es ist sinnvoll seine Amplituden auf eine möglichst geringe Anzahl von Spinstrukturen
zu reduzieren und die Rechnung dadurch allgemeiner und diagrammunspezifisch
zu halten.
Folgende Spinstrukturen wurden für den \textit{t}-Kanal bestimmt, 
wobei für \(\gamma_6 = (\mathds{1}+\gamma_5)\) und \(\gamma_7 = (\mathds{1}-\gamma_5)\) gilt:
\begin{align}
 S_1 & = \overline{u}_{\text{t}}(p_4)\*\gamma_7\*u_{\text{b}}(p_2)
         \overline{u}_{\text{d}}(p_3)\*\gamma_6\*\slashed{p}_4\*u_{\text{u}}(p_1)\,, \nonumber\\         
 S_2 & = \overline{u}_{\text{t}}(p_4)\*\gamma_6\*\slashed{p}_1\*u_{\text{b}}(p_2)
         \overline{u}_{\text{d}}(p_3)\*\gamma_6\*\slashed{p}_4\*u_{\text{u}}(p_1)\,,\nonumber\\     
 S_3 & = \overline{u}_{\text{t}}(p_4)\*\gamma_6\*\gamma_{\mu_1}\*u_{\text{b}}(p_2)
         \overline{u}_{\text{d}}(p_3)\*\gamma_6\*\gamma_{\mu_1}\*u_{\text{u}}(p_1)\,,\nonumber\\           
 S_4 & = \overline{u}_{\text{t}}(p_4)\*\gamma_7\*\gamma_{\mu_1}\*\slashed{p}_1\*u_{\text{b}}(p_2)
         \overline{u}_{\text{d}}(p_3)\*\gamma_6\*\gamma_{\mu_1}\*u_{\text{u}}(p_1)\,,\nonumber\\
 S_5 & = \overline{u}_{\text{t}}(p_4)\*\gamma_7\*\gamma_{\mu_1}\*\gamma_{\mu_2}\*u_{\text{b}}(p_2)
         \overline{u}_{\text{d}}(p_3)\*\gamma_6\*\gamma_{\mu_1}\*\gamma_{\mu_2}\*\slashed{p}_4\*u_{\text{u}}(p_1)\,,\nonumber\\                
 S_6 & = \overline{u}_{\text{t}}(p_4)\*\gamma_6\*\gamma_{\mu_1}\*\gamma_{\mu_2}\*\slashed{p}_1\*u_{\text{b}}(p_2)
         \overline{u}_{\text{d}}(p_3)\*\gamma_6\*\gamma_{\mu_1}\*\gamma_{\mu_2}\*\slashed{p}_4\*u_{\text{u}}(p_1)\,,\nonumber\\        
 S_7 & = \overline{u}_{\text{t}}(p_4)\*\gamma_6\*\gamma_{\mu_1}\*\gamma_{\mu_2}\*\gamma_{\mu_3}\*u_{\text{b}}(p_2)
         \overline{u}_{\text{d}}(p_3)\*\gamma_6\*\gamma_{\mu_1}\*\gamma_{\mu_2}\*\gamma_{\mu_3}\*u_{\text{u}}(p_1)\,.  
\label{spinstrukturen}  
\end{align}

Benutzt man die Passarino-Veltman-Reduktion ergibt sich für Gl. \eqref{looptchannel}:
\begin{align}
\ampli^{\text{\tiny 1. Vertex}}  = & {\alpha\*\alpha_s\*V_{\text{tb}}\*V_{\text{ud}}^*
                                     \over 8\*\sin^2 \theta_W\*(t-m_{\text{W}}^2)}
                                   S_3\*\left[ (d -7)\*B_0(2,3) - 2\*t\*C_0(1,2,3)\right]\,.
\label{looptchannel2}
\end{align}
Die Argumente der Koeffizienten \(B_0\) und \(C_0\) geben an, welche Propagatoren
in den Integralen auftreten. Für die hier betrachtete Vertexkorrektur des \textit{t}-Kanals
sind die Argumente wie folgt definiert:
\begin{align}
 1 &\rightarrow k^2 + i\epsilon \,,\nonumber\\
 2 &\rightarrow (k + p_1)^2 + i\epsilon \,,\nonumber\\
 3 &\rightarrow (k + p_1 + p_2 - p_4)^2 + i\epsilon \,.
 \label{argumentsMasters}
\end{align}
Die Masterintegrale in der Gl. \eqref{looptchannel2} sind nicht unabhängig voneinander,
denn man kann das eine durch das andere Integral ausdrücken. Da beide Integrale unterschiedliche 
Arten von Divergenzen aufweisen, ist es sinnvoll die Trennung nach Divergenzen zunächst 
beizubehalten.
Möchte man später das Amplitudenquadrat auswerten, ist es praktisch, die Anzahl
der skalaren Masterintegrale auf ein Minimum zu begrenzen und Beziehungen 
zwischen den Masterintegralen wie die folgende zu verwenden \cite{smirnov}:
\begin{align}
C_0(1,2,3) = B_0(2,3) \*{1\over t}\* {(d-3)\over (2-{d\over2})}\,.
\label{eq:smirnov}
\end{align}
Die Amplitude für das hier betrachtete Diagramm ergibt sich unter Ausnutzung der
Gl. \eqref{eq:smirnov} zu:
\begin{align}
\ampli^{\text{\tiny 1. Vertex}} = & {\alpha\*\alpha_s\*V_{\text{tb}}\*V_{\text{ud}}^*
                                  \over 8\*\sin^2 \theta_W\*(t-m_{\text{W}}^2)}
                                  {(16 + (-7 + d) d)\cdot S_3\cdot B_0(2,3)\over (-4 + d)}\,.
\label{gehrmannVgl}                     
\end{align}
Der in der Gl. \eqref{gehrmannVgl} bestimmte Faktor
\(\left((16 + (-7 + d) d)\cdot B_0(2,3)\over (-4 + d)\right)\) stimmt mit 
dem Formfaktor aus Ref. \cite[Gl. (8)]{Gehrmann} überein.
In der Gl. \eqref{gehrmannVgl} treten noch Divergenzen auf, die in der
bisherigen Diskussion noch nicht berücksichtigt wurden. Um die
Behandlung der Divergenzen soll es im nächsten Abschnitt gehen.

	\subsection{Renormierung}
\label{Kap:Renorm}

In Gl. \eqref{looptchannel} wurde das zu lösende Tensorintegral
bestimmt. Wie im Kapitel \ref{Kap:Vtxkorrekturen} erwähnt,
divergiert dieses Integral in vier Dimensionen.
Aus diesem Grund wurde das Integral konventionell dimensional 
regularisiert. 
Für \(d \neq 4\) ist es somit kein divergentes Integral.
Die Idee ist, dass man das Integral für \(\text{Re}(d) \not= 4\) berechnet, 
ein Ergebnis als eine Funktion in Abhängigkeit von \(d\) erhält und 
es analytisch fortsetzt.

Für die Divergenzen treten Pole an der Stelle \(d=4\) auf. Für die UV-Divergenzen
kann man sogenannte "`Counterterme"' nutzen, die bei \(d=4\) ebenfalls Pole aufweisen,
jedoch mit entgegengesetzten Vorzeichen.

Der Notwendigkeit der Renormierung liegt allgemein die Tatsache zu Grunde, 
dass die "`nackte"' (unrenormierte) Lagrangedichte \(\mathcal{L}_{\text{n}}\) keine 
physikalischen Kopplungen und Felder aufweist und divergent sein kann.
Um die Kopplungen (Massen sind im weiteren Sinn auch unter Kopplungen zu zählen)
und die UV-Divergenz zu beheben, fügt man einen Renormierungsfaktor \(\sqrt{Z_\Psi}\) für
die nackten Felder in der Lagrangedichte hinzu:
\begin{align*}
 \Psi_0 = \sqrt{Z_\Psi} \Psi_R\,.
\end{align*}
Für die Kopplungen in der nackten Lagrangedichte führt man entsprechende
Renormierungsfaktoren ein:
\begin{align*}
 g_0 &= \sqrt{Z_g} g_R\,, \\
 m_0 &= \sqrt{Z_m} m_R\,.
\end{align*}
Die Renormierungsfaktoren beinhalten neben endlichen Termen auch die
divergenten Anteile der nackten Theorie.
Durch die Substitution der nackten Felder und Kopplungen erhält man
eine Lagrangedichte in Abhängigkeit der Renormierungsfaktoren und renormierten
Felder und Kopplungen\footnote{Diese Herangehensweise nennt man renormierte Störungstheorie.}. 
Durch Umschreiben der nackten Lagrangedichte erhält man zwei Summanden:
\begin{align}
 \mathcal{L}(\Psi_0,g_0,m_0) = \mathcal{L}(\Psi_R,g_R,m_R) + \mathcal{L}_{\text{CT}}\,,
\label{eq:LagrangeRen} 
\end{align}
der zusätzliche Lagrangian \(\mathcal{L}_{\text{CT}}\) in Gl. \eqref{eq:LagrangeRen}
wird "`Counterterm"' genannt. Der erste Summand \(\mathcal{L}(\Psi_R,g_R,m_R)\) 
in Gl. \eqref{eq:LagrangeRen} hat dieselbe Form wie der nackte Lagrangian, hängt jedoch
von den renormierten Größen ab. Stellt man nach diesem renormierten Lagrangian um,
sieht man, dass sich dieser aus der Summe des nackten Lagrangians und des Counterterms
ergibt. Durch Hinzufügen des Counterterms zur nackten Lagrangedichte erhält man
eine endliche Lagrangedichte. Der Counterterm hebt also gerade die Divergenzen
der nackten Theorie auf.
%

Die IR-Divergenzen bleiben in dieser Rechnung bestehen.
Nimmt man die reellen Korrekturen derselben Ordnung hinzu, 
kompensieren sich gegenseitig die IR-Divergenzen (nach dem
Kinoshita-Lee-Nauenberg-Theorem \cite{Kinoshita,LeeNauenberg}).
Die reellen Korrekturen treten bei der Phasenraumintegration in den 
Regionen auf, in denen kollineare Teilchen vom Detektor nicht auflösbar sind.
Diese Divergenzen weisen die gleiche Pol-Struktur wie die virtuellen
Korrekturen auf, jedoch wieder mit entgegengesetzten Vorzeichen. In der Summe
heben sich die IR-Divergenzen einer Ordnung konsistent weg.
Dieses Theorem gilt jedoch nur für IR-sichere Observable, also Observable,
die nicht sensitiv auf kollineare und softe Divergenzen sind.

Im \textit{t}-Kanal weisen nur die beiden Vertexkorrekturen UV-Divergenzen auf
(in dieser Eichung).
Um ein UV-endliches Ergebnis zu erhalten, werden also die UV-divergenten
Anteile in beiden Diagrammen mittels Counterterm renormiert.
Die Quark"=Feld"=Re\-nor\-mie\-rungs\-kon\-stan\-te (auf Einschleifengenauigkeit) im \(\overline{\text{MS}}\)-Schema
lautet nach Ref. \cite[Gl. (4.15)]{Zfactor}, wie folgt:

\begin{align}
 Z_q = 1 - C_F\*a\*{\alpha_s \over 4\*\piup\*\epsilon} + \mathcal{O}(\alpha_s^2) \,,
 \label{ZFaktor}
\end{align}
wobei für die Feynman-ähnliche Eichung der Eichparameter \(a = 1\) gesetzt wird 
und \(\epsilon\) läuft gegen Null 
(\(\epsilon \rightarrow 0\), \(d = 4-2\epsilon\)).
Für die vollständige Rechnung der \nnlo n Ordnung muss
der Renormierungsfaktor \(Z\) bis einschließlich der Ordnung \(\mathcal{O}(\alpha_s^2)\)
berücksichtigt werden.
Der in dieser Arbeit bestimmte Beitrag der Einschleifenamplituden quadriert, weist 
lediglich UV-Divergenzen auf, die mittels der Ordnung \(\mathcal{O}(\alpha_s)\)
renormiert werden können. Aus diesem Grund wird im Weiteren nur 
der Beitrag, \(\delta Z = - C_F\*{\alpha_s \over 4\*\piup\*\epsilon}\), zur 
Renormierung der hier auftretenden UV-Divergenzen benutzt.

Die Vertexkorrektur auf der leichten Seite wurde mittels der 
Passarino"=Veltman"=Ko\-effi\-zien\-ten
auf die Form in Gl. \eqref{eq:Vtx1PV} gebracht.
Nur der \(C_{24}\)-Koeffizient, der Koeffizient vom metrischen Tensor \(g_{\mu\nu}\)
(siehe Gl. \eqref{PVtensors}), weist UV-Divergenzen auf.
Betrachtet man nur den UV-Anteil in Gl. \eqref{eq:Vtx1PV}
ergibt sich:
\begin{align}
 \mathcal{T}_{fi | \text{\tiny UV}}^{\text{\tiny 1. Vertex}}  
                         = & {\alpha\*\alpha_s\*V_{\text{tb}}\*V_{\text{ud}}^*
                           \over 8\*\sin^2 \theta_W\*(t-m_{\text{W}}^2)}
                          S_3\*\left[(d - 2)^2\*C_{24}\right]\,.
\label{eq:UVvtx1}                          
\end{align}
Nach Ref. \cite{renorm} sieht die Struktur
der UV-Divergenz für \(C_{24}\) \(( \hat{=} \,\,C_{00}\text{ in Ref. \cite{renorm} })\) so aus:
\begin{align}
 C_{24} =  \overline{C}_{24} + {1\over 4\*\epsilon} \,,
\label{eq:C24} 
\end{align}
so dass in der Rechnung \(C_{24}\) getrennt in einen endlichen \(\overline{C}_{24}\) 
und divergenten Anteil \(\tfrac{1}{4\*\epsilon}\) aufgeschlüsselt werden kann, aus Gl. \eqref{eq:UVvtx1} wird dann:
\begin{align}
\tilde{C_1}\*\mathcal{T}_{fi | \text{\tiny UV}}^{\text{\tiny 1. Vertex}}  
                         = & {\alpha\*\alpha_s\*V_{\text{tb}}\*V_{\text{ud}}^*
                           \over 8\*\sin^2 \theta_W\*(t-m_{\text{W}}^2)}\*\tilde{C_1}
                          S_3\*\left[(4-8\epsilon +4\epsilon^2)\cdot
                            \left(\overline{C}_{24} + {1\over 4 \epsilon}\right)\right]\nonumber \\
                         = & {\alpha\*\alpha_s\*V_{\text{tb}}\*V_{\text{ud}}^*
                           \over 8\*\sin^2 \theta_W\*(t-m_{\text{W}}^2)}\*{C_F\*C_1\over N}
                          S_3\*\left[4\*\overline{C}_{24} - 2 + {1\over \epsilon} 
                              + \mathcal{O}(\epsilon)\right]\,.
\label{eq:Vtx1UVC24}                          
\end{align}
Für die Renormierung auf dem Einschleifenniveau benötigt man nur den 
\(\mathcal{O}(\alpha_s)\)-Anteil von \(Z_q\) (Gl. \eqref{ZFaktor}).
Das zugehörige Counterdiagramm ist proportional zur Bornamplitude
des betrachteten Prozesses:
\begin{align*}
 \mathcal{T}_{\text{\tiny CT}} = {\piup\*\alpha\*V_{\text{tb}}\*V_{\text{ud}}^*
                                 \over 2\*\sin^2 \theta_W\*(t-m_{\text{W}}^2)}\*
                                 {C_1 \over N}\*S_3\*
                                 \left(- C_F\*{\alpha_s \over 4\*\piup\*\epsilon}\right)\,,
\end{align*}
zusammengefasst, vereinfacht sich der Term zu:
\begin{align}
 \mathcal{T}_{\text{\tiny CT}} = -{\alpha\*\alpha_s\*V_{\text{tb}}\*V_{\text{ud}}^*
                                 \over 8\*\sin^2 \theta_W\*(t-m_{\text{W}}^2)}\*S_3\*
                                 \left( {C_F\*C_1 \over N\*\epsilon}\right)\,.
\label{eq:CT}                                 
\end{align}
Die betrachtete Vertexkorrektur und das Counterdiagramm weisen denselben
Farbfaktor auf (vgl. Gl. \eqref{C1} mit Gl. \eqref{eq:CT}).
Durch Addition des Counterterms (Gl. \eqref{eq:CT}) von Gl. \eqref{eq:Vtx1UVC24}
erhält man ein endliches Resultat:
\begin{align}
 \tilde{C_1}\*\mathcal{T}_{fi | \text{\tiny UV}}^{\text{\tiny 1. Vertex}} 
 + \mathcal{T}_{\text{\tiny CT}} =& 
 {\alpha\*\alpha_s\*V_{\text{tb}}\*V_{\text{ud}}^*
 \over 8\*\sin^2 \theta_W\*(t-m_{\text{W}}^2)}\*S_3\*
 \left[4\*\overline{C}_{24} - 2 \right]\,.
\label{eq:vtx1renormiert}                                 
\end{align}
Analog zu dieser Vorgehensweise ist es möglich ausgehend von Gl. \eqref{looptchannel2}
die UV"=Di\-ver\-gen\-zen zu extrahieren und mittels des Counterterms zu entfernen.

Für die Vertexkorrektur auf der schweren Seite (Abb. \ref{vtx2}) erfolgt
die Behandlung der UV-Divergenzen analog.

In beiden Amplituden wurde im Zuge der Renormierung ein endlicher Beitrag produziert, der
unabhängig vom Schleifenintegral ist.
Nach der Renormierung wird das \(d\) in den \(d\)-abhängigen Termen, die aus der 
Dirac-Algebra stammen, zu \(d = 4\) gesetzt 
(FDH-ähnliches Schema, "`Four Dimensional Helicity Scheme"').

Für die UV-endlichen Vertexdiagramme ergeben sich folgende Ausdrücke:
\begin{align}
 \ampli^{\text{\tiny 1. Vertex}} =& C\left[-2\*S_3 - 3\*S_3 \overline{B}_{0}(v_1,2,3) 
                                    - 2\*S_3\*t\*C_{0}(v_1,1,2,3)\right], \nonumber \\
 \ampli^{\text{\tiny 2. Vertex}} =& C\left[-2\*S_3 + {2\*S_1\*\overline{A}(v_2,3)
                                    \over m_{\text{t}}^3 - m_{\text{t}}\*t} 
                                    - {(2\*(2\*m_{\text{t}}\*S_1 + S_3\*t)\*
                                    \overline{B}_{0}(v_2,1,3)\over m_{\text{t}}^2 - t}\right. \nonumber \\     
                                  & \left.+ {(2\*m_{\text{t}}\*S_1 - 
                                  m_{\text{t}}^2\*S_3 + 3\*S_3\*t)\*
                                  \overline{B}_{0}(v_2,2,3) \over m_{\text{t}}^2 - t}
                                  + 2\*S_3\*(m_{\text{t}}^2 - t)\*C_{0}(v_2,1,2,3)\right]\,,
\end{align}
wobei folgendes gilt:
\begin{align}
 B_0(v_i,a,b) &= \overline{B}_0(v_i,a,b) + {1\over \epsilon}, \nonumber \\
 A(v_i,a) &= \overline{A}(v_i,a) + {m^2 \over \epsilon}, \qquad \text{ mit } i = \{1,2\}\,.
\label{eq:NotationMasters} 
\end{align}
Die \(\overline{B}_0(v_i,a,b)\) und \(\overline{A}(v_i,a)\) stellen die
endlichen Beiträge der Masterintegrale für das erste (\(v_1\)) und das zweite (\(v_2\))
Vertexdiagramm dar, \(a\) und \(b\) sind die Argumente für die Propagatoren der
Masterintegrale.

Der Proportionalitätsfaktor \(C\) ist in beiden Fällen gleich:
\begin{align}
 C = {1 \over (4\* \piup)^2}\*{e^2\*g_s^2\*V_{\text{tb}}\*V_{\text{ud}}^* \over 8 \sin^2 \theta_W\*(t-m_{\text{W}}^2)} 
   = {1 \over 4\* \piup}\*{e^2\*\alpha_s\*V_{\text{tb}}\*V_{\text{ud}}^* \over 8 \sin^2 \theta_W\*(t-m_{\text{W}}^2)} 
   = {\alpha\*\alpha_s\*V_{\text{tb}}\*V_{\text{ud}}^* \over 8\*\sin^2 \theta_W\*(t-m_{\text{W}}^2)}\,. 
\label{eq:Vorfaktor}
\end{align}

Das zweite Vertexdiagramm weist neben der Bornspinstruktur auch eine weitere 
Spinstruktur \(S_1\) auf.
Diese Spinstruktur hat keinen IR-divergenten Anteil und sollte daher nach der UV-Renormierung 
nur endliche Terme und Terme in höheren Ordnungen von \(\epsilon\) aufweisen (auf 
Amplitudenniveau).
Diese Tatsache konnte für das Amplitudenniveau auch numerisch\footnote{Zur Berechnung der
verschiedenen \(\epsilon\)-Ordnungen der Masterintegrale wurde {\tt QCDLoop}
\cite{QCDLoop} verwendet.} verifiziert werden:
Für die Summe der Vertexdiagramme  auf Amplitudenniveau ist die Ordnung \(\epsilon^{-1}\)
um eine Größenordnung von \(10^{-15}\) kleiner als die einzelnen Beiträge der Ordnung 
\(\epsilon^{-1}\), wenn man die \(\mathcal{O}(\epsilon^{-1})\) mit einem künstlich hohem
Faktor (z.B. \(10^{50}\)) multipliziert.

In diesem Abschnitt wurde im Rahmen der konventionellen dimensionalen Regularisierung
ein antikommutierendes Schema für die \(\gamma_5\)-Matrix verwendet.
Wie bereits erwähnt, ist es in diesem Prozess gerechtfertigt dieses naive Schema 
zu verwenden. Im nächsten Abschnitt wird ein anderes mögliches Schema der 
\(\gamma_5\)-Matrix beschrieben.
%

	\subsection[Die \ensuremath{\gamma_5}-Matrix in \textit{d}-Dimensionen]
        {Die \ensuremath{\boldsymbol{\gamma_{\text{5}}}}-Matrix in \textit{d}-Dimensionen}
\label{Kap:Larin}        

In diesem Kapitel soll es um ein bekanntes Problem 
(z.B. in den Ref. \cite{Jegerlehner,KreimerGmma5}) im Rahmen der
konventionellen dimensionalen Regularisierung gehen: 
Die \(\gamma_5\)-Matrix in \(d\) Dimensionen.

Die \(\gamma_5\)-Matrix ist aufgrund ihrer Definition 
(Gl. \eqref{eq:gmma5def}) ein rein vierdimensionales
Objekt, sie in \(d\) Dimensionen fortzusetzen ist nicht trivial:
Es ergeben sich verschiedene Fragen aus der Behandlung von 
\(\gamma_5\) im Rahmen der dimensionalen Regularisierung.
Zum Beispiel ist es streng genommen nicht möglich, \(\gamma_5\) als antikommutierend
in CDR anzunehmen.
Ebenso stellt sich die Frage 
wie man die Spur der Spinstrukturen in \(d\) Dimensionen berechnet, 
die eine \(\gamma_5\)-Matrix beinhalten,
da bei der Verwendung der dimensionalen Regularisierung die Spur
in \(d = 4-2\epsilon\) Dimensionen zu bestimmen ist.

Bei den bisher ausgeführten Rechnungen wurde stets angenommen, dass die 
\(d\)-di\-men\-sio\-na\-len Gamma-Matrizen mit der \(\gamma_5\)-Matrix 
antikommutieren.
Diese naive Annahme ist gerechtfertigt, solange keine Anomalie
in der Rechnung zu erwarten ist.
Das Auftreten dieser ABJ-Anomalie (Adler-Bell-Jackiw-Anomalie \cite{Adler,BellJackiw})
bedeutet, dass der Axialvektorstrom und der Vektorstrom nicht gleichzeitig 
erhalten sind und dass die jeweils zugehörige Ward-Identität
durch das Auftreten eines anomalen Terms verletzt wird. 
Charakteristisch dafür ist das Auftreten der \(\gamma_5\)-Matrix
in einer Fermionschleife \cite{Bertlmann}.
Die ABJ-Anomalie benötigt daher eine andere als die naive Behandlung der \(\gamma_5\) 
in \(d\)-Dimensionen. Eine Möglichkeit ist das Schema nach Larin \cite{Larin}.
Das Besondere an diesem Schema ist die Tatsache, dass obwohl die \(\gamma_5\)-Matrix
ein vierdimensionales Objekt darstellt, im Prozess 
der dimensionalen Regularisierung dennoch in \(d\) Dimensionen 
nach Ref. \cite[Gl. (2)]{Larin} berücksichtigt wird:
\begin{align}
\gamma_5 = {\tn{i}\over 4 !}\cdot \epsilon_{\mu_1 \mu_2 \mu_3 \mu_4}
              \gamma^{\mu_1} \gamma^{\mu_2} \gamma^{\mu_3} \gamma^{\mu_4}\,.
\label{eq:gmma5def}              
\end{align}              
Die Annahme nach Larin ist, dass die Indizes \(\mu_1,...,\mu_4\) 
in \(d\) Dimensionen definiert sind.
Beim Prozess der dimensionellen Regularisierung wird der Epsilontensor,
der formal ein vierdimensionales Objekt darstellt, zunächst ignoriert.
Der Epsilontensor wird vor das in vier Dimensionen divergierende Integral gezogen
und erst am Ende der Regularisierung wieder berücksichtigt.
Die Kontraktion der \(d\)-dimensionalen Indizes in Gl. \eqref{eq:gmma5def} 
findet in vier Dimensionen statt.
Das Schema, dass \(\gamma_5\) in beschriebener Form in \(d\) Dimensionen 
verwendet, beruht auf der Verletzung der Ward-Identitäten 
(bzw. Slavnov-Taylor-Identitäten), 
indem als Folge der Verwendung von Gl. \eqref{eq:gmma5def} für die \(\gamma_5\)-Matrix 
nicht mehr mit den \(d\)-dimensionalen Gamma-Matrizen antikommutiert:
\begin{align}
\{\gamma_\mu,\gamma_5\} = \gamma_\mu\gamma_5 + \gamma_5\gamma_\mu \neq 0\,.
\label{eq:antikommut}
\end{align}
Man kann die Form der \(\gamma_5\)-Matrix nach Larin in Kombination mit 
einer Gamma-Matrix (Axialvektor-Struktur) vereinfacht zusammenfassen und dadurch einen 
kompakteren Ausdruck für den axialen Strom in einer Rechnung erhalten:
\begin{align}
 \gamma_\mu \gamma_5 = {i\over 3!}\cdot \epsilon_{\mu \mu_1 \mu_2 \mu_3} 
        \gamma^{\mu_1} \gamma^{\mu_2} \gamma^{\mu_3}\,.
\label{eq:Larinkurz}        
\end{align}
Diese kompaktere Schreibweise (Gl. \eqref{eq:Larinkurz}) hat den Vorteil, 
dass nach Ref. \cite{Larin} die Rechenzeit aufgrund der geringeren Anzahl
von Gamma-Matrizen wesentlich verkürzt ist.

Die Ward-Identitäten sind erhalten, wenn angenommen wird, dass \(\gamma_5\) mit 
den \(d\)-di\-men\-sio\-na\-len Gamma-Matrizen \(\gamma_\mu\) antikommutiert (naives Schema). 
In diesem Fall wird jedoch auf eine der dimensionellen Regularisierung angepasste
Behandlung der \(\gamma_5\) verzichtet und es gilt schon wie in vier Dimensionen
folgende Beziehung:
\begin{align}
\{ \gamma_\mu,\gamma_5\ \} = 0\,.
\end{align}
Der Unterschied in beiden Betrachtungsweisen ist die Kontraktion 
\(d\)-dimensionaler Indizes über \(\gamma_5\) hinweg.
Da IR-Divergenzen auf dieselben \(d\)-dimensionalen Strukturen 
in reellen und virtuellen Korrekturen führen,
ist es hinsichtlich der IR-Divergenzen
nicht essentiell mit welcher Annahme 
\(\gamma_5\) in \(d\)-Dimensionen betrachtet wird.
In der Summe mit den reellen Korrekturen werden die virtuellen IR-Divergenzen
derselben Ordnung verschwinden. 

Es ist nur notwendig und möglich einen Konsistenzcheck 
innerhalb der UV-divergenten Terme vorzunehmen. 
Wenn nun durch die Verwendung eines anderen Schemas (hier das Schema nach Larin) 
aus ursprünglich UV-divergenten Termen endliche Terme generiert werden,
müssen diese überflüssig produzierten, endlichen Terme wieder entfernt
werden. Im Fall des Schemas nach Larin bedeutet das, dass ein zusätzlicher 
Counterterm bei der UV-Renormierung berücksichtigt werden muss. Diesen Counterterm erhält
man aus der Differenz der beiden Schemen (naiv, Larin).

Der folgende Term, aus dem der universelle Counterterm für
UV-Divergenzen im Schema nach Larin folgt, folgt aus der Spinstruktur,
die proportional zum Passarino-Veltman-Koeffizient \(C_{24}\) ist.
Denn wie im vorherigen Abschnitt beschrieben, weist nur dieser 
Koeffizient \(C_{24}\) UV-Divergenzen auf. 
Und die Notwendigkeit der besonderen Behandlung der \(\gamma_5\)-Matrix
ist wie eben erläutert nur gegeben, wenn es sich um UV-Divergenzen handelt.
Aus folgender Struktur folgt der Counterterm für das Schema nach Larin:
\begin{align}
\gamma_\rho \gamma_\alpha \gamma_{\mu} (\mathds{1} -\gamma_5)
        \gamma_\beta \gamma_\rho\,.
\label{eq:Larinspinstruktur}        
\end{align}        
In der Annahme des naiven Schemas, dass \(\{\gamma_\mu,\gamma_5\} = 0\)
gilt, folgt:
\begin{align*}
\gamma_\rho \gamma_\alpha \gamma_\mu \gamma_5 \gamma_\beta \gamma_\rho
& = (2 g_{\rho\alpha} - \gamma_\alpha \gamma_\rho) \gamma_\mu \gamma_5
    (2 g_{\beta\rho} - \gamma_\rho \gamma_\beta)\\
& = 4g_{\alpha\beta}\gamma_\mu\gamma_5
      - 2\gamma_\mu\gamma_5\gamma_\alpha\gamma_\beta 
      - 2\gamma_\alpha\gamma_\beta\gamma_\mu\gamma_5 
      + \gamma_\alpha\gamma_\rho \gamma_\mu \gamma_5\gamma_\rho\gamma_\beta\\
& = 4g_{\alpha\beta}\gamma_\mu\gamma_5
      - 2\gamma_\mu\gamma_5\gamma_\alpha\gamma_\beta 
      - 2\gamma_\alpha\gamma_\beta\gamma_\mu\gamma_5 
      + {\color{red}(-(2-d)) \gamma_\mu \gamma_5}\,.
\end{align*}
Schreibt man \(\gamma_5\) nun nach Larin um und nutzt 
Gl. \eqref{eq:Larinkurz}, folgt für die Struktur der Gamma-Matrizen:      
\begin{align*}
\gamma_\rho \gamma_\alpha \gamma_\mu \gamma_5 \gamma_\beta \gamma_\rho
& = (2 g_{\rho\alpha} - \gamma_\alpha \gamma_\rho) \gamma_\mu \gamma_5
     (2 g_{\beta\rho} - \gamma_\rho \gamma_\beta)\\
& = 4g_{\alpha\beta}\gamma_\mu\gamma_5
      - 2\gamma_\mu\gamma_5\gamma_\alpha\gamma_\beta 
      - 2\gamma_\alpha\gamma_\beta\gamma_\mu\gamma_5 
      + \gamma_\alpha\gamma_\rho 
      \left({\color{red} \epsilon_{\nu_1\nu_2\nu_3} 
      \gamma_{\nu_1}\gamma_{\nu_2}\gamma_{\nu_3}}\right)
      \gamma_\rho\gamma_\beta\\
& = 4g_{\alpha\beta}\gamma_\mu\gamma_5
      - 2\gamma_\mu\gamma_5\gamma_\alpha\gamma_\beta 
      - 2\gamma_\alpha\gamma_\beta\gamma_\mu\gamma_5 
      + {\color{red}(6-d) \gamma_\mu \gamma_5}\,.
\end{align*}      
Der Vergleich zwischen beiden Schemen offenbart:
\begin{align}
-(2-d)\gamma_\mu\gamma_5 =(6-d)\gamma_\mu\gamma_5, \, \text{ für }\, d=4,\nonumber\\
 2(1-\epsilon)\gamma_\mu\gamma_5 \not =
2(1+\epsilon)\gamma_\mu\gamma_5, \, \text{ für }\, \epsilon \not = 0\,.
\end{align}
Die Differenz liefert den Counterterm (für die Ordnung \(\mathcal{O}(\alpha_s)\)) 
für UV-Pole \(\left(\tfrac{1}{\epsilon}\right)\) (bis auf konstante Vorfaktoren):
\begin{align}
\tn{Differenz} = {1\over \epsilon }\left(2-2\epsilon 
                  -(2+2\epsilon)\right)\gamma_\mu\gamma_5
               = {\color{red}-4\gamma_\mu\gamma_5}\,.
\label{eq:LarinCounterterm}               
\end{align}
Der in Gl. \eqref{eq:LarinCounterterm} bestimmte Counterterm stimmt für die
Ordnung \(\mathcal{O}(\alpha_s)\) mit dem aus Ref. \cite{Larin} überein.

\subsubsection{Berechnung der Vertexkorrektur des \textit{t}-Kanals in beiden Schemen}

\begin{figure}[t]
\centering
  \begin{fmffile}{tchannellarin}
  \begin{fmfgraph*}(70,35)
  \fmfleftn{i}{2} \fmfrightn{o}{2}
  \fmflabel{$\text{b}(p_2)$}{i1} \fmflabel{$\text{u}(p_1)$}{i2}
  \fmflabel{$\text{t}(p_4)$}{o1} \fmflabel{$\text{d}(p_3)$}{o2}
  \fmf{fermion}{i1,v1}
  \fmf{heavy}{v1,o1}
  \fmf{boson}{v1,v3}
  \fmf{fermion}{i2,v2}
  \fmf{vanilla,label=$\small k+p_1$,label.side=right,label.dist=0.25mm}{v2,v3}
  \fmf{vanilla,label=$\small k+p_3$,label.side=right,label.dist=0.35mm}{v3,v4}
  \fmf{fermion}{v4,o2}
  \fmf{gluon,left=0.5,tension=0,label=$k$,label.side=left}{v2,v4} 
  \fmfv{label=${\color{red}\gamma_\nu \gamma_5}$}{v3}
  \fmfv{label=${\color{red}\gamma_\rho}$,label.angle=68,label.dist=2mm}{v4}
  \fmfv{label=${\color{red}\gamma_\rho}$,label.angle=100,label.dist=2.5mm}{v2}
  \fmfdotn{v}{4}
 \end{fmfgraph*}
 \end{fmffile}
 \vspace{5mm}
 \caption[Vertexkorrektur \textit{t}-Kanal auf der masselosen Seite]
         {Vertexkorrektur \textit{t}-Kanal auf der masselosen Seite.}   
 \label{tchlarin}
 \end{figure}
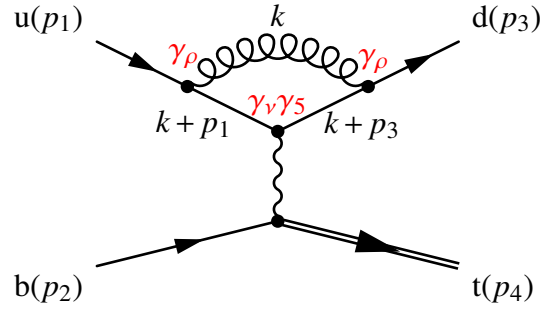

Um zu veranschaulichen, dass die Berechnung in beiden Schemen gleichwertig und
in sich konsistent ist, wird im Folgenden die Vertexkorrektur auf der masselosen
Seite des \textit{t}-Kanals bestimmt.
Es handelt sich um das Diagramm in Abb. \ref{tchlarin}.
\setlength{\unitlength}{0.85mm}
Aufgrund der Feynmanregel für den Gluonpropagator in der Feynman-ähnlichen Eichung
(\(i \delta_{ab} ({-g_{\sigma\rho}\over k^2})\))
erhält man zwei
\(d\)-dimensionale Indizes, die entlang einer Fermionlinie über \(\gamma_5\)
hinweg kontrahiert werden. Für den UV-divergenten Anteil
der betrachteten Vertexkorrektur, der proportional
zum \(C_{24}\)-Koeffizienten ist (vgl. Kapitel  \ref{Kap:Renorm}), 
tritt die oben beschrieben Struktur
(Gl. \eqref{eq:Larinspinstruktur}) mit einem \(g^{\alpha \beta}\) auf:
\begin{align*}
\gamma_\rho \gamma_\alpha \gamma_{\mu} (\mathds{1} -\gamma_5)
        \gamma_\beta \gamma_\rho g^{\alpha \beta} =
\gamma_\rho \gamma_\alpha \gamma_{\mu} (\mathds{1} -\gamma_5)
        \gamma_\alpha \gamma_\rho\,,          
\end{align*} 
so dass insgesamt zwei \(d\)-dimensionale Indizes auftreten,
die über die \(\gamma_5\)-Matrix hinweg kontrahiert werden.
Betrachtet wird im Folgenden nur
der UV-divergente Anteil in beiden Schemen 
Für das naive Schema ergibt sich Gl. \eqref{eq:UVvtx1}:
\begin{align*}
  \mathcal{T}_{\text{\tiny naiv} | \text{\tiny UV}}^{\text{\tiny 1. Vertex}}  
                         = & {\alpha\*\alpha_s\*V_{\text{tb}}\*V_{\text{ud}}^*
                           \over 8\*\sin^2 \theta_W\*(t-m_{\text{W}}^2)}
                          S_3\*\left[(d - 2)^2\*C_{24}\right] \nonumber \\
                        = & {\alpha\*\alpha_s\*V_{\text{tb}}\*V_{\text{ud}}^*
                           \over 8\*\sin^2 \theta_W\*(t-m_{\text{W}}^2)}
                          S_3\*\left[(4 - 8\epsilon + 4\epsilon)\*C_{24}\right]  \,,
\end{align*}
für das Schema nach Larin folgt der UV-divergente Anteil zu:
\begin{align}
 \mathcal{T}_{\text{\tiny Larin} | \text{\tiny UV}}^{\text{\tiny 1. Vertex}}  
                        = & {\alpha\*\alpha_s\*V_{\text{tb}}\*V_{\text{ud}}^*
                          \over 8\*\sin^2 \theta_W\*(t-m_{\text{W}}^2)}
                          S_3\*\left[(6 - d)^2\*C_{24}\right] \nonumber \\
                       = & {\alpha\*\alpha_s\*V_{\text{tb}}\*V_{\text{ud}}^*
                          \over 8\*\sin^2 \theta_W\*(t-m_{\text{W}}^2)}
                          S_3\*\left[(4 + 8\epsilon + 4\epsilon)\*C_{24}\right]\,.                          
\end{align}
Der Unterschied in beiden Schemen besteht lediglich in dem 
\(\mathcal{O}(\epsilon)\)-Term.
Verwendet man für \(C_{24} = \overline{C}_{24} + \tfrac{1}{4\epsilon}\) nach Gl. \eqref{eq:C24}, 
folgt für beide Schemen:
\begin{align}
 \mathcal{T}_{\text{\tiny naiv} | \text{\tiny UV}}^{\text{\tiny 1. Vertex}}  
                        = &  {\alpha\*\alpha_s\*V_{\text{tb}}\*V_{\text{ud}}^*
                           \over 8\*\sin^2 \theta_W\*(t-m_{\text{W}}^2)}
                          S_3\*\left[4\*\overline{C}_{24} + {1\over \epsilon} - 2\right],\nonumber \\
 \mathcal{T}_{\text{\tiny Larin} | \text{\tiny UV}}^{\text{\tiny 1. Vertex}}  
                        = &  {\alpha\*\alpha_s\*V_{\text{tb}}\*V_{\text{ud}}^*
                          \over 8\*\sin^2 \theta_W\*(t-m_{\text{W}}^2)}
                          S_3\*\left[4\*\overline{C}_{24} + {1\over \epsilon} + 2\right] \,.  
\end{align}
In der Differenz der beiden Amplituden bleib übrig:
\begin{align}
 \mathcal{T}_{\text{\tiny naiv} | \text{\tiny UV}}^{\text{\tiny 1. Vertex}} - 
 \mathcal{T}_{\text{\tiny Larin} | \text{\tiny UV}}^{\text{\tiny 1. Vertex}}
 & = -{\alpha\*\alpha_s\*V_{\text{tb}}\*V_{\text{ud}}^*
     \over 8\*\sin^2 \theta_W\*(t-m_{\text{W}}^2)}
     S_3\*4\,.
\label{eq:DifferenzLaNa}     
\end{align}
Der Beitrag des Larin-Counterterms ist proportional zum Bornprozess und der
Proportionalitätsfaktor ist der Larin-Renormierungsfaktor (Gl. \eqref{eq:Z5}) 
in \(\mathcal{O}(\alpha_s)\) \cite{Larin}:
\begin{align}
 \mathcal{L}_{\text{CT}}^{\text{\tiny Larin}} = &
  -{\alpha\*\piup\*V_{\text{tb}}\*V_{\text{ud}}^*
   \over 2\*\sin^2 \theta_W\*(t-m_{\text{W}}^2)}{C_1\over N}
   S_3\* \left({\alpha_s\*C_F \over \piup}\right)\nonumber \\
  = & -{\alpha\*\alpha_s\*V_{\text{tb}}\*V_{\text{ud}}^*
     \over 8\*\sin^2 \theta_W\*(t-m_{\text{W}}^2)} {C_1\*C_F\over N} 4 \,.
\label{eq:LarinCT}     
\end{align}
Gl. \eqref{eq:LarinCT} im Vergleich mit Gl. \eqref{eq:DifferenzLaNa} offenbart
in Gl. \eqref{eq:DifferenzLaNa} nur den nicht explizit berücksichtigten Farbfaktor \(\tilde{C_1}\).
Addiert man den Larin-Counterterm von der Amplitude im Schema nach Larin hinzu
unter der Vernachlässigung der Farbe beim Larin-Counterterm Gl. \eqref{eq:LarinCT}:
\begin{align*}
 \mathcal{T}_{\text{\tiny Larin} | \text{\tiny UV}}^{\text{\tiny 1. Vertex}} \,+ \,
 \left(\tfrac{C_1 C_F}{N}\right)^{-1}\*\mathcal{L}_{\text{CT}}^{\text{\tiny Larin}}  \,=\,
 {\alpha\*\alpha_s\*V_{\text{tb}}\*V_{\text{ud}}^*
 \over 8\*\sin^2 \theta_W\*(t-m_{\text{W}}^2)}
 S_3\*\left[4\*\overline{C}_{24} - {1\over \epsilon} + 2 - 4\right]\,,
\end{align*}
ergibt sich der UV-divergente Anteil im naiven Schema:
\begin{align*}
 \mathcal{T}_{\text{\tiny Larin} | \text{\tiny UV}}^{\text{\tiny 1. Vertex}} \,+ \,
 \left(\tfrac{C_1 C_F}{N}\right)^{-1}\*\mathcal{L}_{\text{CT}}^{\text{\tiny Larin}}  \,=\,
 \mathcal{T}_{\text{\tiny naiv} | \text{\tiny UV}}^{\text{\tiny 1. Vertex}}\,.
\end{align*}
Eine unabhängige Rechnung mittels des Schemas nach Larin ist also möglich.
Im Endergebnis sollten dann die \(\epsilon\)-unabhängigen Terme, also der 
jeweils endliche Beitrag, identisch sein.
Wendet man das Schema nach Larin an, ist dabei der Larin-Renormierungsfaktor 
\(Z_5^{\text{ns}}\) \cite{Larin} in der UV-Renormierung mit zu berücksichtigen:
\begin{align}
 Z_5^{\text{ns}} = 1 - {\alpha_s C_F \over \piup} + \mathcal{O}(\alpha_s^2)\,,
\label{eq:Z5}
\end{align}
um die überflüssig erzeugten endlichen \(\left({\epsilon \over \epsilon}\right)\)-Terme
(im Schema nach Larin) adäquat zu behandeln.

Bei der Berechnung des Amplitudenquadrats tritt in der hier betrachteten Vertexkorrektur
im Fall der naiven Rechnung eine einzige Spinstruktur 
(\(S_3\) siehe Gl. \eqref{spinstrukturen}) auf. 
Für die Larin-Rechnung treten acht unterschiedliche 
Spinstrukturen auf. 
Setzt man die Dimension auf \(d = 4\) fest, stimmen beide Ergebnisse überein. 
Für die Betrachtung \(d \neq 4\) ergeben sich unterschiedliche Ergebnisse. Der
Unterschied für die UV-Divergenzen
lässt sich auf den theoretisch ermittelten Counterterm zurückführen.
Da in der hier betrachteten Produktion einzelner \topq s keine ABJ-Anomalie
zu erwarten ist (da keine reinen Fermionschleifen auftreten), 
ist es vorteilhaft die Rechnung im naiven Schema vorzunehmen.
Der Vorteil besteht darin, dass eine geringere Anzahl von unterschiedlichen
Spinstrukturen auftreten wird, so dass die Rechnung kompakter und schneller
durchgeführt werden kann. Außerdem wird durch die Verwendung einer antikommutierenden
\(\gamma_5\)-Matrix keine Ward-Identität verletzt.

	\subsection[Virtuelle Beiträge in \nlo r Ordnung im \textit{t}-Kanal]
           {Virtuelle Beiträge in \nlo r Ordnung im \textit{t}-Kanal}
\label{kap:nlo}        

\setlength{\unitlength}{0.70mm}
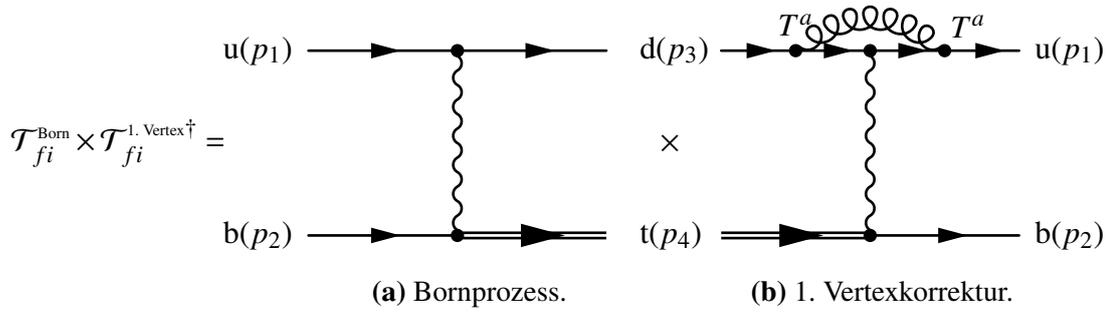
\begin{figure}[t!]
\centering
\hspace{2.5cm}
 \begin{subfigure}[t]{0.4\textwidth}
  \centering
  \begin{fmffile}{tchannelborn}
  \begin{fmfgraph*}(70,35)
  \fmfleft{i1,i3,i2}
  \fmfright{o1,o3,o2}
  \fmfv{lab=$\text{b}(p_2)$,label.angle=180}{i1}
  \fmfv{lab=$\text{u}(p_1)$,label.angle=180}{i2}
  \fmfv{lab=$\text{d}(p_3)$,lab.dist=0.09w,label.angle=0}{o2}
  \fmfv{lab=$\text{t}(p_4)$,lab.dist=0.09w,label.angle=0}{o1}
  \fmf{fermion}{i1,v1}
  \fmf{heavy}{v1,o1}
  \fmf{boson,tension=0}{v1,v3}
  \fmf{fermion}{i2,v3,o2}
  \fmfdot{v1,v3}
  \fmfv{lab=$\ampli^{\text{\tiny Born}}\times \ampli^{\text{{\tiny 1. Vertex}}\dagger} =$,lab.dist=6mm}{i3}
  \fmfv{lab=$\times$}{o3}
 \end{fmfgraph*}
 \end{fmffile}
 \vspace{3mm}
 \caption{Bornprozess.}
\end{subfigure}
\hspace{-9mm}
\begin{subfigure}[t]{0.4\textwidth}
  \centering
  \begin{fmffile}{tchannel1loopnlo}
  \begin{fmfgraph*}(70,35)
  \fmfleft{i1,i3,i2} \fmfrightn{o}{2}
  \fmfv{lab=$\text{b}(p_2)$,label.angle=0}{o1}
  \fmfv{lab=$\text{u}(p_1)$,label.angle=0}{o2}
  \fmfv{lab=$T^a$,lab.dist=0.04w,label.angle=90}{v2}
  \fmfv{lab=$T^a$,lab.dist=0.04w,label.angle=67}{v4}
  \fmf{heavy}{i1,v1}
  \fmf{fermion}{v1,o1}
  \fmf{boson,tension=0}{v1,v3}
  \fmf{fermion}{i2,v2}
  \fmf{fermion}{v4,o2}
  \fmf{fermion}{v2,v3,v4}
  \fmf{gluon,left=0.5,tension=0}{v2,v4} 
  \fmfdotn{v}{4}
 \end{fmfgraph*}
 \end{fmffile}
 \vspace{3mm}
 \caption{1. Vertexkorrektur.}
\end{subfigure}
 \caption[Interferenz zwischen Born- und Vertexdiagramm]
         {Interferenz zwischen Born- und Vertexdiagramm.}
 \label{interfvtxborn}
\end{figure}
\begin{figure}[t!]
\centering
\hspace{2.5cm}
 \begin{subfigure}[t]{0.4\textwidth}
  \centering
  \begin{fmffile}{tchannelborn2}
  \begin{fmfgraph*}(70,35)
  \fmfleft{i1,i3,i2}
  \fmfright{o1,o3,o2}
  \fmfv{lab=$\text{b}(p_2)$,label.angle=180}{i1}
  \fmfv{lab=$\text{u}(p_1)$,label.angle=180}{i2}
  \fmfv{lab=$\text{d}(p_3)$,lab.dist=0.09w,label.angle=0}{o2}
  \fmfv{lab=$\text{t}(p_4)$,lab.dist=0.09w,label.angle=0}{o1}
  \fmf{fermion}{i1,v1}
  \fmf{heavy}{v1,o1}
  \fmf{boson,tension=0}{v1,v3}
  \fmf{fermion}{i2,v3,o2}
  \fmfdot{v1,v3}
  \fmfv{lab=$\ampli^{\text{\tiny Born}}\times \ampli^{\text{{\tiny 1. Box}}\dagger} =$,lab.dist=6mm}{i3}
  \fmfv{lab=$\times$}{o3}
 \end{fmfgraph*}
 \end{fmffile}
 \vspace{3mm}
 \caption{Bornprozess.}
\end{subfigure}
\hspace{-9mm}
\begin{subfigure}[t]{0.4\textwidth}
  \centering
  \begin{fmffile}{intbornbox}
  \begin{fmfgraph*}(70,35)
  \fmfleft{i1,i3,i2} \fmfrightn{o}{2}
  \fmfv{lab=$\text{b}(p_2)$,label.angle=0}{o1}
  \fmfv{lab=$\text{u}(p_1)$,label.angle=0}{o2}
  \fmfv{lab={$T^a$},lab.dist=0.035w,label.angle=-80}{v3}
  \fmfv{lab=$T^a$,lab.dist=0.04w,label.angle=61}{v4}
  \fmf{heavy,straight}{i1,v1}
  \fmf{vanilla}{v1,v3}
  \fmf{fermion,straight}{v3,o1}
  \fmf{boson,straight,tension=0,label.side=left}{v1,v2}
  \fmf{vanilla}{v2,v4}
  \fmf{fermion,straight}{i2,v2}
  \fmf{fermion,straight}{v4,o2}
  \fmf{gluon,straight,tension=0,label.side=right}{v3,v4} 
  \fmfdotn{v}{4}
 \end{fmfgraph*}
 \end{fmffile}
 \vspace{3mm}
 \caption{1. Box.}
\end{subfigure}
 \caption[Interferenz zwischen Born- und Boxdiagramm] 
         {Interferenz zwischen Born- und Boxdiagramm.}
 \label{interfbornbox}
\end{figure}
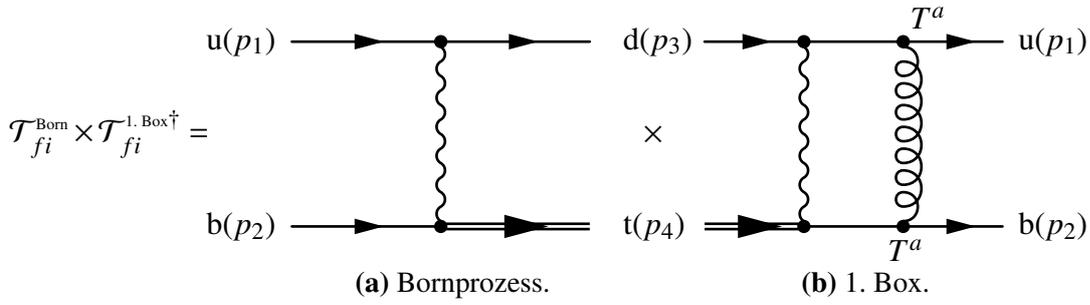
In diesem Abschnitt werden die Beiträge der Einschleifenamplituden
zur \nlo n Ordnung bestimmt (NLO).
Bei dieser Ordnung interferieren die Einschleifendiagramme mit 
der führenden Ordnung (Bornprozess\index{Bornprozess}).
Durch die einfache Struktur werden nicht alle Interferenzen von
Einschleifendiagrammen mit dem Borndiagramm  
ein von Null verschiedenes Ergebnis produzieren.
Einen Beitrag zu den virtuellen Korrekturen liefern 
die Interferenzen zwischen den 
Vertexkorrekturen und der führenden Ordnung, wie in
Abb. \ref{interfvtxborn} zu sehen.
Die Interferenzen zwischen den Boxdiagrammen und dem Bornprozess
liefern hingegen keinen Beitrag.
In Abb. \ref{interfbornbox} ist zu erkennen, dass bei
einer Spurbildung über beide Fermionlinien 
die Spur nur über einen Farbgenerator gebildet wird.
Da alle Farbgeneratoren jedoch spurlos sind, wird
bei der Interferenz mit dem Bornprozess insgesamt kein Beitrag in 
dieser Ordnung generiert (analog zur Diskussion in Kapitel \ref{einschleifen}).
Zur Bestimmung der Beiträge zur \nlo n Ordnung
im \textit{t}-Kanal tragen also nur die Vertexkorrekturen der 
Einschleifendiagramme bei.
Der Farbanteil der Rechnung kann wieder separat vom restlichen
Anteil bestimmt werden.
Der Farbfaktor der beiden Vertexkorrekturen ist, siehe Gl. \eqref{C1}:
\[\tilde{C_1} = {1\over N}C_F\delta_{ij}\delta_{mo}.\]
Der Farbfaktor des Bornprozesses ergibt sich analog dazu:
\[{1\over N}\delta_{ij}\delta_{mo}.\]
Beide Farbfaktoren miteinander multipliziert ergeben:
\begin{align}
{1\over N^2}\delta_{ij}\delta_{mo}\*C_F\delta_{ij}\delta_{mo} = C_F\,.
\label{eq:nloFarbe} 
\end{align}
Der Farbfaktor in Gl. \eqref{eq:nloFarbe} muss beim Endergebnis
mitberücksichtigt werden.

Bei der Berechnung der virtuellen und reellen Beiträge können verschiedene
Schemen verwendet werden. Wichtig für jede Rechnung ist, dass
in jeder betrachteten Ordnung ein konsistentes Schema
für beide Beiträge verwendet wird.
In dieser Arbeit wird einheitlich das gleiche Schema verwendet, welches auch für
die Beiträge der Einschleifenamplituden quadriert in der 
\nnlo n Ordnung verwendet wurde.
Für die aufgestellten Amplituden werden 
Polynome in \(d\) nach der UV-Renormierung, die aus der Gamma-Algebra stammen, in
\(d = 4\) Dimensionen gesetzt, da die zu bildenden Spuren über
die Gamma-Matrizen ebenfalls in vier Dimensionen berechnet werden.
Die Berechnung der Spur in vier Dimensionen wurde
gewählt, um Probleme mit der \(\gamma_5\)-Matrix im Rahmen der
dimensionalen Regularisierung zu vermeiden (siehe Kapitel \ref{Kap:Larin}).
Die notwendige Reduktion der Tensorintegrale (mittels der 
Passarino-Veltman-Reduktion) bleibt in \(d\) Dimensionen erhalten.
Dieses verwendete Schema hat vor allem technische Vorteile, es ist
dem FDH Schema ähnlich. 

Für die in Abb. \ref{interfvtxborn} dargestellte Interferenz ergibt sich:
\begin{align*}
 {1\over 4}\sum_{\text{\tiny{Spins}}} \ampli^{\text{\tiny Born}} \times 
 \ampli^{\text{\tiny 1. Vertex}\dagger} \,= \,  
 C_F\*(4\*\piup)^{-2}\*\alpha_s\*\left[2 + 3\*\overline{B}_{0}(v_1,2,3) + 2\*tC_{0}(v_1,1,2,3) \right]
   \cdot\vert \ampli^{\text{\tiny Born}} \vert^2\,. 
\end{align*}
Man sieht, dass der infrarot-divergente Term \(\left(\sim C_{0}(v1,1,2,3)\right)\) in das 
Bornamplitudenquadrat faktorisiert, der Faktor, \(2\*t\), ist \(d\)-unabhängig.
Bei der Interferenz des Bornprozesses mit der Vertexkorrektur auf der schweren Seite,
sieht man ein äquivalentes Bild:
\begin{align*}
{1\over 4}\sum_{\text{\tiny{Spins}}} \ampli^{\text{\tiny Born}} \times 
 \ampli^{\text{\tiny 2. Vertex}\dagger} \,= \,  
 C_F\*(4\*\piup)^{-2}\*\alpha_s\*\left[-2\*(m_{\text{t}}^2-t)C_{0}(v_2,1,2,3) \right]
   \cdot\vert \ampli^{\text{\tiny Born}}\vert^2 + \,\, ... \,\,\, ,
\end{align*}
im Unterschied zur ersten Vertexkorrektur, faktorisiert der endliche Anteil der
ursprünglich UV-divergenten Anteile nicht
in das Bornamplitudenquadrat, was wegen der komplexeren Struktur
aufgrund der \topq-Masse (im Fermionpropagator) in der zweiten Vertexkorrektur zu
verstehen ist.
Nach Ref. \cite[Gl.(12)]{Catani} erwartet man für Einschleifenrechnungen
diese globale Struktur der IR-Divergenzen.

Der gesamte Beitrag der virtuellen Korrekturen zur \nlo n 
Ordnung ergibt sich zu: 
\begin{align}
 {1\over 4}\sum_{\text{\tiny{Spins}}}
 \ampli^{\text{\tiny Born}} \times (
 \ampli^{\text{\tiny 1. Vertex}\dagger} + \ampli^{\text{\tiny 2. Vertex}\dagger}) = &
   \,(4\*\piup)^{-1}\*\alpha_s\*C_F\*\left[2+3\*\overline{B}_{0}(v_1,2,3) + 2\*tC_{0}(v_1,1,2,3) 
        \right.  \,\,\, +  \nonumber\\ 
   &\left. - 2\*(m_{\text{t}}^2-t)C_{0}(v_2,1,2,3) \right] 
   \cdot\vert \ampli^{\text{\tiny Born}}\vert^2  \,\,\, +  \nonumber\\ 
   & {\piup^2\*\alpha^2\*\vert V_{tb}\vert^2\*\vert V_{ud}\vert ^2\over 
      4\*\sin^4 \theta_W\*(t -m_{\text{W}}^2)^2}(4\*\piup)^{-1}\*\alpha_s\*C_F \,\,\, \times \nonumber \\
 & \left[ {128\*s\*(s\*t + m_{\text{t}}^2\*(s-m_{\text{t}}^2))\over m_{\text{t}}^2-t}\*
    \overline{B}_{0}(v_2,1,3) \,\,\, + \right.\*\nonumber \\
 & \left. {64\*s\*t\*\left(2\*m_{\text{t}}^2 -3\*s\right)\over
    (m_{\text{t}}^2-t)}\*\overline{B}_{0}(v_2,2,3) \,\,\, + \right.\*\nonumber \\
 & \left. 64\*s\*\left(1 + {s\over t-m_{\text{t}}^2}\right)\*\overline{A}(v_2,3) \,\,
    +\,\,  128\*s\*(s-m_{\text{t}}^2)\right]\,,
 \label{nlo}
\end{align}
wobei \(\vert \ampli^{\text{\tiny Born}}\vert^2\) dem Bornamplitudenquadrat aus der
Gl. \eqref{4DimBorn} im Kapitel \ref{Kap:ampliquborn} entspricht.
Die angegebenen Masterintegrale 
\(\overline{B}_{0}(v_i,2,3)\), \(C_{0}(v_i,1,2,3)\), \(\overline{B}_{0}(v_i,1,3)\) und 
\(\overline{A}(v_i,3)\) (für \(i = \{1,2\}\))
können mit Hilfe von {\tt QCDLoop} \cite{QCDLoop} für eine beliebige
Kinematik ausgewertet werden\footnote{Die Argumente der Masterintegrale entsprechen
verschiedenen Propagatoren. Für die erste Vertexkorrektur siehe Gl. \ref{argumentsMasters},
die zweite Vertexkorrektur siehe Anhang Gl. \eqref{MastersInQCDLoop}.}. 
Wobei beachtet werden muss, dass
die UV-divergenten Anteile in diesem Ergebnis (Gl. \eqref{nlo}) nicht weiter 
berücksichtigt werden dürfen.
Die Notation der Masterintegrale entspricht der Notation in Gl. \eqref{eq:NotationMasters}.

Die UV-Divergenzen wurden mittels Countertermen renormiert, so dass das Ergebnis 
in Gl. \eqref{nlo} UV-endlich ist. IR-Divergenzen sind noch vorhanden. Die reellen 
Korrekturen werden im Rahmen dieser Arbeit nicht näher betrachtet.
In der \nlo n Ordnung werden sich die IR-Divergenzen in der Summe mit den
reellen Korrekturen kompensieren (Kinoshita-Lee-Nauenberg-Theorem).
Verwendet man ein anderes Schema als das hier vorgestellte, bleibt die Faktorisierung
der IR-Divergenzen erhalten \cite{Catani2}.
Für manche Schemen wird das Bornamplitudenquadrat in \(d\) Dimensionen benötigt, welches
hier zur Vollständigkeit mit  angegeben wird.
Das \(d\)-dimensionale Amplitudenquadrat des \textit{t}-Kanals 
(\(\text{ub} \rightarrow \text{dt}\)) lautet:
\begin{align}
\vert \ampli^{\text{\tiny Born}}(d) \vert^2 = & 
{\pi^2\*\alpha^2\*\vert V_{tb}\vert^2\*\vert V_{ud}\vert ^2\over \*\sin^4 \theta_W\*(t -m_{\text{W}}^2)^2} \nonumber\\
   & \left\{ (d-4) t\cdot [(-2 + d) mt^2 - 2 (-1 + d) s] \right. \nonumber \\
   & \left. - (d-4)\cdot (-2 + d) t^2 + 4 s (s-mt^2) \right\}\,.
   \label{dDimBorn}
\end{align}

Nachdem nun in diesem und im vorherigen Kapitel die Vertexdiagramme
des \textit{t}-Kanals ausführlich
besprochen wurden, soll es im nächsten Kapitel um die Beiträge der
Boxdiagramme in \nnlo r Ordnung gehen.
Aufgrund der komplexeren Struktur (Auftreten von vier Propagatoren, 
andere Farbstruktur) ist die Behandlung der Boxdiagramme im Vergleich
zu den Vertexdiagrammen aufwendiger.
	\section[Berechnung der Boxdiagramme im \textit{t}-Kanal]
        {Berechnung der Boxdiagramme im \textit{t}-Kanal}
\label{KapitelBox}

In diesem Kapitel werden nun die Beiträge der Boxdiagramme
in \(\text{NLO}^2\) berechnet.

Da ein physikalisches Ergebnis nicht von der Wahl der Eichung
abhängen darf, ist es sinnvoll eine Rechnung mit 
physikalischem Ergebnis in verschiedenen Eichungen
vorzunehmen.

Hat man die unitäre Eichung gewählt, werden in Schleifenrechnungen einzelne Diagramme 
hoch divergent so wie in nicht renormierbaren Theorien (vgl. Kapitel \ref{Kap:ampliquborn})
\cite{Nachtmann}. 
Die Summe solcher hoch divergenter Diagramme ist wieder in einer renormierbaren
Theorie eingebettet. Es ist aber sinnvoll, Einschleifenrechnungen in 
einer Eichung vorzunehmen, in der keine hoch divergenten Diagramme auftreten 
nach Ref. \cite{Nachtmann}. 

Nimmt man die Eichung \(\xi \rightarrow \infty\), die in Ref. \cite{Nachtmann} verwendet wird, kann man
zeigen, dass sich die Beiträge des zusätzlichen Terms im \W-Propagator
paarweise innerhalb der Summe der Boxdiagramme aufheben.

\begin{figure}[t]
\centering
  \vspace{3mm}
 \begin{subfigure}[b]{0.4\textwidth}
  \centering
  \begin{fmffile}{tchannel3_2}
  \begin{fmfgraph*}(70,35)
  \fmfleftn{i}{2} \fmfrightn{o}{2}
  \fmfv{lab=$\text{b}(p_2)$,label.angle=180}{i1}
  \fmfv{lab=$\text{u}(p_1)$,label.angle=180}{i2}
  \fmfv{lab=$\text{d}(p_3)$,label.angle=0}{o2}
  \fmfv{lab=$\text{t}(p_4)$,label.angle=0}{o1}
  \fmf{fermion,straight}{i1,v1}
  \fmf{vanilla,label=$p-p_4$,label.side=right}{v1,v3}
  \fmf{heavy,straight}{v3,o1}
  \fmf{boson,straight,tension=0,label=$p$,label.side=right}{v3,v4}
  \fmf{vanilla,label=$p+p_3$,label.side=left}{v2,v4}
  \fmf{fermion,straight}{i2,v2}
  \fmf{fermion,straight}{v4,o2}
  \fmf{gluon,straight,tension=0,label=$p-p_1+p_3$,label.side=left}{v1,v2} 
  \fmfdotn{v}{4}
 \end{fmfgraph*}
 \end{fmffile}
 \vspace{5mm}
 \caption{1. Box.}
 \end{subfigure}
 \hspace{1cm}
 \begin{subfigure}[b]{0.4\textwidth}
  \centering
  \begin{fmffile}{tchannel5_2}
  \begin{fmfgraph*}(70,35)
  \fmfleftn{i}{2} \fmfrightn{o}{2}
  \fmfv{lab=$\text{b}(p_2)$,label.angle=180}{i1}
  \fmfv{lab=$\text{u}(p_1)$,label.angle=180}{i2}
  \fmfv{lab=$\text{d}(p_3)$,label.angle=0}{o2}
  \fmfv{lab=$\text{t}(p_4)$,label.angle=0}{o1}
  \fmf{fermion,straight}{i1,v1}
  \fmf{vanilla,tension=0.5,label=$p-p_4$,label.side=right}{v1,v3}
  \fmf{heavy}{v3,o1}
  \fmf{boson,tension=0}{v3,v2}
  \fmf{vanilla,tension=0.5,label=$p_1-p$,label.side=left}{v2,v4}
  \fmf{fermion,straight}{i2,v2}
  \fmf{fermion,straight}{v4,o2}
  \fmf{gluon,tension=0}{v1,v4} 
  \fmfdotn{v}{4}
 \end{fmfgraph*}
 \end{fmffile}
 \vspace{5mm}
 \caption{3. Box.}
 \end{subfigure}
 \caption[Summe des 1. und 2. Boxdiagramms des \textit{t}-Kanals] 
         {Summe des 1. und 2. Boxdiagramms des \textit{t}-Kanals.
          Der Schleifenimpuls \(p\) ist für beide Diagramme der Impuls
          des \W s.} 
 \label{sum2boxes}
\end{figure}
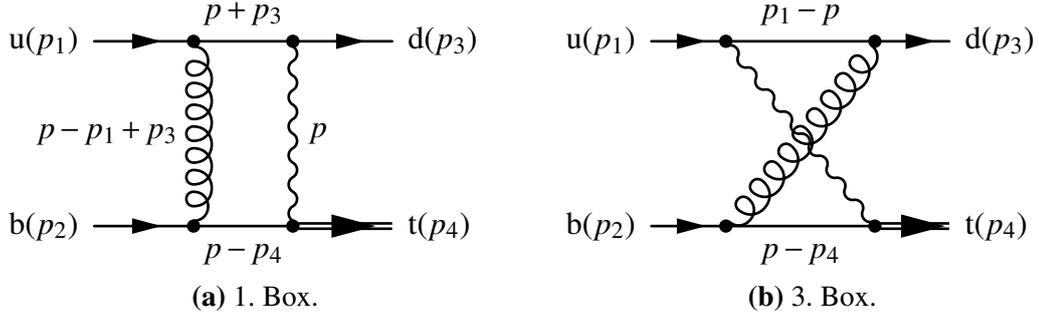

Betrachtet man zunächst die Summe des \(\tfrac{p_\mu\*p_\nu}{m_{\text{W}}^2}\)-Anteils
der 1. und der 3. Box (Abb. \ref{sum2boxes}),
wobei \(\gamma_{6,7} = (\mathds{1} \pm \gamma_5)\) ist und 
der Faktor \(C\) dem aus Gl. \eqref{eq:Vorfaktor} entspricht,
ergibt sich:

\begin{align}
 C\cdot \left(\ampli^{\text{\tiny 1. Box}} + \ampli^{\text{\tiny 3. Box}} \right)_{|{\tiny 
    ({p_\mu\*p_\nu}/{m_{\text{W}}^2})}} = & \,\, C\cdot P\cdot
         \left[ \overline{u}_{\text{t}}(p_4)\*\gamma^\mu\*\gamma_7\*
          [\slashed{p}_4 -\slashed{p}]^{-1}\*\gamma_\alpha\*u_{\text{b}}(p_2)\right] \, 
          {p_\mu\*p_\nu \over m_{\text{W}}^2} \times \nonumber\\
         & \left[ \overline{u}_{\text{d}}(p_3)\*\gamma^\nu\*\gamma_7\*
            [\slashed{p}_3+\slashed{p}]^{-1}\*\gamma^\alpha\*u_{\text{u}}(p_1) + \right. \nonumber \\
         &  \left. \hphantom{[}\overline{u}_{\text{d}}(p_3)\*\gamma^\alpha\*
            [\slashed{p}_1-\slashed{p}]^{-1}\*\gamma^\nu\*\gamma_7\*u_{\text{u}}(p_1)\right] \nonumber\\
         = & \,\,C\cdot P\cdot
         \left[ \overline{u}_{\text{t}}(p_4)\*\gamma^\mu\*\gamma_7\*
          [\slashed{p}_4 -\slashed{p}]^{-1}\*\gamma_\alpha\*u_{\text{b}}(p_2)\right] \,  
          {p_\mu \over m_{\text{W}}^2} \times \nonumber\\
          & \left[ \overline{u}_{\text{d}}(p_3)\*\slashed{p}\*\gamma_7\*
            [\slashed{p}_3+\slashed{p}]^{-1}\*\gamma^\alpha\*u_{\text{u}}(p_1) + \right. \nonumber \\
          & \left.\hphantom{[} \overline{u}_{\text{d}}(p_3)\*\gamma^\alpha\*
            [\slashed{p}_1-\slashed{p}]^{-1}\*\slashed{p}\*\gamma_7\*u_{\text{u}}(p_1)\right]\,,
\label{weichung}
\end{align}
mit dem zusätzlichen Proportionalitätsfaktor \(P\) für die Propagatoren der Eichbosonen:
\begin{align}
 P & = {1 \over [p^2 -m_{\text{W}}^2]\cdot (p+p_3-p_1^2)} \,.
\label{eq:PropP}
\end{align}
Interessant in der Gl. \eqref{weichung} ist der letzte geklammerte Term.
Ersetzt man in dem ersten Summanden 
(\(\slashed{p} \rightarrow \slashed{p} + \slashed{p}_3\))\footnote{Diese Ersetzung wird
nicht für den Propagator \([\slashed{p}_3 +\slashed{p} ]^{-1}\) vorgenommen
(siehe Gl. \eqref{endweichung}).}, hat man lediglich
eine Null eingeführt, da durch Anwendung der Dirac-Gleichung die 
zusätzlichen Terme verschwinden.
Für den zweiten Summanden ersetzt man äquivalent 
(\(\slashed{p} \rightarrow \slashed{p} - \slashed{p}_1\)).
Mit diesen Ausdrücken kann man die Propagatoren kürzen und erhält vereinfachte 
Spinstrukturen:

\begin{align}
 C\cdot \left(\ampli^{\text{\tiny 1. Box}} + \ampli^{\text{\tiny 3. Box}} \right)_{|{\tiny 
    ({p_\mu\*p_\nu}/{m_{\text{W}}^2})}} = & \,\, C\cdot P\cdot
          \left[ \overline{u}_{\text{t}}(p_4)\*\gamma^\mu\*\gamma_7\*
          [\slashed{p}_4 -\slashed{p}]^{-1}\*\gamma_\alpha\*u_{\text{b}}(p_2)\right] \,
          {p_\mu \over m_{\text{W}}^2} \times \nonumber\\
         & \left[ \overline{u}_{\text{d}}(p_3)\*(\slashed{p}+ \slashed{p}_3)\*\gamma_7\*
           [\slashed{p}_3+\slashed{p}]^{-1}\*\gamma^\alpha\*u_{\text{u}}(p_1) + \right. \nonumber \\
         & \left. \hphantom{[} \overline{u}_{\text{d}}(p_3)\*\gamma^\alpha\*
           [\slashed{p}_1-\slashed{p}]^{-1}\*(\slashed{p}-\slashed{p}_1)\*\gamma_7\*u_{\text{u}}(p_1)\right] \nonumber\\
       = & \,\,C\cdot P\cdot
          \left[ \overline{u}_{\text{t}}(p_4)\*\gamma^\mu\*\gamma_7\*
          [\slashed{p}_4 -\slashed{p}]^{-1}\*\gamma_\alpha\*u_{\text{b}}(p_2)\right] \, 
          {p_\mu \over m_{\text{W}}^2} \times \nonumber\\
         & \left[ \overline{u}_{\text{d}}(p_3)\*\gamma_7\*(-\slashed{p}-\slashed{p}_3)\*
           [\slashed{p}_3+\slashed{p}]^{-1}\*\gamma^\alpha\*u_{\text{u}}(p_1) + \right. \nonumber \\
         & \left. \hphantom{[} \overline{u}_{\text{d}}(p_3)\*\gamma^\alpha\*
           [\slashed{p}_1-\slashed{p}]^{-1}\*(\slashed{p}-\slashed{p}_1)\*\gamma_7\*u_{\text{u}}(p_1)\right] \nonumber\\
       = & \,\, C\cdot P\cdot
          \left[ \overline{u}_{\text{t}}(p_4)\*\gamma^\mu\*\gamma_7\*
          [\slashed{p}_4 -\slashed{p}]^{-1}\*\gamma_\alpha\*u_{\text{b}}(p_2)\right] \, 
          {p_\mu \over m_{\text{W}}^2} \times \nonumber\\
         & \left[- \overline{u}_{\text{d}}(p_3)\*\gamma_7\*\gamma^\alpha\*u_{\text{u}}(p_1) -
           \overline{u}_{\text{d}}(p_3)\*\gamma^\alpha\*\gamma_7\*u_{\text{u}}(p_1)\right] \nonumber\\                        
       = & \,\,C\cdot P\cdot
          \left[ \overline{u}_{\text{t}}(p_4)\*\gamma^\mu\*\gamma_7\*
          [\slashed{p}_4 -\slashed{p}]^{-1}\*\gamma_\alpha\*u_{\text{b}}(p_2)\right] \, 
          {p_\mu \over m_{\text{W}}^2} \times \nonumber\\
         & \left[ \overline{u}_{\text{d}}(p_3)\*\gamma^\alpha\*\gamma_7\*u_{\text{u}}(p_1) -
           \overline{u}_{\text{d}}(p_3)\*\gamma^\alpha\*\gamma_7\*u_{\text{u}}(p_1)\right] \nonumber\\                                    
       = & \,\, 0 \, .            
 \label{endweichung}
\end{align}
Wie man in Gl. \eqref{endweichung} sieht, verschwinden
die Beiträge des \(\tfrac{p_\mu\*p_\nu}{m_{\text{W}}^2}\)-Terms des \W-Propagators
in der Summe des ersten und dritten Boxdiagramms.
Analoges ergibt sich bei der Betrachtung der verbleibenden zwei Boxen (Abb. \ref{4boxes}).

Die Rechnung ist unabhängig von der Wahl der Eichung. Es
ist, wie oben erwähnt, jedoch für die Ausführung der Rechnung
effizienter die Feynman-ähnliche Eichung zu verwenden. 
Gleichzeitig bietet die Eichfreiheit die Möglichkeit für einen 
Konsistenzcheck der Rechnung.

Mit einer ähnlichen Argumentation für das Verschwinden des 
\(\tfrac{p_\mu\*p_\nu}{m_{\text{W}}^2}\)-Terms des \W-Propagators, kann
man argumentieren, dass kollineare Divergenzen in der Summe der Box-Amplituden
verschwinden.
Kollineare Divergenzen finden sich in den \(\tfrac{1}{\epsilon^2}\)-Termen
wieder. Zusätzlich zu einer soften Divergenz hat man hier noch eine kollineare Divergenz,
die aufgrund des zusätzlichen Freiheitsgrads des Winkels zwischen zwei Teilchen
auftritt.
Heben sich diese kollinearen Divergenzen in der Summe der Box-Amplituden 
weg, wird der \(\tfrac{1}{\epsilon^2}\)-Term für die Summe verschwinden.

Schreibt man zunächst die Summe der Boxen \(\ampli^{\text{\tiny Boxen}}\)
(Abb. \ref{4boxes}) auf Amplitudenniveau
nach den verschiedenen Spinorlinien, erhält man folgendes:
\begin{align}
 \ampli^{\text{\tiny Boxen}}  = C\cdot \tilde{P}\cdot M_{24}^{\alpha_1 \beta_1}(p_4,p_2,k)\cdot 
                                      M_{13}^{\alpha_2 \beta_2}(p_3,p_1,k)\,,
\end{align}
wobei \(\tilde{P}\) die Propagatoren der Eichbosonen beinhaltet und: 
\begin{align*}
 M_{24}^{\alpha_1 \beta_1}(p_4,p_2,k) = & \,\, 
 \overline{u}_{\text{t}}(p_4)\*\gamma^{\alpha_1}\*\gamma_7\*
   [\slashed{p}_2 - \slashed{k}]^{-1}\*\gamma^{\beta_1}\*u_{\text{b}}(p_2) +  \\
 & \,\,\overline{u}_{\text{t}}(p_4)\*\gamma^{\beta_1}\*[\slashed{p}_4 + \slashed{k} - m_{\text{t}}]^{-1}\*
   \gamma^{\alpha_1}\*\gamma_7\*u_{\text{b}}(p_2),\\
 M_{13}^{\alpha_2 \beta_2}(p_3,p_1,k)  = & \,\,
 \overline{u}_{\text{d}}(p_3)\*\gamma^{\alpha_2}\*\gamma_7\*
   [\slashed{p}_1 + \slashed{k}]^{-1}\*\gamma^{\beta_2}\*u_{\text{u}}(p_1) +  \\
  &\,\, \overline{u}_{\text{d}}(p_3)\*\gamma^{\beta_2}\*[\slashed{p}_3 - \slashed{k}]^{-1}\*
   \gamma^{\alpha_2}\*\gamma_7\*u_{\text{u}}(p_1)   \,.
\end{align*}
Zu zeigen ist nun für den Schleifenimpuls \(k\) des Gluons\footnote{Zu beachten ist, das
der Impulsfluss der Schleife für diese Rechnung anders gewählt wurde als in Abb. \ref{sum2boxes}.}:
\begin{align}
 M_{24}^{\alpha_1\beta_1}k_{\beta_1} & = 0 \,, \nonumber \\
 M_{13}^{\alpha_2\beta_2}k_{\beta_2} & = 0 \,.
 \label{WardId}
\end{align}
Die explizite Rechnung zeigt, dass Gl. \eqref{WardId} erfüllt ist:
\begin{align*}
 M_{24}\cdot k_{\beta_1} = & \,\,
 \overline{u}_{\text{t}}(p_4)\*\gamma^{\alpha_1}\*\gamma_7\*
   [\slashed{p}_2 - \slashed{k}]^{-1}\*\slashed{k}\*u_{\text{b}}(p_2) +  \\
 & \,\, \overline{u}_{\text{t}}(p_4)\*\slashed{k}\*[\slashed{p}_4 + \slashed{k} - m_{\text{t}}]^{-1}\*
   \gamma^{\alpha_1}\*\gamma_7\*u_{\text{b}}(p_2) \,,
\end{align*}
addiert man eine Null zum Schleifenimpuls \(k\) hinzu (analog zu Gl. \eqref{endweichung}), 
ergibt sich:
\begin{align*}
 M_{24}\cdot k_{\beta_1} = & \,\,
 \overline{u}_{\text{t}}(p_4)\*\gamma^{\alpha_1}\*\gamma_7\*
   [\slashed{p}_2 - \slashed{k}]^{-1}\*(\slashed{p}_2 -(\slashed{p}_2 - \slashed{k}))\*u_{\text{b}}(p_2) +  \\
 & \,\,\overline{u}_{\text{t}}(p_4)\*(\slashed{k}+\slashed{p}_4 -m_{\text{t}} -(\slashed{p}_4 -m_{\text{t}}))\*
   [\slashed{p}_4 + \slashed{k} - m_{\text{t}}]^{-1}\*
   \gamma^{\alpha_1}\*\gamma_7\*u_{\text{b}}(p_2)\\
 = & - \overline{u}_{\text{t}}(p_4)\*\gamma^{\alpha_1}\*\gamma_7\*u_{\text{b}}(p_2) + 
     \overline{u}_{\text{t}}(p_4)\*\gamma^{\alpha_1}\*\gamma_7\*u_{\text{b}}(p_2) \\
 = & \,\, 0\,.
\end{align*}
Ausgenutzt wurde hier die Dirac-Gleichung. Durch das Hinzufügen der Null ist es möglich
die Propagatoren zu kürzen und somit auf die Bornspinstruktur zu kommen.
Die zweite Fermionlinie \(M_{13}\) verhält sich analog und erzeugt die
Gl. \eqref{WardId}.

Nun betrachtet man den kollinearen Limes zum Beispiel in der ersten Box (Abb. \ref{4boxes}),
was \(k || p_2\) bedeutet.
Der Schleifenimpuls \(k\) kann dann in diesem Limes als Vielfaches 
des äußeren Impulses \(p_2\) geschrieben werden:
\begin{align}
 k = & \lambda \cdot p_2, \nonumber\\
 {\slashed{p}_2 - \slashed{k} \over (p_2 - k)^2}\*\gamma^{\beta_1}\*u_{\text{b}}(p_2)  = & 
 {(1-\lambda)\*\slashed{p}_2 \over (p_2 - k)^2}\*\gamma^{\beta_1}\*u_{\text{b}}(p_2) \nonumber\\
 = & {1-\lambda \over (p_2 - k)^2}\*\left(-\gamma^{\beta_1}\*\slashed{p}_2 +
      2\*p_2^{\beta_1} \right)\*u_{\text{b}}(p_2) \nonumber\\
 = & {2\*(1-\lambda) \over \lambda}\cdot {k^{\beta_1}\*u_{\text{b}}(p_2) \over (p_2 - k)^2}\,.
\label{eq:collinear} 
\end{align}
In der vorletzten Zeile von Gl. \eqref{eq:collinear} fällt der erste Summand aufgrund 
der Dirac-Gleichung weg, da die Masse des b Quarks zu Null angenommen wird:
\[\slashed{p}_2\*u_{\text{b}}(p_2) = \mathds{1}\* m_b\*u_{\text{b}}(p_2) = 0.\]
Man kann danach den äußeren Impuls wieder in den Schleifenimpuls zurückschreiben.
Aufgrund des Gluonpropagators \(\left(\sim g_{\beta_1\beta_2}\right)\) wird
bei der Multiplikation von \(M_{24}\) mit \(M_{13}\), \(M_{13}\) mit \(k_{\beta_2}\)
kontrahiert, was nach Gl. \eqref{WardId} ein Null generieren wird.
Der kollineare Grenzfall verschwindet also in der Kombination der verschiedenen
Boxdiagramme.

Die Fälle \(k || p_1,p_3\) sind analog zu dem eben gezeigten Fall \( k || p_2\).
Sobald ein externes Teilchen eine Masse besitzt, wie hier das \topq \, (zugehöriger 
Impuls \(p_4\)), tritt keine kollineare Divergenz auf.
Man kann Teilchen mit einer Masse in ihr Ruhesystem transformieren, so dass für das Skalarprodukt
des Viererimpulses mit dem Schleifenimpuls folgendes geschrieben werden kann:
\begin{align*}
 k\*p_4 &= E_k\*E_{p_4}, \,\, \text{mit  } \overrightarrow{p_4} = \overrightarrow{0} \,.                 
\end{align*}                                                       
Falls ein externes Teilchen Masse besitzt, treten demnach nur
softe Divergenzen auf. Im masselosen Fall hat man zusätzlich noch
kollineare Divergenzen\footnote{Man nennt diese Divergenzen aus diesem Grund auch
Massensingularitäten.} aufgrund des weiteren Freiheitsgrads des Winkels \(\vartheta\):
\begin{align*}
 k\*p_i &= E_k\*E_{p_i}\*(1-\cos\vartheta), \,\, \text{mit  } 
                    E_{p_i}^2 = \vert \overrightarrow{p_i}\vert^2 \,\,\, (i = 1,2,3)\,,
                    E_{k}^2 = \vert \overrightarrow{k}\vert^2  \,.                 
\end{align*}  

Die Summe der Boxdiagramme im \(t\)-Kanal (Amplitudenniveau) sollte keine kollinearen Divergenzen
mehr aufweisen. Numerisch konnte das verifiziert\footnote{Zur Berechnung der
\({1\over \epsilon^2}\)-Terme der Masterintegrale wurde {\tt QCDLoop} 
\cite{QCDLoop} verwendet.} 
werden: Die Größenordnung der Summe der \(\tfrac{1}{\epsilon^2}\)-Beiträge 
ist wesentlich kleiner als die Größenordnung der Einzelterme der Boxdiagramme,
wenn man die \(\mathcal{O}(\epsilon^{-2})\) mit einem künstlich hohem
Faktor (z.B. \(10^{50}\)) multipliziert, weist die Summe der 
\(\mathcal{O}(\epsilon^{-2})\)-Ordnung eine Größenordnung von rund \(10^{35}\) auf.
Die Amplitude der Boxdiagramme \(\ampli^{\text{\tiny Boxen}}\) befindet sich im Anhang \ref{ampli}.

An diesem Punkt sind nun alle notwendigen Beiträge zum \(\text{NLO}^2\)
bestimmt.
Bei den Vertizes treten nur zwei Spinstrukturen (\(S_1,S_3\)) auf. Die Boxdiagramme
weisen alle sieben Spinstrukturen in Gl. \eqref{spinstrukturen} auf.
Es treten in der hier verwendeteten Eichung keine UV-Divergenzen in den Boxdiagrammen auf.
In den Boxdiagrammen vorhandene Tensorintegrale wurden analog zu den Vertexdiagrammen
mit Hilfe der Passarino-Veltman-Reduktion (Kapitel \ref{pv}) zu skalaren Masterintegralen reduziert.

Die für die Berechnung verwendeten Feynmanregeln sind in Ref. \cite{Nachtmann} zu finden, wobei
für den \W- und Gluon-Propagator, wie hier beschrieben, die Feynman-ähnliche Eichung gewählt wurde.
Bei einer unabhängigen Rechnung unter Verwendung der Reduktionsmethode nach 
Davydychev \cite{Davydychev:1991va} für die Reduktion der Tensorintegrale
konnte dasselbe Ergebnis für die Amplitude bestimmt werden (siehe \ref{ampli}). 

	\section{Verwendetes Schema im Rahmen der dimensionalen Regularisierung}

Zur Berechnung des Amplitudenquadrats ist es notwendig festzulegen in welcher Art und Weise
die Spur im Rahmen der dimensionalen Regularisierung berechnet wird.
Wie im vorherigen Kapitel dargelegt, ist für die Form der Amplitude das verwendete
Schema der \(\gamma_5\)-Matrix ausschlaggebend.
Die UV-Divergenzen haben jedoch eine universelle Struktur: Bei der Verwendung
des Schemas nach Larin ist es daher notwendig den berechneten Larin-Counterterm für
die UV-Divergenzen mit einzubeziehen, um die durch die Verwendung des Schemas
erzeugten, überflüssigen, endlichen Terme \(\left({\epsilon \over \epsilon}\right)\) 
zu entfernen.
Die Verwendung eines bestimmten Schemas hat ebenfalls Einfluss auf die Form der 
IR-Divergenzen. Die Behandlung dieser im Rahmen einer Rechnung muss daher einheitlich in
einem Schema sein, damit sich die IR-Divergenzen (reell und virtuell)  einer Ordnung 
kompensieren können.

In dieser Arbeit wurde die Rechnung im folgenden Schema vorgenommen:
Um den Satz einer minimalen Anzahl von Spinstrukturen zu erhalten 
(sieben Stück im \textit{t}-Kanal), 
wurde eine antikommutierende \(\gamma_5\)-Matrix angenommen.
Die auftretenden UV-Divergenzen wurden im \(\overline{\text{MS}}\)-Schema mittels Countertermen
renormiert. Auf Amplitudenniveau im \textit{t}-Kanal 
treten lediglich UV-Divergenzen für die Vertexkorrekturen auf.
Nach der Renormierung wurde der Grenzwert \(d \rightarrow 4\) (bzw. \(\epsilon 
\rightarrow 0\)) gebildet
allerdings nur für \(d\)'s, die aus der \(\gamma\)-Algebra entstanden sind
(FDH ähnliches Schema).
Die \(d\)-Abhängigkeit aus der Reduktion der Tensorintegrale 
(nach Passarino und Veltman) wurde beibehalten.
Bei der Bildung des Amplitudenquadrats wurden die auftretenden Spuren in vier Dimensionen
berechnet, um keine Inkonsistenzen mit der \(\gamma_5\)-Matrix in \(d\) Dimensionen zu erhalten.
Ein Problem stellt dabei der Verlust der Zyklizität der Spur in \(d \neq 4\) dar \cite{Kreimer}. 
Eine weitere
Möglichkeit als die in dieser Arbeit vorgestellte Methode, ist die Methode nach Körner, Kreimer
und Schilcher \cite{Kreimer}. Diese soll hier nur erwähnt und nicht näher 
betrachtet werden.

Aufgrund der Farbe (wie oben beschrieben) mischen die Beiträge der Vertexkorrekturen nicht
mit denen aus den Korrekturen der Boxdiagramme.
Eine getrennte Berechnung der Beiträge des Amplitudenquadrats bietet sich daher an.

Auf Amplitudenniveau wurden die Spinstrukturen und die Masterintegrale
aus der Reduktion vom Rest separiert, so dass die Amplitude aus einer Koeffizientenmatrix
besteht, die mit einem Vektor der Spinstrukturen und einem Vektor der Integrale multipliziert
wird. Die Einträge des Integralvektors werden für die Ordnungen 
\(\epsilon^{-2}\), \(\epsilon^{-1}\) und \(\epsilon^0\) mit {\tt QCDLoop} 
\cite{QCDLoop} bestimmt,
höhere Ordnungen in \(\epsilon\) sind zunächst auf eins gesetzt.
Die Amplitude der Vertexdiagramme wird in folgender Weise strukturiert:
\begin{align}
\ampli^{\text{\tiny Vertizes}} 
= \ampli^{\text{\tiny 1. Vertex}} + \ampli^{\text{\tiny 2. Vertex}}
= f_{ijk} \cdot S_{j} \cdot I_{ik} ,
\label{eq:ProgrammAmpli} 
\end{align}
wobei \(i\) über die Anzahl der auftretenden Masterintegrale (für die Vertizes acht Integrale und
für die Boxdiagramme sind es 32) läuft, endliche Beiträge,
die im Zuge der Renormierung entstanden sind, werden formal auch als ein Integral betrachtet, welches
für die endliche Ordnung \(\epsilon^0\) immer den Wert eins und für alle anderen Ordnungen den
Wert null hat.
Der Index \(j\) läuft für die Vertexdiagramme über die zwei möglichen Spinstrukturen,
für die Boxdiagramme hingegen über die sieben möglichen Spinstrukturen.
Der Index  \(k\) gibt die Ordnung in \(\epsilon\) an.
Da die Spinstrukturen nach der Renormierung in vier Dimensionen betrachtet werden, 
hat \(S_j\) keine \(k\)-Abhängigkeit.

Das Amplitudenquadrat ergibt sich dann aus einer Matrixmultiplikation:
\begin{align}
\vert \ampli^{\text{\tiny Vertizes}} \vert^2 
= \left[(f_{ijk} I_{ik}) \times (f_{i'j'k'} I_{i'k'})\right] (S_{j}\times S_{j'}) .
\label{eq:ProgrammAmpliSq} 
\end{align}
Für das  Amplitudenquadrat der Boxen \(\vert \ampli^{\text{\tiny Boxen}} \vert^2 \)
ergibt sich eine analoge Schreibweise zu Gl. \eqref{eq:ProgrammAmpliSq}.

Die vollständige Liste der in dieser Rechnung auftretenden Integrale findet man
im Anhang \ref{MastersInQCDLoop}. Unter der Verwendung der Vertauschung von Impulsen
und Massen erhält man 16 unterschiedliche Masterintegrale. Im Vergleich mit 
der Rechnung der \topq-Paar-Erzeugung \cite{babis} ist das nur ein zusätzliches Integral.
Aufgrund der zusätzlich auftretenden Masse in der Produktion einzelner \topq s erwartet 
man mehr Masterintegrale.
Die Matrix und die Vektoren werden für eine bestimmte Kinematik ausgerechnet (Programm in {\tt C++}).
Mit Hilfe dieses Programms ist die Berechnung des Amplitudenquadrats für einen Phasenraumpunkt möglich.
Am Ende der Rechnung liegt das Ergebnis als Zahlenwert für verschiedene 
Ordnungen in \(\epsilon\) vor, woraus man einen totalen hadronischen Wirkungsquerschnitt
oder Verteilungen bestimmen kann.

Für z.B. folgenden Phasenraumpunkt mit den partonischen Impulsen:
\begin{align}
 p_1 &= (724;0;0;724), \nonumber \\
 p_2 &= (675;0;0;-675), \nonumber \\
 p_3 &= (707,1;-417,85;-94,11;563,55), \nonumber \\
 p_4 &= (691,9;417,85,94,11,-514,55), 
\end{align}
wobei \(p_4\) der Impuls des \topq s ist und alle Werte in GeV angegeben sind,
erhält man mit dem entwickelten {\tt C++}-Programm das Ergebnis
für die verschiedenen Ordnungen von \(\epsilon\) in Tabelle \ref{Tab:Benchmark}.
\begin{table}[h]
	\centering
	\begin{tabular}{cr@{,}l}
	\toprule
		Ordnung in \(\epsilon\) & \multicolumn{2}{c}{\(\tilde{C_1}^2\cdot \vert \ampli^{\text{\tiny Vertizes}} \vert^2 + 
                                          \tilde{C_2}^2\cdot \vert \ampli^{\text{\tiny Boxen}} \vert^2\)} \\
	\midrule
		\(\epsilon^{-4}\)   & \hspace{20mm} 0&2\\
		\(\epsilon^{-3}\)   &  \(-0\)&\(8\)\\
		\(\epsilon^{-2}\)   &  \(2\)&\(4\)\\
		\(\epsilon^{-1}\)   &  \(-8\)&\(7\cdot 10^{4}\)\\
		\(\epsilon^{0}\)    &  \(1\)&\(2\cdot 10^{10}\)\\
	\bottomrule
	\end{tabular}
	\caption
	{Beitrag der Einschleifenamplituden quadriert in \nnlo r Ordnung für den 
	\textit{t}-Kanal für einen Phasenraumpunkt ausgewertet.}        
	\label{Tab:Benchmark}
\end{table}
\section[Der tW-Kanal in \nnlo r Ordnung]
        {Der tW-Kanal in \nnlo r Ordnung}
        
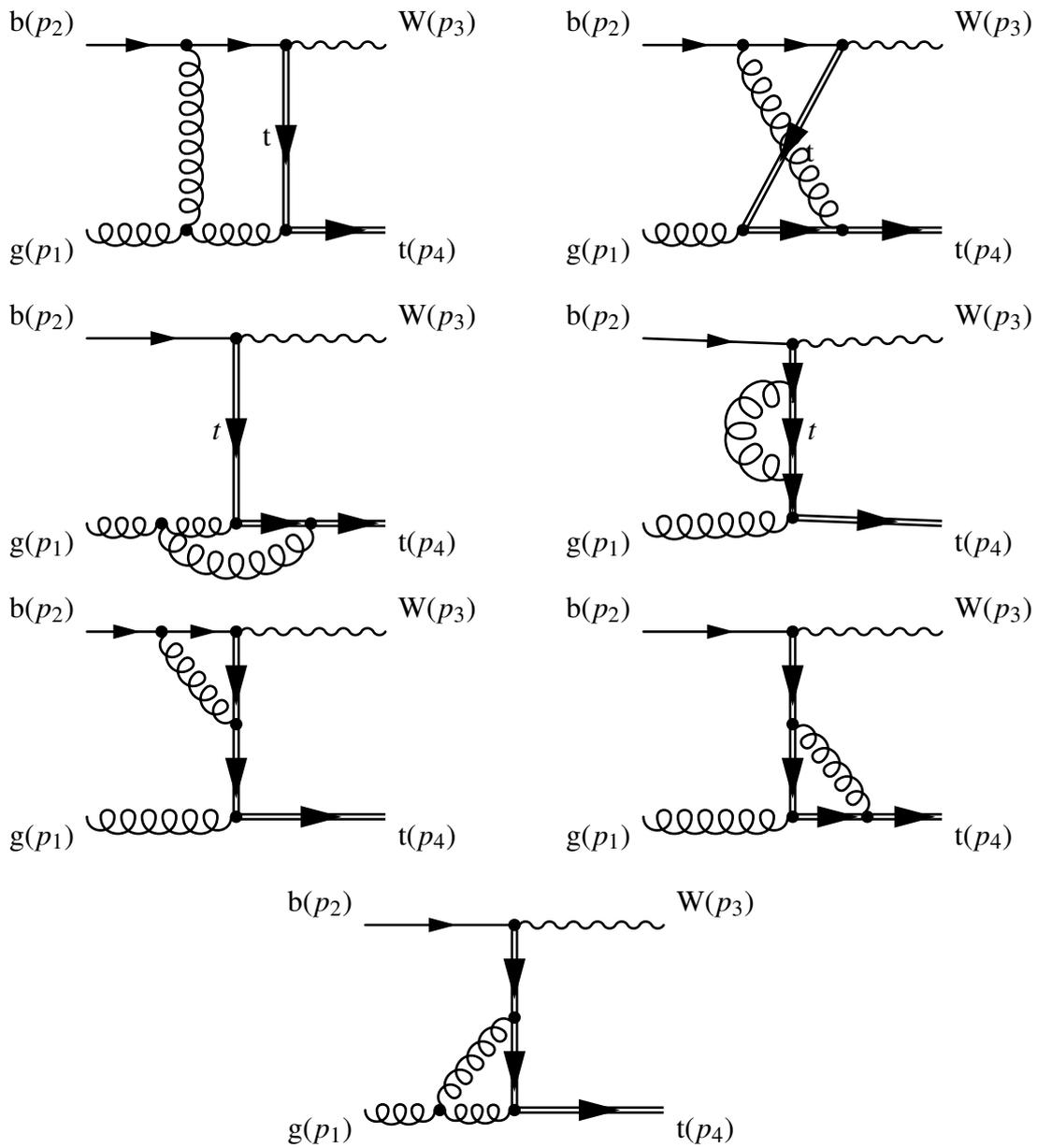
\begin{figure}[b]
\centering
 \begin{subfigure}[b]{0.4\textwidth}
  \centering
  \begin{fmffile}{tW2oneloop1}
  \begin{fmfgraph*}(70,35)
  \fmfleftn{i}{2} \fmfrightn{o}{2}
  \fmf{gluon}{i1,v3,v1}
  \fmf{heavy}{v1,o1}
  \fmf{heavy,tension=0,label=$\text{t}$}{v2,v1}
  \fmf{fermion}{i2,v4,v2}
  \fmf{boson}{v2,o2}
  \fmf{gluon,tension=0}{v3,v4}
  \fmflabel{$\text{g}(p_1)$}{i1} \fmflabel{$\text{b}(p_2)$}{i2}
  \fmflabel{$\text{W}(p_3)$}{o2} \fmflabel{$\text{t}(p_4)$}{o1}
  \fmfdotn{v}{4}
  \end{fmfgraph*}
 \end{fmffile}
 \vspace{15mm}
\end{subfigure}  
  \hspace{15mm}
  \begin{subfigure}[b]{0.4\textwidth}
  \centering
  \begin{fmffile}{tW2oneloop2}
   \begin{fmfgraph*}(70,35)
  \fmfleftn{i}{2} \fmfrightn{o}{2}
  \fmf{gluon}{i1,v1}
  \fmf{heavy}{v1,v4,o1}
  \fmf{heavy,tension=0,label=$\text{t}$}{v2,v1}
  \fmf{fermion}{i2,v3,v2}
  \fmf{boson}{v2,o2}
  \fmf{gluon,tension=0}{v3,v4}
  \fmflabel{$\text{g}(p_1)$}{i1} \fmflabel{$\text{b}(p_2)$}{i2}
  \fmflabel{$\text{W}(p_3)$}{o2} \fmflabel{$\text{t}(p_4)$}{o1}
  \fmfdotn{v}{4}
  \end{fmfgraph*}
 \end{fmffile}
  \vspace{15mm}
\end{subfigure} 
 \begin{subfigure}[b]{0.4\textwidth}
 \centering
 \begin{fmffile}{tW2oneloop3}
  \begin{fmfgraph*}(70,35)
  \fmfleftn{i}{2} \fmfrightn{o}{2}
  \fmf{gluon}{i1,v3,v1}
  \fmf{heavy}{v1,v4,o1}
  \fmf{heavy,tension=0,label=$t$}{v2,v1}
  \fmf{fermion}{i2,v2}
  \fmf{boson}{v2,o2}
  \fmf{gluon,left=-0.5,tension=0}{v3,v4}
  \fmflabel{$\text{g}(p_1)$}{i1} \fmflabel{$\text{b}(p_2)$}{i2}
  \fmflabel{$\text{W}(p_3)$}{o2} \fmflabel{$\text{t}(p_4)$}{o1}
  \fmfdotn{v}{4}
  \end{fmfgraph*}
 \end{fmffile}
 \vspace{15mm}
\end{subfigure} 
  \hspace{15mm}
  \begin{subfigure}[b]{0.4\textwidth}
  \centering
  \begin{fmffile}{tW2oneloop4}
   \begin{fmfgraph*}(70,35)
  \fmfleftn{i}{2} \fmfrightn{o}{2}
  \fmf{gluon}{i1,v1}
  \fmf{heavy}{v1,o1}
  \fmf{heavy,tension=0.2}{v2,v3}
  \fmf{heavy,tension=0.1,label=$t$}{v3,v4}
  \fmf{heavy,tension=0.2}{v4,v1}
  \fmf{fermion}{i2,v2}
  \fmf{boson}{v2,o2}
  \fmf{gluon,left=-1.5,tension=0.1}{v3,v4}
  \fmflabel{$\text{g}(p_1)$}{i1} \fmflabel{$\text{b}(p_2)$}{i2}
  \fmflabel{$\text{W}(p_3)$}{o2} \fmflabel{$\text{t}(p_4)$}{o1}
  \fmfdotn{v}{2}
  \end{fmfgraph*}
 \end{fmffile}
  \vspace{15mm}
\end{subfigure} 
 \begin{subfigure}[b]{0.4\textwidth}
  \centering
 \begin{fmffile}{tW2oneloop5}
  \begin{fmfgraph*}(70,35)
  \fmfleftn{i}{2} \fmfrightn{o}{2}
  \fmf{gluon}{i1,v1}
  \fmf{heavy}{v1,o1}
  \fmf{phantom}{i2,v2}
  \fmf{phantom,tension=0}{v2,v1}
  \fmf{boson}{v2,o2}
  \fmffreeze
  \fmf{fermion}{i2,v3,v2}
  \fmf{heavy}{v2,v4,v1}
  \fmffreeze
  \fmf{gluon}{v3,v4}
  \fmflabel{$\text{g}(p_1)$}{i1} \fmflabel{$\text{b}(p_2)$}{i2}
  \fmflabel{$\text{W}(p_3)$}{o2} \fmflabel{$\text{t}(p_4)$}{o1}
  \fmfdotn{v}{4}
  \end{fmfgraph*}
 \end{fmffile}
 \vspace{15mm}
\end{subfigure} 
  \hspace{15mm}
  \begin{subfigure}[b]{0.4\textwidth}
  \centering
  \begin{fmffile}{tW2oneloop6}
   \begin{fmfgraph*}(70,35)
  \fmfleftn{i}{2} \fmfrightn{o}{2}
  \fmf{gluon}{i1,v1}
  \fmf{phantom}{v1,o1}
  \fmf{fermion}{i2,v2}
  \fmf{phantom,tension=0}{v2,v1}
  \fmf{boson}{v2,o2}
  \fmffreeze
  \fmf{heavy}{v1,v3,o1}
  \fmf{heavy}{v2,v4,v1}
  \fmffreeze
  \fmf{gluon}{v3,v4}
  \fmflabel{$\text{g}(p_1)$}{i1} \fmflabel{$\text{b}(p_2)$}{i2}
  \fmflabel{$\text{W}(p_3)$}{o2} \fmflabel{$\text{t}(p_4)$}{o1}
  \fmfdotn{v}{4}
  \end{fmfgraph*}
 \end{fmffile}
 \vspace{15mm}
\end{subfigure} 
\begin{subfigure}[b]{0.4\textwidth}
  \centering
  \begin{fmffile}{tW2oneloop7}
  \begin{fmfgraph*}(70,35)
  \fmfleftn{i}{2} \fmfrightn{o}{2}
  \fmf{phantom}{i1,v1}
  \fmf{heavy}{v1,o1}
  \fmf{fermion}{i2,v2}
  \fmf{phantom,tension=0}{v2,v1}
  \fmf{boson}{v2,o2}
  \fmffreeze
  \fmf{gluon}{i1,v3,v1}
  \fmf{heavy}{v2,v4,v1}
  \fmffreeze
  \fmf{gluon}{v4,v3}
  \fmflabel{$\text{g}(p_1)$}{i1} \fmflabel{$\text{b}(p_2)$}{i2}
  \fmflabel{$\text{W}(p_3)$}{o2} \fmflabel{$\text{t}(p_4)$}{o1}
  \fmfdotn{v}{4}
  \end{fmfgraph*}
 \end{fmffile}
 \vspace{10mm}
\end{subfigure} 
 \caption[Einschleifendiagramme der tW-Produktionen]
         {Einschleifendiagramme der tW-Produktionen.}  
 \label{twNNLO}
\end{figure}

Bei der Bestimmung der quadrierten Einschleifenamplituden 
im tW-Kanal liegt strukturell ein anderes Problem als
in den anderen beiden Produktionskanälen vor.
Es gibt nur eine Fermionlinie und zwei Eichbosonen,
ein Gluon und ein \W, im Anfangs- bzw. Endzustand.
Für jede mögliche Variante der tW-Produktion
(siehe Abb. \ref{twproduktion}), existieren
sieben Einschleifendiagramme, welche für nur
eine Variante in Abb. \ref{twNNLO} dargestellt sind.

\setlength{\unitlength}{0.75mm}
Aufgrund der Struktur dieses Produktionskanals interferieren alle
Diagramme miteinander, so dass eine hohe Kombinatorik von Diagrammen
in der Berechnung zu finden ist.
Auch durch das Auftreten des 3-Gluon-Vertexes ist die Rechnung
im tW-Kanal aufwendiger.

Bei der Beschäftigung mit der tW-Produktion konnten
analog zum \textit{t}-Kanal in einem naiven Schema für 
die \(\gamma_5\)-Matrix unterschiedliche Spinstrukturen
ermittelt werden. Im Fall der tW-Produktion sind es
neun unterschiedliche Spinstrukturen:
\begin{align}
S^{\text{tW}}_1 &= \overline{u}_{\text{t}}(p_4)\*\gamma_6\*\slashed{p}_2\*u_{\text{b}}(p_1)\*\varepsilon_{\text{W}}^{*\mu}\*\varepsilon_{\text{g}}^{a\nu} \nonumber \,, \\
S^{\text{tW}}_2 &= \overline{u}_{\text{t}}(p_4)\*\gamma_6\*\gamma_\mu\*\gamma_\nu\*\slashed{p_2}\*u_{\text{b}}(p_1)\*\varepsilon_{\text{W}}^{*\mu}\*\varepsilon_{\text{g}}^{a\nu} \nonumber \,, \\ 
S^{\text{tW}}_3 &= \overline{u}_{\text{t}}(p_4)\*\gamma_6\*\gamma_\mu\*u_{\text{b}}(p_1)\*\varepsilon_{\text{W}}^{*\mu}\*\varepsilon_{\text{g}}^{a\nu} \nonumber \,, \\
S^{\text{tW}}_4 &= \overline{u}_{\text{t}}(p_4)\*\gamma_7\*u_{\text{b}}(p_1)\*\varepsilon_{\text{W}}^{*\mu}\*\varepsilon_{\text{g}}^{a\nu} \nonumber  \,, \\ 
S^{\text{tW}}_5 &=\overline{u}_{\text{t}}(p_4)\*\gamma_7\*\gamma_\mu\*\slashed{p}_2\*u_{\text{b}}(p_1)\*\varepsilon_{\text{W}}^{*\mu}\*\varepsilon_{\text{g}}^{a\nu} \nonumber \,, \\
S^{\text{tW}}_6 &= \overline{u}_{\text{t}}(p_4)\*\gamma_7\*\gamma_\mu\*\gamma_\nu\*u_{\text{b}}(p_1)\*\varepsilon_{\text{W}}^{*\mu}\*\varepsilon_{\text{g}}^{a\nu} \nonumber \,, \\
S^{\text{tW}}_7 &= \overline{u}_{\text{t}}(p_4)\*\gamma_6\*\gamma_\nu\*u_{\text{b}}(p_1)\*\varepsilon_{\text{W}}^{*\mu}\*\varepsilon_{\text{g}}^{a\nu} \nonumber \,, \\
S^{\text{tW}}_8 &= \overline{u}_{\text{t}}(p_4)\*\gamma_6\*\gamma_\nu\*\gamma_\mu\*\slashed{p}_2\*u_{\text{b}}(p_1)\*\varepsilon_{\text{W}}^{*\mu}\*\varepsilon_{\text{g}}^{a\nu} \nonumber \,, \\
S^{\text{tW}}_9 &= \overline{u}_{\text{t}}(p_4)\*\gamma_7\*\gamma_\nu\*\slashed{p}_2\*u_{\text{b}}(p_1)\*\varepsilon_{\text{W}}^{*\mu}\*\varepsilon_{\text{g}}^{a\nu} \,.
\end{align}
Der \textit{t}-Kanal stellt den größten Anteil am
gesamten Wirkungsquerschnitt, wie man in
der Abb. \ref{wqcms} und \ref{ratio} erkennen kann.
Zur vollständigen Beschreibung der Produktion einzelner \topq s
ist die Bestimmung der tW-Produktion natürlich unerlässlich.
Die Bestimmung des Anteils des \textit{t}-Kanals hat dem gegenüber aber
die höhere Priorität.

	\chapter{Diskussion der erzielten Ergebnisse}
 
 \begin{table}[b!]
	\centering
	\begin{tabular}{cc}
        \toprule
		\(\sqrt{s_{\text{\tiny Had}}}\) [TeV] &
		\(\sigmaup_{\text{\tiny Had}}\) [\(10^{10}\)pb] \\
                \midrule
		\(1{,}96\) & \(0{,}004\) \\
		\(7{,}0\)  & \(0{,}318\) \\
		\(8{,}0\)  & \(0{,}461\) \\
		\(13{,}0\) & \(1{,}626\) \\
		\(14{,}0\) & \(1{,}957\) \\
	\bottomrule
        \end{tabular}
	\caption
	{Hadronische Wirkungsquerschnitte der Produktion einzelner \topq s
	im \textit{t}-Kanal für den Beitrag der \nnlo n
	Ordnung (endlicher Anteil \(\mathcal{O}(\epsilon^0)\)) in Abhängigkeit von 
	verschiedenen Schwerpunktenergien \(\sqrt{s_{\text{\tiny Had}}}\)
	für pp-Kollisionen bei einer 
	\topq-Masse von \(m_{\text{t}} = \unit{172{,}5}{GeV}\),
	PDF-Set: MRST2004nnlo (\(\mu_F = \mu_R = m_{\text{t}}\)),
	Statistischer Fehler (Standardabweichung) liegt in der letzten 
	angegebenen Stelle.}        
	\label{Tab:nlosq}
\end{table}
In dieser Arbeit wurde der Beitrag der Einschleifenamplituden quadriert
in der Produktion einzelner \topq s in \nnlo r 
Ordnung im \textit{t}-Kanal bestimmt.

Zu Beginn wurden die 
führenden Ordnungen für die drei möglichen Produktionskanäle berechnet. 
Partonische und hadronische
Wirkungsquerschnitte wurden ebenfalls bestimmt und diskutiert.
Ergebnisse aus Ref. \cite{Kidonakis} konnten reproduziert werden.
Auch experimentelle Werte von ATLAS \cite{atlasWert} und CMS \cite{cmsWert}
stimmen im Rahmen ihrer Messfehler mit den hier berechneten hadronischen
Wirkungsquerschnitten überein.

Im darauffolgenden Kapitel wurden die virtuellen Beiträge zur
\nlo n Ordnung im \textit{t}-Kanal bestimmt.
Alle Konsistenzüberprüfungen dieser Rechnung waren erfolgreich.
Ein weiterer Vergleich für die virtuellen Beiträge in \nlo r Ordnung
ist noch in Bearbeitung.

Bei der Berechnung der quadrierten Einschleifenamplituden im Haupteil der Arbeit wurden 
neben der Passarino-Veltman-Reduktion  auch zwei weitere Methoden verwendet.
Die drei verschiedenen Rechnungen liefern identische Ergebnisse \cite{mohammad}.

Auf Amplitudenniveau wurde ein Vergleich mit einer zweiten Rechnung
im naiven Schema vorgenommen. Die hier bestimmte, 
gesamte Amplitude mit Vertex- und Boxdiagrammen
stimmt mit der unabhängigen Rechnung überein.
Theoretisch erwartetes Verhalten einzelner Ordnungen in \(\epsilon\)
konnte numerisch überprüft werden.
Teilergebnisse sind mit Ergebnissen aus der Literatur verglichen worden, wenn
diese vorhanden waren.

Durch die Anwendung der Clifford-Algebra und der Dirac-Gleichung in dem
naiven Schema für die \(\gamma_5\)-Matrix in \(d\) Dimensionen, erhält man sieben unterschiedliche
Spinstrukturen im \textit{t}-Kanal.
Die Möglichkeit einer unabhängigen Rechnung nach der Arbeit von Larin
ist möglich und wurde hier vorgestellt.
Der Vergleich beider Schemen auf Amplitudenquadratniveau für die endlichen
Beiträge ist noch in Arbeit
\cite{mohammad}.

Für die Berechnung der benötigten Integrale wurde {\tt QCDLoop} \cite{QCDLoop}
verwendet, obwohl diese Bibliothek nicht alle benötigten Ordnungen in \(\epsilon\)
bereitstellt. Die fehlenden Ordnungen der Masterintegrale wurden in dieser Arbeit 
zu eins gesetzt.
Ziel ist, die fehlenden Ordnungen in \(\epsilon\) der Masterintegrale zu bestimmen, 
so dass {\tt QCDLoop} durch eine um die fehlenden Ordnungen in \(\epsilon\)
erweiterte Bibliothek ersetzt werden kann.

Um ein numerisches Resultat zu erzeugen, was sehr relevant für den Vergleich
mit dem Experiment ist, wurde ein Programm in {\tt C++} entwickelt.
Dieses Programm berechnet zu einer bestimmten 
Kinematik den Beitrag der Einschleifenamplituden quadriert
im \textit{t}-Kanal. Das Amplitudenquadrat ist sortiert nach 
Ordnungen in \(\epsilon\).
Mit dem erzeugten Beitrag des Matrixelements ist es auch möglich, 
den Beitrag zum hadronischen Wirkungsquerschnitt (Tabelle \ref{Tab:nlosq}) zu bestimmen.
Die Prozedur verläuft analog zur Bestimmung hadronischer
Wirkungsquerschnitte in führender Ordnung (siehe Tabelle \ref{Tab:WqCms}).
In Tabelle \ref{Tab:nlosq} wurden für verschiedene Schwerpunktenergien
die endlichen Beiträge\footnote{Die IR-divergenten Anteile werden je nach verwendetem Schema
unterschiedliche Vorfaktoren aufweisen. Ein Vergleich bei der Verwendung verschiedener
Schemata ist somit nur für die endlichen Beiträge möglich.}
(\(\mathcal{O}(\epsilon^0)\)) der \nnlo n Ordnung am 
hadronischen Wirkungsquerschnitt bestimmt. Dabei wurden alle Ordnungen in \(\epsilon\)
der Masterintegrale ab \(\mathcal{O}(\epsilon^1)\) zu eins gesetzt.
Die bestimmten Werte
des hadronischen Wirkungsquerschnitts stellen somit 
keine realistischen Werte dar, demonstrieren jedoch die Möglichkeit zur
Berechnung der physikalischen Werte.
Die Werte der verwendeteten Konstanten im Programm, befinden sich im
Anhang \ref{A:numerik}, Gl. \eqref{eq:Konstanten}. 

Die Arbeit stellt damit unter anderem einen Beitrag zur \nnlo n 
Ordnung dar und ist somit in das Projekt\footnote{An diesem Projekt
sind noch folgende Personen beteiligt: Mohammad Assadsolimani, Dr. Philipp Kant,
Christoph Meyer, Sascha Peitzsch, 
Stefan Mölbitz, Tilmann Sult, Dr. Bas Tausk, Prof. Dr. Peter Uwer.} 
der Berechnung der gesamten \nnlo n Ordnung
der Produktion einzelner \topq s eingebettet.

Zum Schluss der Arbeit wurden noch die anderen Produktionskanäle besprochen:
In der Arbeit erfolgte die Berechnung des \textit{t}-Kanals für den LHC.
Durch Vertauschen von kinematischen Größen erhält man auch den
\textit{t}-Kanal mit Antiquarks sowie den \textit{s}-Kanal.
Der Beitrag des dritten Produktionkanals (tW-Produktion) wurde im Rahmen
dieser Arbeit nicht bestimmt, jedoch wurden die Grundlagen für
diese Berechnung gelegt.

	\appendix
	
	\chapter{Anhang}

\section{Alle verwendeten Größen}
\label{A:groessen}

Liste aller verwendeter Größen:
\begin{align*}
p_i &= \text{Teilchenimpuls des Teilchens $i$}\,,\\
k &= \text{übertragener Impuls, Impuls des Propagators, Schleifenimpuls}\,,\\
g^{\mu\nu} &= \text{metrischer Tensor mit}\, \text{diag}(1,-1,-1,-1)\,,\\
\gamma^\mu &= \text{Gamma-Matrix in Dirac-Basis}\,,\\
\gamma^5 &= \text{i}\gamma^0\gamma^1\gamma^2\gamma^3 \, \text{in Dirac-Basis}\,,\\
u_j(p_i) &= \text{Spinor des Teilchens $j$ mit Impuls $p_i$}\,,\\
\overline{u}_j(p_i) &= u^\dagger_j(p_i)\*\gamma_0\,,\\
v_j(p_i) &= \text{Spinor des Antiteilchens $j$ mit Impuls $p_i$}\,,\\
\overline{v}_j(p_i) &= v^\dagger_j(p_i)\*\gamma_0\,,\\
\varepsilon^*_\mu &= \text{Polarisationsvektor des \W s}\,,\\
\varepsilon^a_\nu &= \text{Polarisationsvektor des Gluons}\,,\\
\slashed{p}_i &= p_i^\mu\*\gamma_\mu\,,\\
V_{ij} &= \text{Element der Cabbibo-Kobayashi-Maskawa-Matrix}\,,\\
e &= \text{Elementarladung}\,,\\
\alpha &= {e^2 \over 4\*\pi}\,,\\
\alpha_s &= \text{Kopplungskonstante der QCD}\,,\\
\theta_W &= \text{Weinbergwinkel}\,,\\
g_s &= \sqrt{\alpha_s\*4\*\piup}\,,\\
m_{\text{t}} &= \text{Masse des \topq s}\,,\\
m_{\text{W}} &= \text{Masse des \W s}\,,\\
T^a &= \text{Generator der $\text{SU}(N)_{\text{Farbe}}$-Gruppe}\,,\\
a,b,c &= \text{Farbindizes}\,,\\
N &= \text{Dimension der Symmetriegruppe}\,,\\
\delta^{ab} = \delta_{ab} &= \text{Kroneckerdelta}\,,\\
C_F & = {N^2 - 1\over 2\*N} \,.
\end{align*}

\section{Mandelstam-Variablen}
\label{A:mandelstam}

Die folgenden Zusammenhänge zwischen den Teilchenimpulsen bezeichnet man
im Allgemeinen als Mandelstam-Variablen.
\begin{align*}
 s &= (p_1 + p_2)^2 = (p_3 + p_4)^2\,,\\
 t &= (p_1 - p_3)^2 = (p_2 - p_4)^2\,,\\
 u &= (p_1 - p_4)^2 = (p_2 - p_3)^2\,.
\end{align*}
Schreibt man diese aus, erhält man Ausdrücke für die Skalarprodukte der
Impulse der Teilchen in Abhängigkeit von den Teilchenmassen und den Mandelstam-Variablen.
\begin{align*}
 s &= (p_1 + p_2)^2 = p_1^2 + p_2^2 + 2\*p_1\*p_2\\
 &= m_1^2 + m_2^2 + 2\*p_1\*p_2\\
 \rightarrow p_1\*p_2 &= {1\over2}\*(s - m_1^2 - m_2^2)\,,\\
 s &= m_3^2 + m_4^2 + 2\*p_3\*p_4\\
 \rightarrow p_3\*p_4 &= {1\over2}\*(s - m_3^2 - m_4^2)\,.
\end{align*}
Entsprechend ergeben sich aus den anderen beiden Mandelstam-Variablen folgende
Zusammenhänge:
\begin{align*}
 t &= (p_1 - p_3)^2 = (p_2 - p_4)^2\\
 \rightarrow p_1\*p_3 &= -{1\over2}\*(t - m_1^2 - m_3^2)\\
 \rightarrow p_2\*p_4 &= -{1\over2}\*(t - m_2^2 - m_4^2)\,,\\
 u &= (p_1 - p_4)^2 = (p_2 - p_3)^2\\
 \rightarrow p_1\*p_4 &= -{1\over2}\*(u - m_1^2 - m_4^2)\\
 \rightarrow p_2\*p_3 &= -{1\over2}\*(u - m_2^2 - m_3^2)\,.
\end{align*}
\newpage
\enlargethispage{15cm}
\section[Hadronische Wirkungsquerschnitte für einen \ensuremath{\text{p} \bar{\text{p}}}-Beschleuniger]
        {Hadronische Wirkungsquerschnitte für einen \ensuremath{\mathbf{\text{p} \bar{\text{p}}}}-Beschleuniger}
\label{TevWQ}

\begin{figure}[h!]
\centering
\includegraphics[trim=0.8cm 0.5cm 0.5cm 0.1cm,width=0.65\textwidth,angle=0]{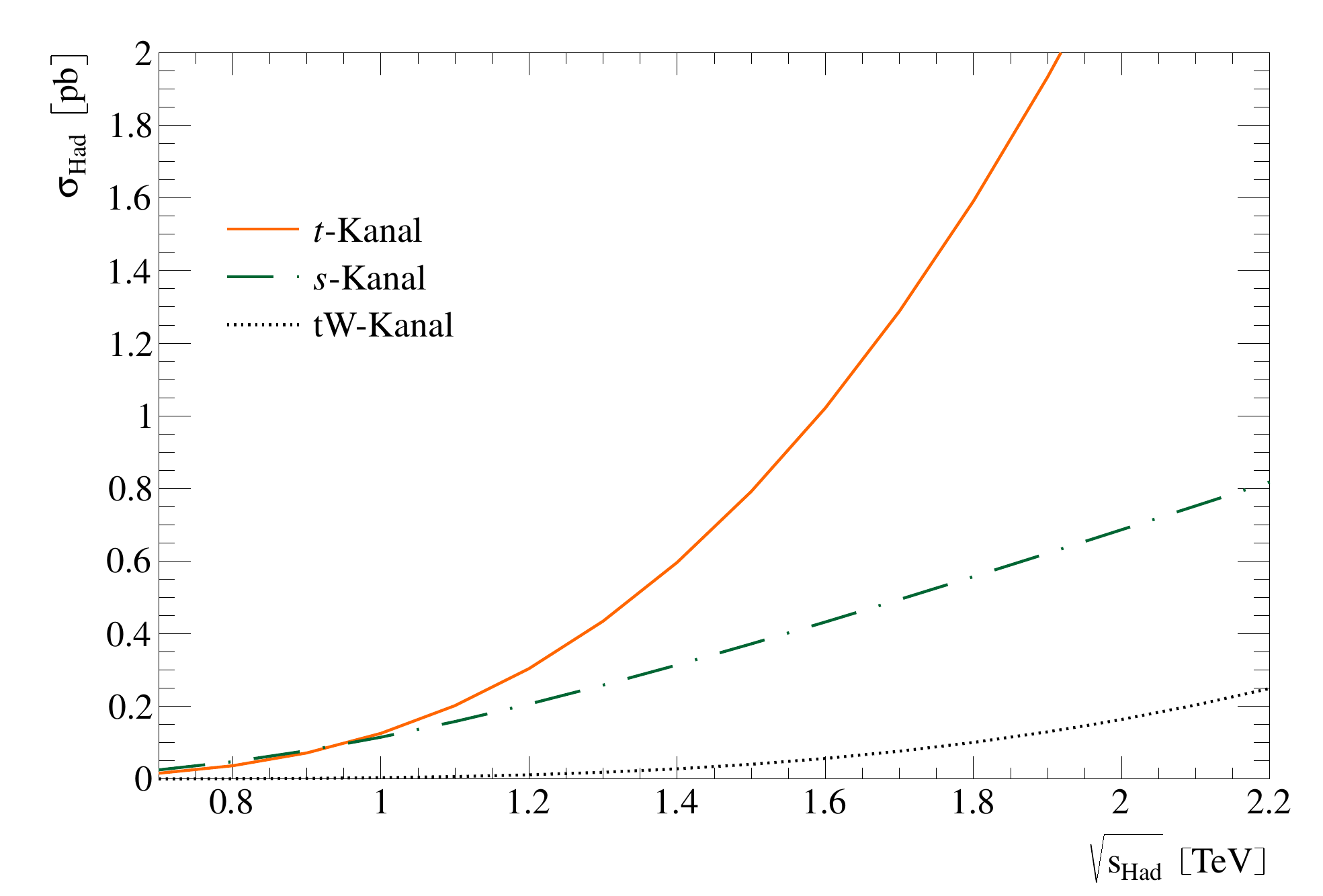}
\caption[Hadronischer Wirkungsquerschnitt in Abhängigkeit von der 
        Schwerpunktenergie für einen p$\bar{\text{p}}$-Beschleuniger]
        {Hadronischer Wirkungsquerschnitt in führender Ordnung in Abhängigkeit 
        von der Schwerpunktenergie bei einer angenommenen \topq-Masse 
        von $m_{\text{t}} = \unit{172{,}5}{GeV}$ für einen 
        p$\bar{\text{p}}$-Beschleuniger (Tevatron).
        Als PDF-Set wurde MRST2004nnlo genutzt (\(\mu_F = \mu_R = m_{\text{t}}\)).}
\label{Twqcms}
\end{figure}

\begin{figure}[h!]
\centering
\includegraphics[trim=0.8cm 0.5cm 0.5cm 0.1cm,width=0.65\textwidth,angle=0]{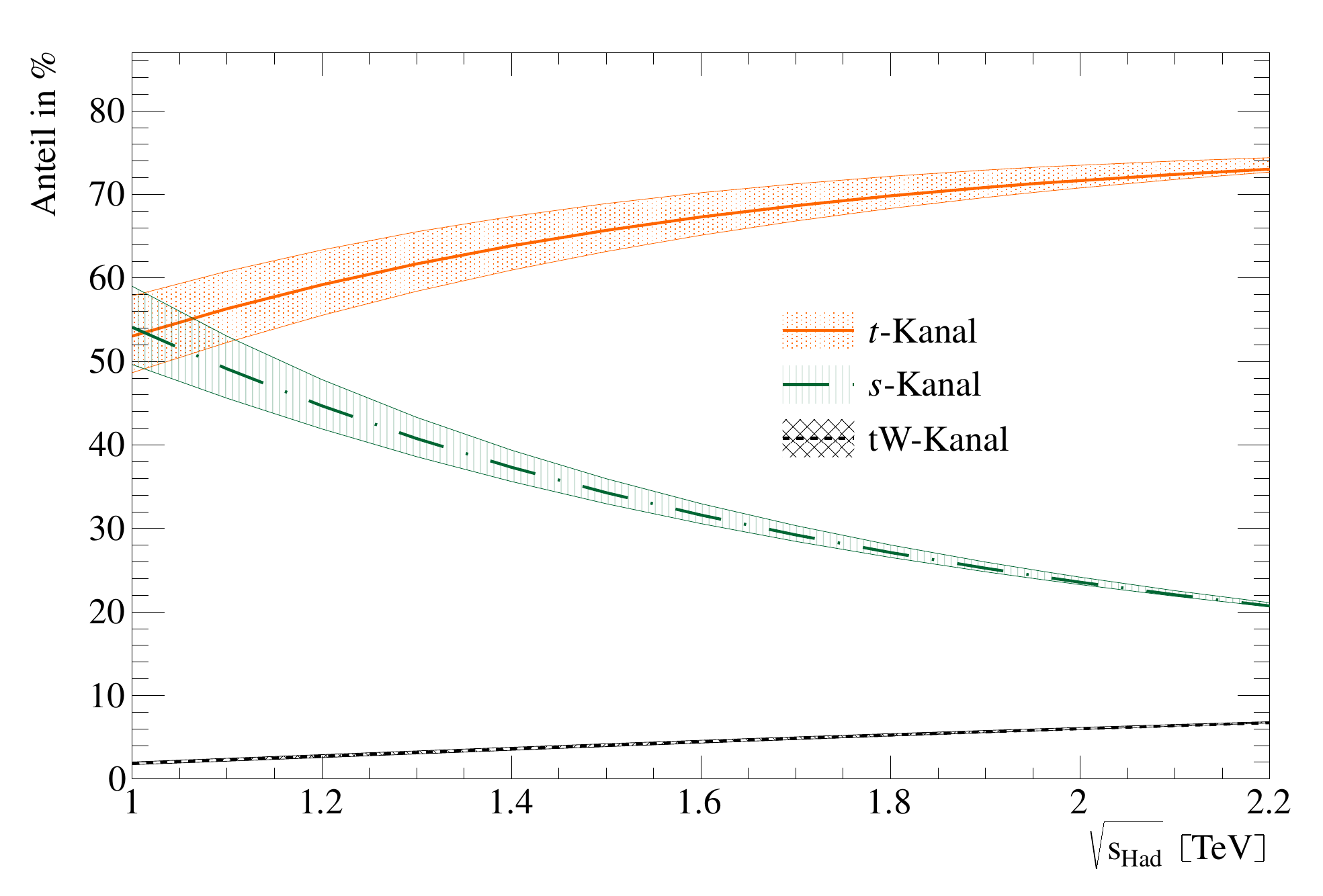}
\caption[Anteil der verschiedenen Produktionskanäle am gesamten hadronischen Wirkungsquerschnitt
         für einen p$\bar{\text{p}}$-Beschleuniger]
        {Anteil der verschiedenen Produktionskanäle am gesamten hadronischen Wirkungsquerschnitt
         in führender Ordnung abhängig von der Schwerpunktenergie bei einer angenommenen 
         \topq-Masse von $m_{\text{t}} = \unit{172{,}5}{GeV}$ für einen p$\bar{\text{p}}$-Beschleuniger 
         (Tevatron), zusätzlich ist die Standardabweichung der Werte (als Fehlerband) eingezeichnet.
         Als PDF-Set wurde MRST2004nnlo genutzt (\(\mu_F = \mu_R = m_{\text{t}}\)).}
\label{Tratio}
\end{figure}

\begin{figure}[h!]
\centering
\includegraphics[trim=0.8cm 0.5cm 0.5cm 0.1cm,width=0.65\textwidth,angle=0]{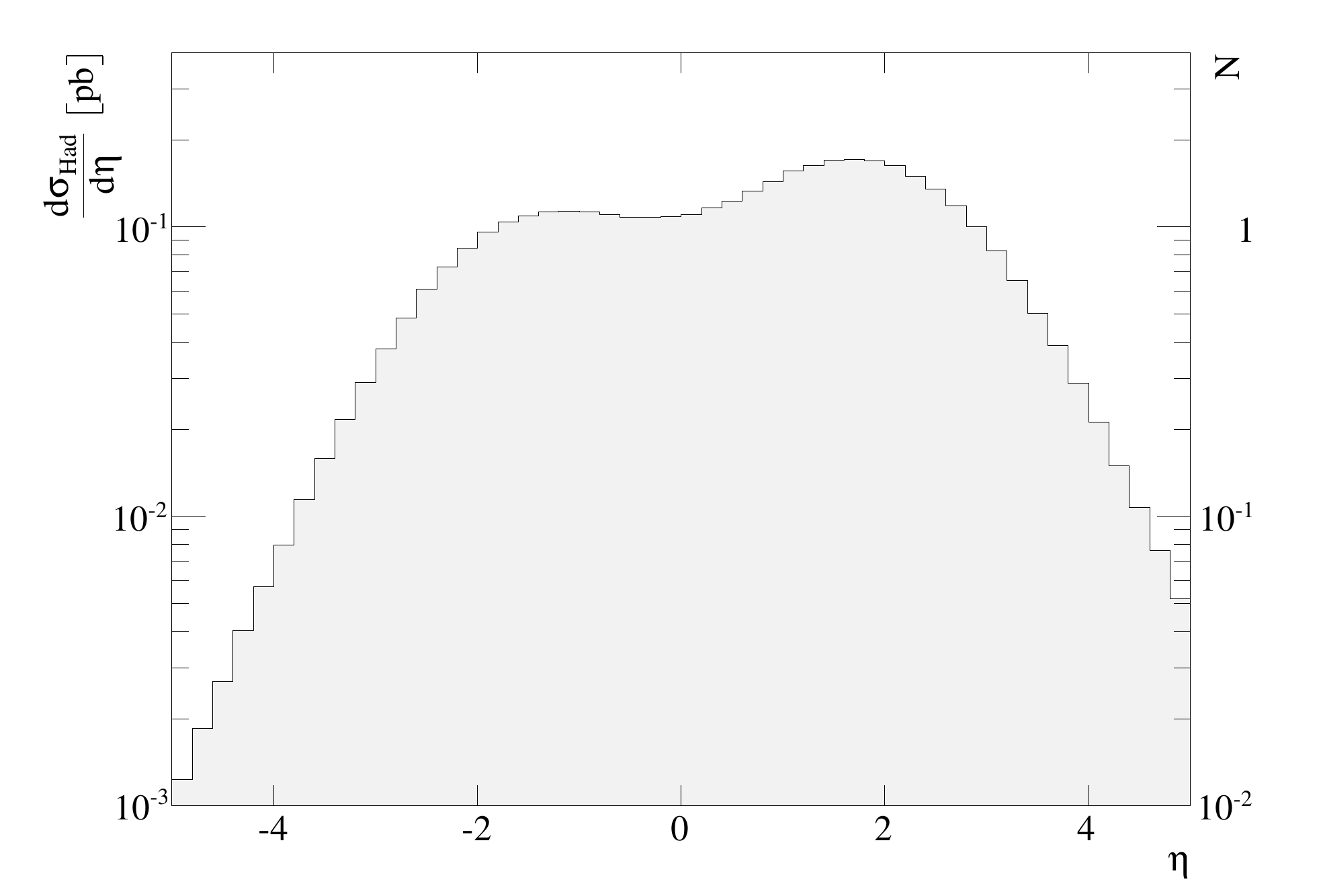}
\caption[Pseudorapidität des \topq s im \textit{t}-Kanal für einen p$\bar{\text{p}}$-Beschleuniger]
        {Pseudorapidität des \topq s im \textit{t}-Kanal bei einer
         Schwerpunktenergie von \(\sqrt{s_{\text{\tiny Had}}} = \unit{1{,}96}{TeV}\)
         für einen p$\bar{\text{p}}$-Beschleuniger (Tevatron).
         Auf der linken Ordinate ist der 
         differentielle Wirkungsquerschnitt, auf der rechten die erwartete Anzahl
         von Ereignissen bei einer angenommenen integrierten Luminosität von 
         \(\unit{100}{fb}^{-1}\) dargestellt.
         Als PDF-Set wurde MRST2004nnlo genutzt (\(\mu_F = \mu_R = m_{\text{t}}\)).}
\label{TevatronPseudorap}
\end{figure}

Die Werte der Abb. \ref{Twqcms} und \ref{Tratio} wurden mit \textsc{HatHor} \cite{hathor} 
bestimmt und die Abbildungen wurden mit {\tt ROOT} \cite{ROOT} erstellt (ebenso Abb. \ref{TevatronPseudorap}).
\newpage

\section{Lagrangedichte der QCD}

Hier steht die Lagrangedichte der QCD nach Ref. \cite{Nachtmann} aus
denen die verwendeten Feynmanregeln (aus \cite{Nachtmann})
abgeleitet wurden:
\begin{align}
\mathcal{L}_{\text{QCD}}(x) &= -{1\over 4}\*G^a_{\lambda\rho}(x)\*G^{\lambda\rho a}(x) + 
                              \sum_{j=1}^f \overline{q}^j\*\left[ 
                              \text{i}\gamma^\lambda\left(
                                \partial_\lambda + \text{i}g_s\*G_\lambda^a(x)\*{\lambda_a\over 2}
                              \right)-m_j\right]\*
                              q^j(x)\,, \nonumber\\
\text{mit } a &= 1,...,8,                              
\label{eq:LagrangeQCD}
\end{align}
wobei \(G^a_{\lambda\rho}(x)\) den Komponenten des Gluon-Feldstärke-Tensors
entspricht und \(\lambda_a\) den Gell-Mann-Matrizen entspricht,
\(f\) den verschiedenen Flavour-Quantenzahlen zu zuordnen ist, \(a\) die Kennzeichnung
der acht unterschiedlichen Gluonen darstellt und \(q_j\) die betrachteten Quarkfelder sind.
Der Lagrangian, der zur Ableitung der Feynmanregeln in unitärer Eichung für das Standardmodell
genutzt wurde, kann \cite[Glg. (22-123)]{Nachtmann} entnommen werden.

\section[Die Masterintegrale in der Notation von {\tt QCDLoop}]
        {Die Masterintegrale in der Notation von {\tt \textbf{QCDLoop}}}
\label{MastersInQCDLoop}

Die auftretenden Masterintegrale wurden mittels {\tt QCDLoop} \cite{QCDLoop} bis einschließlich
der endlichen Ordnung benutzt.
Die Notation nach Ref. \cite[Glg. (2.1)]{QCDLoop} ist die folgende (der Schleifenimpuls \(k\) wird
hier in \(l\) umbenannt):

\begin{align}
 I_1^D(m_1^2) &= \intmas {1\over (l^2 - m_1^2 +\text{i}\epsilon)} \,,\nonumber \\
 I_2^D(p_1^2;m_1^2,m_2^2) &= \intmas {1\over (l^2 - m_1^2 +\text{i}\epsilon)\*
                            \left((l+q_1)^2-m_2^2+\text{i}\epsilon\right)}\,, \nonumber \\
 I_3^D(p_1^2,p_2^2,p_3^2;m_1^2,m_2^2,m_3^2) &= \intmas  \,\,\,\times \nonumber  \\
                    & {1\over (l^2 - m_1^2 +\text{i}\epsilon)\*
                     \left((l+q_1)^2-m_2^2+\text{i}\epsilon\right)
                     \left((l+q_2)^2-m_3^2+\text{i}\epsilon\right)}\,, \nonumber \\   
 I_4^D(p_1^2,p_2^2,p_3^2,p_4^2;s_{12},S_{23};m_1^2,m_2^2,m_3^2,m_4^2) &= \intmas 
		      {1\over (l^2 - m_1^2 +\text{i}\epsilon)\* 
                     \left((l+q_1)^2-m_2^2+\text{i}\epsilon\right)}\,\,\, \times \nonumber \\
                     & {1 \over \left((l+q_2)^2-m_3^2+\text{i}\epsilon\right)
                     \left((l+q_3)^2-m_4^2+\text{i}\epsilon\right)}\,,                         
\end{align}
wobei \(q_n = \sum_{i=1}^n p_i\), \(s_{12} = (p_1+p_2)^2 \) und \(s_{23} = (p_2+p_3)^2\) sind.

Für den betrachteten Prozess der einzelnen \topq-Produktion im \(t\)-Kanal 
gilt folgende Kinematik:
\begin{align}
p_1+p_2+p_3+p_4 &= 0 \,, \nonumber\\
p_1^2 = p_2^2 = p_3^2 &= 0 \,,\nonumber\\
p_4^2 &= m_{\text{t}}^2 \,,\nonumber\\
(p_1+p_2)^2 = (p_3+p_4)^2 &= s \,,\nonumber\\
(p_1+p_3)^2 = (p_2+p_4)^2 &= t \,,\nonumber\\
(p_1+p_4)^2 = (p_2+p_3)^2 &= u \,.
\end{align}
1. Vertex:
\begin{align*}
B_0(2,3) &= I_2^D[(p_2+p_4)^2,0,0]\,,\\
C_0(1,2,3) &= I_3^D[p_1^2,(p_2+p_4)^2,p_3^2,0,0,0]\,.
\end{align*}
2. Vertex:
\begin{align*}
B_0(2,3) &= I_2^D[(p_1+p_3)^2,0,m_{\text{t}}^2]\,,\\
B_0(1,3) &= I_2^D[(p_1+p_2+p_3)^2 = (-p_4)^2,0,m_{\text{t}}^2]\,,\\
C_0(1,2,3) &= I_3^D[p_2^2,(p_1+p_3)^2,p_4^2,0,0,m_{\text{t}}^2]\,,\\
A(3) &= I_1^D[m_{\text{t}}^2]\,.
\end{align*}
1. Box:
\begin{align*}
B_0(1,3) &= I_2^D[(p_1+p_3)^2,0,m_{\text{W}}^2]\,,\\
B_0(2,4) &= I_2^D[(-p_1-p_2)^2,0,0]\,,\\
B_0(3,4) &= I_2^D[(-p_1-p_2-p_3)^2= p_4^2,m_{\text{W}}^2,0]\,,\\
C_0(1,2,3) &= I_3^D[p_1^2,p_3^2,(p_2+p_4)^2,0,0,m_{\text{W}}^2]\,,\\
C_0(1,2,4) &= I_3^D[p_1^2,(p_3+p_4)^2,p_2^2,0,0,0]\,,\\
C_0(1,3,4) &= I_3^D[(p_1+p_3)^2,p_4^2,p_2^2,0,m_{\text{W}}^2,0]\,,\\
C_0(2,3,4) &= I_3^D[p_3^2,p_4^2,(p_1+p_2)^2,0,m_{\text{W}}^2,0]\,,\\
D_0(1,2,3,4) &= I_4^D[p_1^2,p_3^2,p_4^2,(-p_2)^2,0,0,m_{\text{W}}^2,0]\,.
\end{align*}
2. Box:
\begin{align*}
B_0(1,3) &= I_2^D[(p_3+p_1)^2,0,m_{\text{W}}^2]\,,\\
B_0(1,4) &= I_2^D[(p_3+p_1+p_2)^2=(-p_4)^2,0,m_{\text{t}}^2]\,,\\
B_0(2,4) &= I_2^D[(p_1+p_2)^2,0,m_{\text{t}}^2]\,,\\
C_0(1,2,3) &= I_3^D[p_3^2,p_1^2,(p_2+p_4)^2,0,0,m_{\text{W}}^2]\,,\\
C_0(1,2,4) &= I_3^D[p_3^2,(p_1+p_2)^2,p_4^2,0,0,m_{\text{t}}^2]\,,\\
C_0(1,3,4) &= I_3^D[(p_3+p_1)^2,p_2^2,p_4^2,0,m_{\text{W}}^2,m_{\text{t}}^2]\,,\\
C_0(2,3,4) &= I_3^D[p_1^2,p_2^2,(p_3+p_4)^2,0,m_{\text{W}}^2,m_{\text{t}}^2]\,,\\
D_0(1,2,3,4) &= I_4^D[p_3^2,p_1^2,p_2^2,p_4^2,0,0,m_{\text{W}}^2,m_{\text{t}}^2]\,.
\end{align*}
3. Box:
\begin{align*}
B_0(1,3) &= I_2^D[(p_3+p_1)^2,0,m_{\text{W}}^2]\,,\\
B_0(2,4) &= I_2^D[(p_1+p_4)^2,0,0]\,,\\
B_0(3,4) &= I_2^D[p_4^2,m_{\text{W}}^2,0]\,,\\
C_0(1,2,3) &= I_3^D[p_3^2,p_1^2,(p_2+p_4)^2,0,0,m_{\text{W}}^2]\,,\\
C_0(1,2,4) &= I_3^D[p_3^2,(p_1+p_4)^2,p_2^2,0,0,0]\,,\\
C_0(1,3,4) &= I_3^D[(p_3+p_1)^2,p_4^2,p_2^2,0,m_{\text{W}}^2,0]\,,\\
C_0(2,3,4) &= I_3^D[p_1^2,p_4^2,(p_2+p_3)^2,0,m_{\text{W}}^2,0]\,,\\
D_0(1,2,3,4) &= I_4^D[p_3^2,p_1^2,p_4^2,p_2^2,0,0,m_{\text{W}}^2,0]\,.
\end{align*}
4. Box:
\begin{align}
B_0(1,3) &= I_2^D[(p_1+p_3)^2,0,m_{\text{W}}^2]\,,\nonumber\\
B_0(1,4) &= I_2^D[(p_1+p_3+p_2)^2=(-p_4)^2,0,m_{\text{t}}^2]\,,\nonumber\\
B_0(2,4) &= I_2^D[(p_3+p_2)^2,0,m_{\text{t}}^2]\,,\nonumber\\
C_0(1,2,3) &= I_3^D[p_1^2,p_3^2,(p_2+p_4)^2,0,0,m_{\text{W}}^2]\,,\nonumber\\
C_0(1,2,4) &= I_3^D[p_1^2,(p_3+p_2)^2,p_4^2,0,0,m_{\text{t}}^2]\,,\nonumber\\
C_0(1,3,4) &= I_3^D[(p_1+p_3)^2,p_2^2,p_4^2,0,m_{\text{W}}^2,m_{\text{t}}^2]\,,\nonumber\\
C_0(2,3,4) &= I_3^D[p_3^2,p_2^2,(p_1+p_4)^2,0,m_{\text{W}}^2,m_{\text{t}}^2]\,,\nonumber\\
D_0(1,2,3,4) &= I_4^D[p_1^2,p_3^2,p_2^2,p_4^2,0,0,m_{\text{W}}^2,m_{\text{t}}^2]\,.
\end{align}

\section[Werte der verwendeten Konstanten]{Werte der verwendeten Konstanten}
\label{A:numerik}

Zur Berechnung der Werte in den Tabellen \ref{Tab:WqCms} und \ref{Tab:nlosq} 
wurden in dem (im Rahmen dieser Arbeit entstandenen) {\tt C++}-Programm folgende
Werte der Konstanten verwendet:
\begin{align}
V_{ud} &= 0{,}97383\,,\nonumber \\
V_{us} &= 0{,}22720\,,\nonumber \\
V_{ub} &= 0{,}00396\,,\nonumber \\
V_{cd} &= 0{,}22710\,,\nonumber \\
V_{cs} &= 0{,}97296\,,\nonumber \\
V_{cb} &= 0{,}04221\,,\nonumber \\
V_{td} &= 0{,}00814\,,\nonumber \\
V_{ts} &= 0{,}04161\,,\nonumber \\
V_{tb} &= 0{,}99910\,,\nonumber \\
m_{\text{t}} &= \unit{172{,}5}{GeV}\,, \nonumber \\
m_{\text{W}} &= \unit{80{,}385}{GeV}\,, \nonumber \\
\text{Umrechnung }1\text{GeV}^{-2}\text{ in pb} &= 0.38937911\cdot 10^9\,
      \tfrac{\text{pb}}{\text{GeV}^{-2}}\,, \nonumber \\
\text{Elementarladung } e &=  \unit{1.602176487\cdot 10^{-19}}{C}\,, \nonumber \\
\alpha &= \frac{1}{132.2332298}\,, \nonumber \\
\alpha_s\,(\text{ bei }m_{\text{t}})\, & = 0{,}106461\,,\nonumber \\
g_s &= \sqrt{\alpha_s 4 \piup} = 1{,}15664\,.
\label{eq:Konstanten}
\end{align}

\section[Amplitude der Boxdiagramme für \ensuremath{\text{NLO}^2}-Beiträge im \textit{t}-Kanal]
        {Amplitude der Boxdiagramme für \ensuremath{\mathbf{\text{NLO}^2}}-Beiträge im \textit{t}-Kanal}
\label{ampli}
        
Hier steht die gesamte Amplitude der Boxdiagramme, wobei
\(\tilde{C_2} = \tfrac{1}{2\*N}C_2 - \tfrac{1}{N^2} C_2\) 
den Farbfaktoren der Boxen entspricht (\(C_1\) und \(C_2\) siehe
Gl. \eqref{C1} und \eqref{C2}). Die Koeffizienten 
\(S_1\), \(S_2\), \(S_3\), \(S_4\), \(S_5\), \(S_6\), \(S_7\) sind 
die Spinstrukturen aus Gl. \eqref{spinstrukturen}. 
Die Masterintegrale haben hier zusätzlich zu der
Nummerierung ihrer Propagatoren noch die Angabe aus welchem Diagramm sie ursprünglich
stammen (\(b_1 \hat{=} \text{ 1. Box}\), \(b_2 \hat{=} \text{ 2. Box}\),
\(b_3 \hat{=} \text{ 3. Box}\), \(b_4 \hat{=} \text{ 4, Box}\)). 
Die Definitionen der Masterintegrale sind im Anhang \ref{MastersInQCDLoop} 
angegeben. Die Variablen \(s\) und \(t\) entsprechen den Mandelstam-Variablen
(siehe Anhang \ref{A:mandelstam}), \(m_{\text{W}}\) ist die \W-Masse und \(m_{\text{t}}\) entspricht
der \topq-Masse.
Formal ergibt sich die Amplitude der Boxdiagramme aus der Summe der Amplituden
der einzelnen Boxdiagramme:
\begin{align*}
\tilde{C_2}\ampli^{\text{\tiny Boxen}}  = \tilde{C_2} {\alpha\*\alpha_s\*V_{\text{tb}}\*V_{\text{ud}}^* 
                                \over 8\*\sin^2 \theta_W\*(t-m_{\text{W}}^2)}\left[
                                \ampli^{\text{\tiny 1. Box}} + \ampli^{\text{\tiny 2. Box}} + 
                                \ampli^{\text{\tiny 3. Box}} + \ampli^{\text{\tiny 4. Box}}\right]\,.
\end{align*}
Nachfolgend sind die Amplituden der Boxdiagramme einzeln angegeben:
\\ 
\\
{\scriptsize
\begin{math}
\ampli^{\text{\tiny 1. Box}} =  
D_0(b_1)\*(((-2 + d)\*m_{\text{t}}\*s\*S_5\*(m_{\text{W}}^2 - t)^2)/(4\*(-3 + d)\*t\*(-m_{\text{t}}^2 + s + t)^2) + 
   (s\*S_7\*(m_{\text{W}}^2 - t)^2)/(4\*(-3 + d)\*t\*(-m_{\text{t}}^2 + s + t)) + 
   (S_6\*(m_{\text{W}}^2 - t)^2\*(-((-4 + d)\*m_{\text{t}}^2) - 2\*s + (-4 + d)\*t))/
    (4\*(-3 + d)\*t\*(-m_{\text{t}}^2 + s + t)^2) - 
   (m_{\text{t}}\*s\*S_1\*(m_{\text{W}}^2 - t)\*(2\*(-3 + d)\*m_{\text{t}}^2 + (-2 + d)\*m_{\text{W}}^2 + 6\*s - 2\*d\*s + 8\*t - 
      3\*d\*t))/((-3 + d)\*t\*(-m_{\text{t}}^2 + s + t)^2) - 
   (2\*S_2\*(m_{\text{W}}^2 - t)\*((-4 + d)\*(s - t)\*t + m_{\text{W}}^2\*((-2 + d)\*s + (-4 + d)\*t)))/
    ((-3 + d)\*t^2\*(-m_{\text{t}}^2 + s + t)) + 
   (m_{\text{t}}\*S_4\*(m_{\text{W}}^2 - t)\*(-(t\*(s + (-4 + d)\*t)) + 
      m_{\text{W}}^2\*((-2 + d)\*s + (-4 + d)\*t)))/((-3 + d)\*t^2\*(-m_{\text{t}}^2 + s + t)) + 
   (s\*S_3\*((-3 + d)\*m_{\text{t}}^4\*t\*(m_{\text{W}}^2 + t) + (s + t)\*(2\*m_{\text{W}}^2\*t\*(-s + t) + 
        m_{\text{W}}^4\*((-2 + d)\*s + (-4 + d)\*t) + t^2\*((-2 + d)\*s + (-4 + d)\*t)) - 
      m_{\text{t}}^2\*(m_{\text{W}}^4\*((-2 + d)\*s - 2\*t) + t^2\*((-5 + 2\*d)\*s + (-5 + d)\*t) + 
        m_{\text{W}}^2\*t\*((-5 + d)\*s + (-5 + 3\*d)\*t))))/
    ((-3 + d)\*t^2\*(-m_{\text{t}}^2 + s + t)^2)) + 
 ((4\*m_{\text{t}}\*S_1)/((m_{\text{t}}^2 - t)\*(m_{\text{t}}^2 - s - t)) - 
   (m_{\text{t}}\*S_5)/((m_{\text{t}}^2 - t)\*(m_{\text{t}}^2 - s - t)) - (4\*S_3\*(s + t))/
    ((m_{\text{t}}^2 - t)\*(m_{\text{t}}^2 - s - t)) - S_6/(s\*(-m_{\text{t}}^2 + s + t)) - 
   (8\*S_2)/(m_{\text{t}}^2\*s - s\*t) + (4\*m_{\text{t}}\*S_4)/(m_{\text{t}}^2\*s - s\*t))\*B_0(b_1,1,3) + 
 ((4\*m_{\text{t}}\*S_1)/((m_{\text{t}}^2 - s)\*(m_{\text{t}}^2 - s - t)) - 
   (m_{\text{t}}\*S_5)/((m_{\text{t}}^2 - s)\*(m_{\text{t}}^2 - s - t)) - (4\*S_3\*(s + t))/((m_{\text{t}}^2 - s - t)\*t) + 
   (8\*S_2)/(m_{\text{t}}^2\*t - s\*t) - (4\*m_{\text{t}}\*S_4)/(m_{\text{t}}^2\*t - s\*t) + 
   S_6/(m_{\text{t}}^4 - 2\*m_{\text{t}}^2\*s + s^2 - m_{\text{t}}^2\*t + s\*t))\*B_0(b_1,2,4) + 
 (-((m_{\text{t}}^2\*S_6)/((m_{\text{t}}^2 - s)\*s\*(m_{\text{t}}^2 - s - t))) + (m_{\text{t}}\*S_5\*(2\*m_{\text{t}}^2 - s - t))/
    ((m_{\text{t}}^2 - s)\*(m_{\text{t}}^2 - t)\*(m_{\text{t}}^2 - s - t)) + (4\*m_{\text{t}}^3\*S_4\*(s - t))/
    ((m_{\text{t}}^2 - s)\*s\*(m_{\text{t}}^2 - t)\*t) + (8\*m_{\text{t}}^2\*S_2\*(s - t))/
    ((m_{\text{t}}^2 - s)\*s\*t\*(-m_{\text{t}}^2 + t)) + (4\*m_{\text{t}}^2\*S_3\*(s + t))/
    ((m_{\text{t}}^2 - t)\*(m_{\text{t}}^2 - s - t)\*t) + (4\*m_{\text{t}}\*S_1\*(-2\*m_{\text{t}}^2 + s + t))/
    ((m_{\text{t}}^2 - s)\*(m_{\text{t}}^2 - t)\*(m_{\text{t}}^2 - s - t)))\*B_0(b_1,3,4) + 
 ((S_7\*(-m_{\text{W}}^2 + t))/(4\*(-3 + d)\*(m_{\text{t}}^2 - s - t)) + ((-2 + d)\*m_{\text{t}}\*S_5\*(m_{\text{W}}^2 - t))/
    (4\*(-3 + d)\*(-m_{\text{t}}^2 + s + t)^2) + 
   (S_6\*(m_{\text{W}}^2 - t)\*(-((-4 + d)\*m_{\text{t}}^2) - 2\*s + (-4 + d)\*t))/
    (4\*(-3 + d)\*s\*(-m_{\text{t}}^2 + s + t)^2) - 
   (m_{\text{t}}\*S_1\*(2\*(-3 + d)\*m_{\text{t}}^2 + (-2 + d)\*m_{\text{W}}^2 + 6\*s - 2\*d\*s + 8\*t - 3\*d\*t))/
    ((-3 + d)\*(-m_{\text{t}}^2 + s + t)^2) - 
   (2\*S_2\*((-4 + d)\*(s - t)\*t + m_{\text{W}}^2\*((-2 + d)\*s + (-4 + d)\*t)))/
    ((-3 + d)\*s\*t\*(-m_{\text{t}}^2 + s + t)) + 
   (m_{\text{t}}\*S_4\*(-(t\*(s + (-4 + d)\*t)) + m_{\text{W}}^2\*((-2 + d)\*s + (-4 + d)\*t)))/
    ((-3 + d)\*s\*t\*(-m_{\text{t}}^2 + s + t)) + 
   (S_3\*(-((-3 + d)\*m_{\text{t}}^4\*t) + (m_{\text{W}}^2 - t)\*(s + t)\*((-2 + d)\*s + (-4 + d)\*t) + 
      m_{\text{t}}^2\*(m_{\text{W}}^2\*(-((-2 + d)\*s) + 2\*t) + t\*((-5 + 2\*d)\*s + (-5 + d)\*t))))/
    ((-3 + d)\*t\*(-m_{\text{t}}^2 + s + t)^2))\*C_0(b_1,1,2,3) + 
 (((-2 + d)\*m_{\text{t}}\*s\*S_5\*(m_{\text{W}}^2 - t))/(4\*(-3 + d)\*t\*(-m_{\text{t}}^2 + s + t)^2) - 
   (s\*S_7\*(-m_{\text{W}}^2 + t))/(4\*(-3 + d)\*t\*(-m_{\text{t}}^2 + s + t)) + 
   (S_6\*(m_{\text{W}}^2 - t)\*(-((-4 + d)\*m_{\text{t}}^2) - 2\*s + (-4 + d)\*t))/
    (4\*(-3 + d)\*t\*(-m_{\text{t}}^2 + s + t)^2) - 
   (m_{\text{t}}\*s\*S_1\*(2\*(-3 + d)\*m_{\text{t}}^2 + (-2 + d)\*m_{\text{W}}^2 + 6\*s - 2\*d\*s + 8\*t - 3\*d\*t))/
    ((-3 + d)\*t\*(-m_{\text{t}}^2 + s + t)^2) - 
   (2\*S_2\*((-4 + d)\*(s - t)\*t + m_{\text{W}}^2\*((-2 + d)\*s + (-4 + d)\*t)))/
    ((-3 + d)\*t^2\*(-m_{\text{t}}^2 + s + t)) + 
   (m_{\text{t}}\*S_4\*(-(t\*(s + (-4 + d)\*t)) + m_{\text{W}}^2\*((-2 + d)\*s + (-4 + d)\*t)))/
    ((-3 + d)\*t^2\*(-m_{\text{t}}^2 + s + t)) + 
   (s\*S_3\*((-3 + d)\*m_{\text{t}}^4\*t + (s + t)\*(m_{\text{W}}^2\*((-2 + d)\*s + (-4 + d)\*t) + 
        t\*((-4 + d)\*s + (-2 + d)\*t)) + m_{\text{t}}^2\*(m_{\text{W}}^2\*(2\*s - d\*s + 2\*t) + 
        t\*(7\*s - 2\*d\*s + 7\*t - 3\*d\*t))))/((-3 + d)\*t^2\*(-m_{\text{t}}^2 + s + t)^2))\*
  C_0(b_1,1,2,4) + (((-2 + d)\*m_{\text{t}}\*S_5\*(m_{\text{t}}^2 - t)\*(-m_{\text{W}}^2 + t))/
    (4\*(-3 + d)\*t\*(-m_{\text{t}}^2 + s + t)^2) + (S_7\*(m_{\text{t}}^2 - t)\*(-m_{\text{W}}^2 + t))/
    (4\*(-3 + d)\*t\*(-m_{\text{t}}^2 + s + t)) + 
   (S_6\*(m_{\text{t}}^2 - t)\*(m_{\text{W}}^2 - t)\*((-4 + d)\*m_{\text{t}}^2 + 2\*s - (-4 + d)\*t))/
    (4\*(-3 + d)\*s\*t\*(-m_{\text{t}}^2 + s + t)^2) + 
   (m_{\text{t}}\*S_1\*(m_{\text{t}}^2 - t)\*(2\*(-3 + d)\*m_{\text{t}}^2 + (-2 + d)\*m_{\text{W}}^2 + 6\*s - 2\*d\*s + 8\*t - 
      3\*d\*t))/((-3 + d)\*t\*(-m_{\text{t}}^2 + s + t)^2) + 
   (2\*S_2\*(m_{\text{t}}^2 - t)\*((-4 + d)\*(s - t)\*t + m_{\text{W}}^2\*((-2 + d)\*s + (-4 + d)\*t)))/
    ((-3 + d)\*s\*t^2\*(-m_{\text{t}}^2 + s + t)) - 
   (m_{\text{t}}\*S_4\*(m_{\text{t}}^2 - t)\*(-(t\*(s + (-4 + d)\*t)) + 
      m_{\text{W}}^2\*((-2 + d)\*s + (-4 + d)\*t)))/((-3 + d)\*s\*t^2\*(-m_{\text{t}}^2 + s + t)) - 
   (S_3\*((-3 + d)\*m_{\text{t}}^6\*t - (m_{\text{W}}^2 - t)\*t\*(s + t)\*((-2 + d)\*s + (-4 + d)\*t) + 
      m_{\text{t}}^4\*(m_{\text{W}}^2\*(-((-2 + d)\*s) + 2\*t) + t\*((7 - 2\*d)\*s - 2\*(-2 + d)\*t)) + 
      m_{\text{t}}^2\*(t\*((-4 + d)\*s^2 - s\*t + 3\*t^2) + 
        m_{\text{W}}^2\*((-2 + d)\*s^2 + (-8 + 3\*d)\*s\*t + (-6 + d)\*t^2))))/
    ((-3 + d)\*t^2\*(-m_{\text{t}}^2 + s + t)^2))\*C_0(b_1,1,3,4) + 
 ((S_7\*(s\*t - m_{\text{t}}^2\*(m_{\text{W}}^2 + t) + m_{\text{W}}^2\*(s + 2\*t)))/(4\*(-3 + d)\*(m_{\text{t}}^2 - s - t)\*
     t) + (m_{\text{t}}\*S_5\*((-2 + d)\*s^2\*t + (-2 + d)\*m_{\text{t}}^4\*(m_{\text{W}}^2 + t) + 
      m_{\text{W}}^2\*((-2 + d)\*s^2 + 2\*(-8 + 3\*d)\*s\*t + 4\*(-3 + d)\*t^2) - 
      2\*m_{\text{t}}^2\*((-2 + d)\*s\*t + m_{\text{W}}^2\*((-2 + d)\*s + (-8 + 3\*d)\*t))))/
    (4\*(-3 + d)\*(m_{\text{t}}^2 - s)\*t\*(-m_{\text{t}}^2 + s + t)^2) - 
   (m_{\text{t}}\*S_1\*(2\*(-3 + d)\*m_{\text{t}}^6 + m_{\text{t}}^4\*((-2 + d)\*m_{\text{W}}^2 - 6\*(-3 + d)\*s + 
        (16 - 5\*d)\*t) + s\*(-2\*(-3 + d)\*s^2 + (16 - 5\*d)\*s\*t - 
        4\*(-3 + d)\*t^2) + m_{\text{W}}^2\*((-2 + d)\*s^2 + 2\*(-8 + 3\*d)\*s\*t + 
        4\*(-3 + d)\*t^2) - 2\*m_{\text{t}}^2\*(-3\*(-3 + d)\*s^2 + (16 - 5\*d)\*s\*t - 
        2\*(-3 + d)\*t^2 + m_{\text{W}}^2\*((-2 + d)\*s + (-8 + 3\*d)\*t))))/
    ((-3 + d)\*(m_{\text{t}}^2 - s)\*t\*(-m_{\text{t}}^2 + s + t)^2) + 
   (S_6\*(-((-4 + d)\*m_{\text{t}}^6\*(m_{\text{W}}^2 + t)) + m_{\text{t}}^4\*(t\*(2\*(-5 + d)\*s + (-4 + d)\*t) + 
        m_{\text{W}}^2\*(2\*(-5 + d)\*s + 3\*(-4 + d)\*t)) + 
      m_{\text{t}}^2\*(s\*t\*(-((-8 + d)\*s) - 2\*(-4 + d)\*t) + 
        m_{\text{W}}^2\*(-((-8 + d)\*s^2) + 8\*s\*t - 2\*(-4 + d)\*t^2)) + 
      s\*(s\*t\*(-2\*s + (-4 + d)\*t) + m_{\text{W}}^2\*(-2\*s^2 + (4 - 3\*d)\*s\*t - 
          2\*(-2 + d)\*t^2))))/(4\*(-3 + d)\*(m_{\text{t}}^2 - s)\*s\*t\*(-m_{\text{t}}^2 + s + t)^2) + 
   (2\*S_2\*(m_{\text{t}}^4\*(-((-4 + d)\*(s - t)\*t) + m_{\text{W}}^2\*((-2 + d)\*s + (-4 + d)\*t)) - 
      2\*m_{\text{t}}^2\*(-((-4 + d)\*s\*(s - t)\*t) + m_{\text{W}}^2\*((-2 + d)\*s^2 + (-4 + d)\*t^2)) + 
      s\*(-((-4 + d)\*s\*(s - t)\*t) + m_{\text{W}}^2\*((-2 + d)\*s^2 - (-4 + d)\*s\*t - 
          2\*(-2 + d)\*t^2))))/((-3 + d)\*(m_{\text{t}}^2 - s)\*s\*(m_{\text{t}}^2 - s - t)\*t^2) - 
   (m_{\text{t}}\*S_4\*(m_{\text{t}}^4\*(t\*((7 - 2\*d)\*s + (-4 + d)\*t) + 
        m_{\text{W}}^2\*((-2 + d)\*s + (-4 + d)\*t)) - 2\*m_{\text{t}}^2\*(-(s\*t\*((-7 + 2\*d)\*s + t)) + 
        m_{\text{W}}^2\*((-2 + d)\*s^2 + (-4 + d)\*t^2)) + 
      s\*(s\*t\*((7 - 2\*d)\*s - (-2 + d)\*t) + m_{\text{W}}^2\*((-2 + d)\*s^2 - (-4 + d)\*s\*t - 
          2\*(-2 + d)\*t^2))))/((-3 + d)\*(m_{\text{t}}^2 - s)\*s\*(m_{\text{t}}^2 - s - t)\*t^2) + 
   (S_3\*((-3 + d)\*m_{\text{t}}^6\*t - m_{\text{t}}^4\*(m_{\text{W}}^2\*((-2 + d)\*s - 2\*t) + 
        t\*((-2 + d)\*s + (-5 + d)\*t)) + 
      (s + t)\*(s\*t\*((-4 + d)\*s + (-2 + d)\*t) + 
        m_{\text{W}}^2\*(-((-2 + d)\*s^2) + (-4 + d)\*s\*t + 2\*(-2 + d)\*t^2)) + 
      m_{\text{t}}^2\*(m_{\text{W}}^2\*(2\*(-2 + d)\*s^2 + (4 - 3\*d)\*t^2) + 
        t\*(-((-5 + d)\*s^2) + (-4 + d)\*t^2 + s\*(t - d\*t)))))/
    ((-3 + d)\*t^2\*(-m_{\text{t}}^2 + s + t)^2))\*C_0(b_1,2,3,4)\,,
\end{math}
}
\\
\\
{\scriptsize
\begin{math}
\ampli^{\text{\tiny 2. Box}} =  
D_0(b_2)\*(((m_{\text{t}}^2 - s)^2\*S_7\*(m_{\text{W}}^2 - t)^2)/(4\*(-3 + d)\*s\*t\*(-m_{\text{t}}^2 + s + t)) - 
   (m_{\text{t}}\*(m_{\text{t}}^2 - s)\*S_5\*(m_{\text{W}}^2 - t)^2\*((-4 + d)\*m_{\text{t}}^2 - (-4 + d)\*s - 
      2\*(-3 + d)\*t))/(4\*(-3 + d)\*s\*t\*(-m_{\text{t}}^2 + s + t)^2) + 
   (m_{\text{t}}\*(m_{\text{t}}^2 - s)\*S_1\*(m_{\text{W}}^2 - t)\*(m_{\text{t}}^2\*((-4 + d)\*m_{\text{W}}^2 + (10 - 3\*d)\*t) + 
      m_{\text{W}}^2\*(-((-4 + d)\*s) - 2\*(-3 + d)\*t) + 
      t\*((-10 + 3\*d)\*s + 4\*(-3 + d)\*t)))/((-3 + d)\*s\*t\*(-m_{\text{t}}^2 + s + t)^2) + 
   ((m_{\text{t}}^2 - s)\*S_6\*(m_{\text{W}}^2 - t)^2\*((-2 + d)\*m_{\text{t}}^4 + s\*(2\*s - (-4 + d)\*t) - 
      m_{\text{t}}^2\*(-2\*t + d\*(s + t))))/(4\*(-3 + d)\*s^2\*t\*(-m_{\text{t}}^2 + s + t)^2) + 
   (m_{\text{t}}\*(m_{\text{t}}^2 - s)\*S_4\*(m_{\text{W}}^2 - t)\*
     (-(s\*(t\*(-s + t) + m_{\text{W}}^2\*((-2 + d)\*s + (-4 + d)\*t))) + 
      m_{\text{t}}^2\*((-2 + d)\*m_{\text{W}}^2\*(s - t) + t\*(-s + (-2 + d)\*t))))/
    ((-3 + d)\*s^2\*t^2\*(-m_{\text{t}}^2 + s + t)) - 
   (2\*(m_{\text{t}}^2 - s)\*S_2\*(m_{\text{W}}^2 - t)\*
     (-(s\*((-4 + d)\*(s - t)\*t + m_{\text{W}}^2\*((-2 + d)\*s + (-4 + d)\*t))) + 
      m_{\text{t}}^2\*((-2 + d)\*m_{\text{W}}^2\*(s - t) + t\*((-4 + d)\*s + (-2 + d)\*t))))/
    ((-3 + d)\*s^2\*t^2\*(-m_{\text{t}}^2 + s + t)) - 
   ((m_{\text{t}}^2 - s)\*S_3\*(m_{\text{t}}^4\*(m_{\text{W}}^4\*((-2 + d)\*s - 2\*t) - 2\*m_{\text{W}}^2\*(s - 2\*t)\*t + 
        ((-2 + d)\*s - 2\*t)\*t^2) + s\*(s + t)\*(2\*m_{\text{W}}^2\*t\*(-s + t) + 
        m_{\text{W}}^4\*((-2 + d)\*s + (-4 + d)\*t) + t^2\*((-2 + d)\*s + (-4 + d)\*t)) - 
      m_{\text{t}}^2\*(-2\*m_{\text{W}}^2\*t\*(2\*s^2 - 2\*s\*t + (-4 + d)\*t^2) + 
        m_{\text{W}}^4\*(2\*(-2 + d)\*s^2 + 2\*(-4 + d)\*s\*t + (-4 + d)\*t^2) + 
        t^2\*(2\*(-2 + d)\*s^2 + 2\*(-4 + d)\*s\*t + (-4 + d)\*t^2))))/
    ((-3 + d)\*s\*t^2\*(-m_{\text{t}}^2 + s + t)^2)) + 
 ((4\*m_{\text{t}}\*S_1)/((m_{\text{t}}^2 - t)\*(m_{\text{t}}^2 - s - t)) - 
   (m_{\text{t}}\*S_5)/((m_{\text{t}}^2 - t)\*(m_{\text{t}}^2 - s - t)) - (4\*S_3\*(s + t))/
    ((m_{\text{t}}^2 - t)\*(m_{\text{t}}^2 - s - t)) - S_6/(s\*(-m_{\text{t}}^2 + s + t)) - 
   (8\*S_2)/(m_{\text{t}}^2\*s - s\*t) + (4\*m_{\text{t}}\*S_4)/(m_{\text{t}}^2\*s - s\*t))\*B_0(b_2,1,3) + 
 ((m_{\text{t}}\*S_5\*(2\*m_{\text{t}}^2 - s - t))/((m_{\text{t}}^2 - s)\*(m_{\text{t}}^2 - t)\*(m_{\text{t}}^2 - s - t)) + 
   (4\*m_{\text{t}}^2\*S_3\*(s + t))/((m_{\text{t}}^2 - t)\*(m_{\text{t}}^2 - s - t)\*t) + 
   (4\*m_{\text{t}}\*S_1\*(-2\*m_{\text{t}}^2 + s + t))/((m_{\text{t}}^2 - s)\*(m_{\text{t}}^2 - t)\*(m_{\text{t}}^2 - s - t)) - 
   (m_{\text{t}}^2\*S_6)/(s\*(-m_{\text{t}}^2 + s)\*(-m_{\text{t}}^2 + s + t)) + 
   S_2\*(8/(m_{\text{t}}^2\*s - s\*t) - 8/(m_{\text{t}}^2\*t - s\*t)) + 
   4\*m_{\text{t}}\*S_4\*(-(m_{\text{t}}^2\*s - s\*t)^(-1) + (m_{\text{t}}^2\*t - s\*t)^(-1)))\*B_0(b_2,1,4) + 
 ((4\*m_{\text{t}}\*S_1)/((m_{\text{t}}^2 - s)\*(m_{\text{t}}^2 - s - t)) - 
   (m_{\text{t}}\*S_5)/((m_{\text{t}}^2 - s)\*(m_{\text{t}}^2 - s - t)) + (4\*S_3\*(s + t))/(t\*(-m_{\text{t}}^2 + s + t)) + 
   (8\*S_2)/(m_{\text{t}}^2\*t - s\*t) - (4\*m_{\text{t}}\*S_4)/(m_{\text{t}}^2\*t - s\*t) + 
   S_6/(m_{\text{t}}^4 - 2\*m_{\text{t}}^2\*s + s^2 - m_{\text{t}}^2\*t + s\*t))\*B_0(b_2,2,4) + 
 (-((m_{\text{t}}^2 - s)\*S_7\*(m_{\text{W}}^2 - t))/(4\*(-3 + d)\*s\*(-m_{\text{t}}^2 + s + t)) + 
   (m_{\text{t}}\*S_5\*(m_{\text{W}}^2 - t)\*((-4 + d)\*m_{\text{t}}^2 - (-4 + d)\*s - 2\*(-3 + d)\*t))/
    (4\*(-3 + d)\*s\*(-m_{\text{t}}^2 + s + t)^2) + 
   ((m_{\text{t}}^2 - s)\*S_3\*(m_{\text{W}}^2 - t)\*(m_{\text{t}}^2\*((-2 + d)\*s - 2\*t) - 
      (s + t)\*((-2 + d)\*s + (-4 + d)\*t)))/((-3 + d)\*s\*t\*(-m_{\text{t}}^2 + s + t)^2) + 
   (m_{\text{t}}\*S_1\*(t\*((10 - 3\*d)\*s - 4\*(-3 + d)\*t) + 
      m_{\text{W}}^2\*((-4 + d)\*s + 2\*(-3 + d)\*t) + m_{\text{t}}^2\*(-((-4 + d)\*m_{\text{W}}^2) + 
        (-10 + 3\*d)\*t)))/((-3 + d)\*s\*(-m_{\text{t}}^2 + s + t)^2) + 
   (S_6\*(m_{\text{W}}^2 - t)\*(-((-2 + d)\*m_{\text{t}}^4) + s\*(-2\*s + (-4 + d)\*t) + 
      m_{\text{t}}^2\*(-2\*t + d\*(s + t))))/(4\*(-3 + d)\*s^2\*(-m_{\text{t}}^2 + s + t)^2) - 
   (m_{\text{t}}\*S_4\*(-(s\*(t\*(-s + t) + m_{\text{W}}^2\*((-2 + d)\*s + (-4 + d)\*t))) + 
      m_{\text{t}}^2\*((-2 + d)\*m_{\text{W}}^2\*(s - t) + t\*(-s + (-2 + d)\*t))))/
    ((-3 + d)\*s^2\*t\*(-m_{\text{t}}^2 + s + t)) - 
   (2\*S_2\*(s\*((-4 + d)\*(s - t)\*t + m_{\text{W}}^2\*((-2 + d)\*s + (-4 + d)\*t)) - 
      m_{\text{t}}^2\*((-2 + d)\*m_{\text{W}}^2\*(s - t) + t\*((-4 + d)\*s + (-2 + d)\*t))))/
    ((-3 + d)\*s^2\*t\*(-m_{\text{t}}^2 + s + t)))\*C_0(b_2,1,2,3) + 
 (((m_{\text{t}}^2 - s)^2\*S_7\*(m_{\text{W}}^2 - t))/(4\*(-3 + d)\*s\*t\*(-m_{\text{t}}^2 + s + t)) - 
   (m_{\text{t}}\*(m_{\text{t}}^2 - s)\*S_5\*(m_{\text{W}}^2 - t)\*((-4 + d)\*m_{\text{t}}^2 - (-4 + d)\*s - 2\*(-3 + d)\*t))/
    (4\*(-3 + d)\*s\*t\*(-m_{\text{t}}^2 + s + t)^2) + 
   (m_{\text{t}}\*(m_{\text{t}}^2 - s)\*S_1\*(m_{\text{t}}^2\*((-4 + d)\*m_{\text{W}}^2 + (10 - 3\*d)\*t) + 
      m_{\text{W}}^2\*(-((-4 + d)\*s) - 2\*(-3 + d)\*t) + 
      t\*((-10 + 3\*d)\*s + 4\*(-3 + d)\*t)))/((-3 + d)\*s\*t\*(-m_{\text{t}}^2 + s + t)^2) + 
   ((m_{\text{t}}^2 - s)\*S_6\*(m_{\text{W}}^2 - t)\*((-2 + d)\*m_{\text{t}}^4 + s\*(2\*s - (-4 + d)\*t) - 
      m_{\text{t}}^2\*(-2\*t + d\*(s + t))))/(4\*(-3 + d)\*s^2\*t\*(-m_{\text{t}}^2 + s + t)^2) + 
   (m_{\text{t}}\*(m_{\text{t}}^2 - s)\*S_4\*(-(s\*(t\*(-s + t) + m_{\text{W}}^2\*((-2 + d)\*s + (-4 + d)\*t))) + 
      m_{\text{t}}^2\*((-2 + d)\*m_{\text{W}}^2\*(s - t) + t\*(-s + (-2 + d)\*t))))/
    ((-3 + d)\*s^2\*t^2\*(-m_{\text{t}}^2 + s + t)) - 
   (2\*(m_{\text{t}}^2 - s)\*S_2\*(-(s\*((-4 + d)\*(s - t)\*t + 
         m_{\text{W}}^2\*((-2 + d)\*s + (-4 + d)\*t))) + 
      m_{\text{t}}^2\*((-2 + d)\*m_{\text{W}}^2\*(s - t) + t\*((-4 + d)\*s + (-2 + d)\*t))))/
    ((-3 + d)\*s^2\*t^2\*(-m_{\text{t}}^2 + s + t)) - 
   ((m_{\text{t}}^2 - s)\*S_3\*(m_{\text{t}}^4\*(m_{\text{W}}^2\*((-2 + d)\*s - 2\*t) + t\*((-4 + d)\*s + 2\*t)) + 
      s\*(s + t)\*(m_{\text{W}}^2\*((-2 + d)\*s + (-4 + d)\*t) + 
        t\*((-4 + d)\*s + (-2 + d)\*t)) + 
      m_{\text{t}}^2\*(m_{\text{W}}^2\*(-2\*(-2 + d)\*s^2 - 2\*(-4 + d)\*s\*t - (-4 + d)\*t^2) + 
        t\*(-2\*(-4 + d)\*s^2 - 2\*(-2 + d)\*s\*t + (-4 + d)\*t^2))))/
    ((-3 + d)\*s\*t^2\*(-m_{\text{t}}^2 + s + t)^2))\*C_0(b_2,1,2,4) + 
 ((S_7\*(m_{\text{W}}^2 - t)\*(-m_{\text{t}}^4 + s\*t + m_{\text{t}}^2\*(s + t)))/
    (4\*(-3 + d)\*s\*t\*(-m_{\text{t}}^2 + s + t)) + 
   (m_{\text{t}}\*S_5\*(m_{\text{W}}^2 - t)\*((-4 + d)\*m_{\text{t}}^6 + m_{\text{t}}^4\*(-((-4 + d)\*s) + 2\*(7 - 2\*d)\*t) + 
      m_{\text{t}}^2\*t\*(8\*(-3 + d)\*s + (-16 + 5\*d)\*t) + 
      t\*(-4\*(-3 + d)\*s^2 + (20 - 7\*d)\*s\*t - 2\*(-3 + d)\*t^2)))/
    (4\*(-3 + d)\*s\*(m_{\text{t}}^2 - t)\*t\*(-m_{\text{t}}^2 + s + t)^2) - 
   (S_6\*(m_{\text{W}}^2 - t)\*((-2 + d)\*m_{\text{t}}^6 + s\*t\*(2\*s - (-4 + d)\*t) + 
      m_{\text{t}}^2\*(2\*s^2 + 2\*(-2 + d)\*s\*t + (-2 + d)\*t^2) - 
      m_{\text{t}}^4\*(-4\*t + d\*(s + 2\*t))))/(4\*(-3 + d)\*s^2\*t\*(-m_{\text{t}}^2 + s + t)^2) + 
   (m_{\text{t}}\*S_1\*(m_{\text{t}}^6\*(-((-4 + d)\*m_{\text{W}}^2) + (-10 + 3\*d)\*t) + 
      m_{\text{t}}^2\*t\*(4\*(-3 + d)\*s^2 + 20\*(-3 + d)\*s\*t + (-34 + 11\*d)\*t^2 + 
        m_{\text{W}}^2\*(-8\*(-3 + d)\*s + (16 - 5\*d)\*t)) + 
      m_{\text{t}}^4\*(t\*((22 - 7\*d)\*s + 2\*(16 - 5\*d)\*t) + 
        m_{\text{W}}^2\*((-4 + d)\*s + 2\*(-7 + 2\*d)\*t)) + 
      t\*(t\*(-8\*(-3 + d)\*s^2 + (38 - 13\*d)\*s\*t - 4\*(-3 + d)\*t^2) + 
        m_{\text{W}}^2\*(4\*(-3 + d)\*s^2 + (-20 + 7\*d)\*s\*t + 2\*(-3 + d)\*t^2))))/
    ((-3 + d)\*s\*(m_{\text{t}}^2 - t)\*t\*(-m_{\text{t}}^2 + s + t)^2) + 
   (m_{\text{t}}\*S_4\*(s\*t^2\*(t\*((5 - 2\*d)\*s + (7 - 2\*d)\*t) + 
        m_{\text{W}}^2\*((-2 + d)\*s + (-4 + d)\*t)) + m_{\text{t}}^6\*((-2 + d)\*m_{\text{W}}^2\*(s - t) + 
        t\*(-s + (-2 + d)\*t)) + m_{\text{t}}^2\*t\*(m_{\text{W}}^2\*(4\*(-3 + d)\*s^2 + (-2 + d)\*s\*t - 
          (-2 + d)\*t^2) + t\*(-2\*(-3 + d)\*s^2 + (-7 + 2\*d)\*s\*t + 
          (-2 + d)\*t^2)) + m_{\text{t}}^4\*(t\*(s^2 + s\*t - 2\*(-2 + d)\*t^2) + 
        m_{\text{W}}^2\*(-((-2 + d)\*s^2) + (8 - 3\*d)\*s\*t + 2\*(-2 + d)\*t^2))))/
    ((-3 + d)\*s^2\*t^2\*(-m_{\text{t}}^2 + t)\*(-m_{\text{t}}^2 + s + t)) + 
   (2\*S_2\*(s\*t^2\*((-4 + d)\*(s - t)\*t + m_{\text{W}}^2\*((-2 + d)\*s + (-4 + d)\*t)) + 
      m_{\text{t}}^6\*((-2 + d)\*m_{\text{W}}^2\*(s - t) + t\*((-4 + d)\*s + (-2 + d)\*t)) + 
      m_{\text{t}}^2\*t\*(m_{\text{W}}^2\*(4\*(-3 + d)\*s^2 + (-2 + d)\*s\*t - (-2 + d)\*t^2) + 
        t\*(-4\*(-3 + d)\*s^2 + (-4 + d)\*s\*t + (-2 + d)\*t^2)) - 
      m_{\text{t}}^4\*(m_{\text{W}}^2\*((-2 + d)\*s^2 + (-8 + 3\*d)\*s\*t - 2\*(-2 + d)\*t^2) + 
        t\*((-4 + d)\*s^2 + (-4 + d)\*s\*t + 2\*(-2 + d)\*t^2))))/
    ((-3 + d)\*s^2\*(m_{\text{t}}^2 - t)\*t^2\*(-m_{\text{t}}^2 + s + t)) + 
   (S_3\*(-(s\*(m_{\text{W}}^2 - t)\*t^2\*(s + t)\*((-2 + d)\*s + (-4 + d)\*t)) + 
      m_{\text{t}}^8\*(m_{\text{W}}^2\*((-2 + d)\*s - 2\*t) + t\*((-4 + d)\*s + 2\*t)) + 
      m_{\text{t}}^6\*(m_{\text{W}}^2\*(-2\*(-2 + d)\*s^2 - 4\*(-3 + d)\*s\*t - (-8 + d)\*t^2) + 
        t\*(-2\*(-4 + d)\*s^2 - 2\*(-3 + d)\*s\*t + (-8 + d)\*t^2)) + 
      m_{\text{t}}^4\*(t\*((-4 + d)\*s^3 - 2\*s^2\*t - 4\*(-3 + d)\*s\*t^2 - 2\*(-5 + d)\*t^3) + 
        m_{\text{W}}^2\*((-2 + d)\*s^3 + 2\*(-11 + 4\*d)\*s^2\*t + 10\*(-3 + d)\*s\*t^2 + 
          2\*(-5 + d)\*t^3)) + m_{\text{t}}^2\*t\*
       (m_{\text{W}}^2\*(-4\*(-3 + d)\*s^3 - 8\*(-3 + d)\*s^2\*t + 2\*(8 - 3\*d)\*s\*t^2 - 
          (-4 + d)\*t^3) + t\*(2\*(-3 + d)\*s^3 + 4\*(-3 + d)\*s^2\*t + 
          2\*(-5 + 2\*d)\*s\*t^2 + (-4 + d)\*t^3))))/((-3 + d)\*s\*(m_{\text{t}}^2 - t)\*t^2\*
     (-m_{\text{t}}^2 + s + t)^2))\*C_0(b_2,1,3,4) + 
 ((S_7\*(s\*t - m_{\text{t}}^2\*(m_{\text{W}}^2 + t) + m_{\text{W}}^2\*(s + 2\*t)))/(4\*(-3 + d)\*(m_{\text{t}}^2 - s - t)\*
     t) + (m_{\text{t}}\*S_5\*(m_{\text{W}}^2\*((-4 + d)\*s - 2\*t) - (-4 + d)\*m_{\text{t}}^2\*(m_{\text{W}}^2 + t) + 
      t\*((-4 + d)\*s + 2\*(-3 + d)\*t)))/(4\*(-3 + d)\*t\*(-m_{\text{t}}^2 + s + t)^2) + 
   (m_{\text{t}}\*S_1\*(m_{\text{W}}^2\*(-((-4 + d)\*s) + 2\*t) + t\*((10 - 3\*d)\*s - 4\*(-3 + d)\*t) + 
      m_{\text{t}}^2\*((-4 + d)\*m_{\text{W}}^2 + (-10 + 3\*d)\*t)))/((-3 + d)\*t\*(-m_{\text{t}}^2 + s + t)^2) - 
   (2\*S_2\*((-4 + d)\*s\*(s - t)\*t + m_{\text{W}}^2\*(-((-2 + d)\*s^2) + (-4 + d)\*s\*t + 
        2\*(-2 + d)\*t^2) + m_{\text{t}}^2\*((-2 + d)\*m_{\text{W}}^2\*(s - t) - 
        t\*((-4 + d)\*s + (-2 + d)\*t))))/((-3 + d)\*s\*t^2\*(-m_{\text{t}}^2 + s + t)) - 
   (m_{\text{t}}\*S_4\*(s\*t\*((7 - 2\*d)\*s + (5 - 2\*d)\*t) + 
      m_{\text{W}}^2\*((-2 + d)\*s^2 - (-4 + d)\*s\*t - 2\*(-2 + d)\*t^2) + 
      m_{\text{t}}^2\*(-((-2 + d)\*m_{\text{W}}^2\*(s - t)) + t\*((-7 + 2\*d)\*s + (-2 + d)\*t))))/
    ((-3 + d)\*s\*t^2\*(-m_{\text{t}}^2 + s + t)) + 
   (S_3\*(m_{\text{t}}^4\*(t\*((-4 + d)\*s + 2\*t) + m_{\text{W}}^2\*(-((-2 + d)\*s) + 2\*t)) + 
      m_{\text{t}}^2\*(m_{\text{W}}^2\*(2\*(-2 + d)\*s^2 + (4 - 3\*d)\*t^2) + 
        t\*(-2\*(-4 + d)\*s^2 - 2\*(-2 + d)\*s\*t + (-4 + d)\*t^2)) + 
      (s + t)\*(s\*t\*((-4 + d)\*s + (-2 + d)\*t) + 
        m_{\text{W}}^2\*(-((-2 + d)\*s^2) + (-4 + d)\*s\*t + 2\*(-2 + d)\*t^2))))/
    ((-3 + d)\*t^2\*(-m_{\text{t}}^2 + s + t)^2) + 
   (S_6\*((-2 + d)\*m_{\text{t}}^4\*(m_{\text{W}}^2 + t) + s\*t\*(2\*s - (-4 + d)\*t) + 
      m_{\text{W}}^2\*(2\*s^2 + (-4 + 3\*d)\*s\*t + 2\*(-2 + d)\*t^2) - 
      m_{\text{t}}^2\*(-2\*t\*(3\*m_{\text{W}}^2 + t) + d\*(t\*(s + t) + m_{\text{W}}^2\*(s + 3\*t)))))/
    (4\*(-3 + d)\*s\*t\*(-m_{\text{t}}^2 + s + t)^2))\*C_0(b_2,2,3,4)\,,
\end{math}
}
\\
\\
{\scriptsize
\begin{math}
\ampli^{\text{\tiny 3. Box}} =  
D_0(b_3)\*(((-4 + d)\*d\*m_{\text{t}}\*S_1\*(m_{\text{W}}^2 - t)^2)/(2\*(-3 + d)\*s\*t) - 
   ((-4 + d)\*m_{\text{t}}\*S_5\*(m_{\text{W}}^2 - t)^2)/(4\*(-3 + d)\*s\*t) + 
   (S_7\*(m_{\text{W}}^2 - t)^2\*(-m_{\text{t}}^2 + s + t))/(4\*(-3 + d)\*s\*t) + 
   (S_6\*(m_{\text{W}}^2 - t)^2\*((-2 + d)\*m_{\text{t}}^2 - 2\*s - (-2 + d)\*t))/(4\*(-3 + d)\*s^2\*t) - 
   (m_{\text{t}}\*S_4\*(m_{\text{W}}^2 - t)\*(-m_{\text{t}}^2 + s + t)\*(-((-2 + d)\*m_{\text{W}}^2\*(s - t)) + 
      t\*(s - (-2 + d)\*t)))/((-3 + d)\*s^2\*t^2) + 
   (S_3\*((s + t)\*((-2 + d)\*m_{\text{W}}^4\*(2\*s - 3\*t) + (-2 + d)\*(2\*s - 3\*t)\*t^2 + 
        2\*m_{\text{W}}^2\*t\*(-2\*s + 3\*(-2 + d)\*t)) + 
      m_{\text{t}}^2\*(-2\*m_{\text{W}}^2\*t\*(-2\*s + (2 + d)\*t) + m_{\text{W}}^4\*(-2\*(-2 + d)\*s + (2 + d)\*t) + 
        t^2\*(-2\*(-2 + d)\*s + (2 + d)\*t))))/(2\*(-3 + d)\*s\*t^2) + 
   (S_2\*(m_{\text{W}}^2 - t)\*((-2 + d)\*m_{\text{W}}^2\*(-4\*s^2 + 2\*s\*t + (2 + d)\*t^2) + 
      t\*(-4\*(-4 + d)\*s^2 + 2\*(14 - 5\*d)\*s\*t - (-4 + d^2)\*t^2) + 
      m_{\text{t}}^2\*(-((-2 + d)\*m_{\text{W}}^2\*(-4\*s + (2 + d)\*t)) + 
        t\*(4\*(-4 + d)\*s + (-4 + d^2)\*t))))/(2\*(-3 + d)\*s^2\*t^2)) + 
 ((-2\*d\*m_{\text{t}}\*S_1)/((m_{\text{t}}^2 - t)\*(m_{\text{t}}^2 - s - t)) + 
   (m_{\text{t}}\*S_5)/((m_{\text{t}}^2 - t)\*(m_{\text{t}}^2 - s - t)) + (4\*S_3\*(s + t))/
    ((m_{\text{t}}^2 - t)\*(m_{\text{t}}^2 - s - t)) + S_6/(s\*(-m_{\text{t}}^2 + s + t)) - 
   (4\*m_{\text{t}}\*S_4)/(m_{\text{t}}^2\*s - s\*t) - (S_2\*(2\*(2 + d)\*m_{\text{t}}^2 - 2\*(4\*s + (2 + d)\*t)))/
    (s\*(m_{\text{t}}^2 - t)\*(-m_{\text{t}}^2 + s + t)))\*B_0(b_3,1,3) + 
 ((-4\*S_3)/t - (4\*m_{\text{t}}\*S_4)/(s\*t) + (S_2\*(8\*s + 2\*(2 + d)\*t))/(s\*t\*(s + t)) - 
   S_6/(s^2 + s\*t))\*B_0(b_3,2,4) + ((2\*d\*m_{\text{t}}\*S_1)/((m_{\text{t}}^2 - t)\*(m_{\text{t}}^2 - s - t)) - 
   (m_{\text{t}}\*S_5)/((m_{\text{t}}^2 - t)\*(m_{\text{t}}^2 - s - t)) + (4\*m_{\text{t}}^2\*S_3\*(m_{\text{t}}^2 - s - 2\*t))/
    (t\*(-m_{\text{t}}^2 + t)\*(-m_{\text{t}}^2 + s + t)) - (m_{\text{t}}^2\*S_6)/(s\*(s + t)\*(-m_{\text{t}}^2 + s + t)) + 
   (4\*m_{\text{t}}^3\*S_4)/(m_{\text{t}}^2\*s\*t - s\*t^2) - 
   (2\*m_{\text{t}}^2\*S_2\*(4\*s^2 + 8\*s\*t + (2 + d)\*t^2 - m_{\text{t}}^2\*(4\*s + (2 + d)\*t)))/
    (s\*(m_{\text{t}}^2 - t)\*t\*(s + t)\*(-m_{\text{t}}^2 + s + t)))\*B_0(b_3,3,4) + 
 ((S_7\*(m_{\text{W}}^2 - t))/(12\*s - 4\*d\*s) - ((-4 + d)\*d\*m_{\text{t}}\*S_1\*(m_{\text{W}}^2 - t))/
    (2\*(-3 + d)\*s\*(-m_{\text{t}}^2 + s + t)) + ((-4 + d)\*m_{\text{t}}\*S_5\*(m_{\text{W}}^2 - t))/
    (4\*(-3 + d)\*s\*(-m_{\text{t}}^2 + s + t)) + 
   (S_6\*(m_{\text{W}}^2 - t)\*(-((-2 + d)\*m_{\text{t}}^2) + 2\*s + (-2 + d)\*t))/
    (4\*(-3 + d)\*s^2\*(-m_{\text{t}}^2 + s + t)) + 
   (m_{\text{t}}\*S_4\*(-((-2 + d)\*m_{\text{W}}^2\*(s - t)) + t\*(s - (-2 + d)\*t)))/((-3 + d)\*s^2\*t) - 
   (S_3\*(m_{\text{W}}^2 - t)\*(m_{\text{t}}^2\*(-2\*(-2 + d)\*s + (2 + d)\*t) + 
      (-2 + d)\*(2\*s^2 - s\*t - 3\*t^2)))/(2\*(-3 + d)\*s\*t\*(-m_{\text{t}}^2 + s + t)) - 
   (S_2\*((-2 + d)\*m_{\text{W}}^2\*(-4\*s^2 + 2\*s\*t + (2 + d)\*t^2) + 
      t\*(-4\*(-4 + d)\*s^2 + 2\*(14 - 5\*d)\*s\*t - (-4 + d^2)\*t^2) + 
      m_{\text{t}}^2\*(-((-2 + d)\*m_{\text{W}}^2\*(-4\*s + (2 + d)\*t)) + 
        t\*(4\*(-4 + d)\*s + (-4 + d^2)\*t))))/(2\*(-3 + d)\*s^2\*t\*
     (-m_{\text{t}}^2 + s + t)))\*C_0(b_2,1,2,3) + 
 (((-4 + d)\*d\*m_{\text{t}}\*S_1\*(m_{\text{W}}^2 - t))/(2\*(-3 + d)\*s\*t) - 
   ((-4 + d)\*m_{\text{t}}\*S_5\*(m_{\text{W}}^2 - t))/(4\*(-3 + d)\*s\*t) + 
   (S_7\*(m_{\text{W}}^2 - t)\*(-m_{\text{t}}^2 + s + t))/(4\*(-3 + d)\*s\*t) + 
   (S_6\*(m_{\text{W}}^2 - t)\*((-2 + d)\*m_{\text{t}}^2 - 2\*s - (-2 + d)\*t))/(4\*(-3 + d)\*s^2\*t) + 
   (m_{\text{t}}\*S_4\*(m_{\text{t}}^2 - s - t)\*(-((-2 + d)\*m_{\text{W}}^2\*(s - t)) + t\*(s - (-2 + d)\*t)))/
    ((-3 + d)\*s^2\*t^2) + 
   (S_3\*((s + t)\*((-2 + d)\*m_{\text{W}}^2\*(2\*s - 3\*t) + 
        t\*(2\*(-4 + d)\*s + 3\*(-2 + d)\*t)) + 
      m_{\text{t}}^2\*(-(t\*(2\*(-4 + d)\*s + (2 + d)\*t)) + 
        m_{\text{W}}^2\*(-2\*(-2 + d)\*s + (2 + d)\*t))))/(2\*(-3 + d)\*s\*t^2) + 
   (S_2\*((-2 + d)\*m_{\text{W}}^2\*(-4\*s^2 + 2\*s\*t + (2 + d)\*t^2) + 
      t\*(-4\*(-4 + d)\*s^2 + 2\*(14 - 5\*d)\*s\*t - (-4 + d^2)\*t^2) + 
      m_{\text{t}}^2\*(-((-2 + d)\*m_{\text{W}}^2\*(-4\*s + (2 + d)\*t)) + 
        t\*(4\*(-4 + d)\*s + (-4 + d^2)\*t))))/(2\*(-3 + d)\*s^2\*t^2))\*
  C_0(b_2,1,2,4) + ((S_7\*(m_{\text{t}}^2 - t)\*(m_{\text{W}}^2 - t))/(4\*(-3 + d)\*s\*t) - 
   ((-4 + d)\*d\*m_{\text{t}}\*S_1\*(m_{\text{t}}^2 - t)\*(-m_{\text{W}}^2 + t))/(2\*(-3 + d)\*s\*t\*
     (-m_{\text{t}}^2 + s + t)) + ((-4 + d)\*m_{\text{t}}\*S_5\*(m_{\text{t}}^2 - t)\*(-m_{\text{W}}^2 + t))/
    (4\*(-3 + d)\*s\*t\*(-m_{\text{t}}^2 + s + t)) - 
   (S_6\*(m_{\text{t}}^2 - t)\*(-m_{\text{W}}^2 + t)\*((-2 + d)\*m_{\text{t}}^2 - 2\*s - (-2 + d)\*t))/
    (4\*(-3 + d)\*s^2\*t\*(-m_{\text{t}}^2 + s + t)) + 
   (m_{\text{t}}\*S_4\*(m_{\text{t}}^2 - t)\*((-2 + d)\*m_{\text{W}}^2\*(s - t) + t\*(-s + (-2 + d)\*t)))/
    ((-3 + d)\*s^2\*t^2) + 
   (S_2\*(m_{\text{t}}^2 - t)\*((-2 + d)\*m_{\text{W}}^2\*(-4\*s^2 + 2\*s\*t + (2 + d)\*t^2) + 
      t\*(-4\*(-4 + d)\*s^2 + 2\*(14 - 5\*d)\*s\*t - (-4 + d^2)\*t^2) + 
      m_{\text{t}}^2\*(-((-2 + d)\*m_{\text{W}}^2\*(-4\*s + (2 + d)\*t)) + 
        t\*(4\*(-4 + d)\*s + (-4 + d^2)\*t))))/(2\*(-3 + d)\*s^2\*t^2\*
     (-m_{\text{t}}^2 + s + t)) - 
   (S_3\*(-((-2 + d)\*(m_{\text{W}}^2 - t)\*t\*(-2\*s^2 + s\*t + 3\*t^2)) + 
      m_{\text{t}}^4\*(m_{\text{W}}^2\*(2\*(-2 + d)\*s - (2 + d)\*t) + t\*(2\*(-4 + d)\*s + (2 + d)\*t)) + 
      m_{\text{t}}^2\*(t\*(-2\*(-4 + d)\*s^2 + (10 - 3\*d)\*s\*t - 4\*(-1 + d)\*t^2) + 
        m_{\text{W}}^2\*(-2\*(-2 + d)\*s^2 - (-2 + d)\*s\*t + 4\*(-1 + d)\*t^2))))/
    (2\*(-3 + d)\*s\*t^2\*(-m_{\text{t}}^2 + s + t)))\*C_0(b_3,1,3,4) + 
 ((S_7\*(m_{\text{W}}^2\*s - m_{\text{W}}^2\*t + s\*t + t^2))/(12\*s\*t - 4\*d\*s\*t) - 
   ((-4 + d)\*d\*m_{\text{t}}\*S_1\*(m_{\text{W}}^2\*(s - t) + t\*(s + t)))/
    (2\*(-3 + d)\*s\*t\*(-m_{\text{t}}^2 + s + t)) + 
   ((-4 + d)\*m_{\text{t}}\*S_5\*(m_{\text{W}}^2\*(s - t) + t\*(s + t)))/
    (4\*(-3 + d)\*s\*t\*(-m_{\text{t}}^2 + s + t)) + 
   (m_{\text{t}}\*S_4\*(-(m_{\text{W}}^2\*((-2 + d)\*s^2 + 2\*(-4 + d)\*s\*t + (-2 + d)\*t^2)) + 
      t\*((-7 + 2\*d)\*s^2 + (-3 + d)\*s\*t + (-2 + d)\*t^2)))/((-3 + d)\*s^2\*t^2) - 
   (S_6\*(-((s + t)\*(t\*(s + t)\*(2\*s + (-2 + d)\*t) + 
         m_{\text{W}}^2\*(2\*s^2 + (8 - 3\*d)\*s\*t - (-2 + d)\*t^2))) + 
      m_{\text{t}}^2\*((-2 + d)\*t\*(s + t)^2 + m_{\text{W}}^2\*((-2 + d)\*s^2 - 4\*(-3 + d)\*s\*t - 
          (-2 + d)\*t^2))))/(4\*(-3 + d)\*s^2\*t\*(s + t)\*(-m_{\text{t}}^2 + s + t)) - 
   (S_3\*((s + t)\*(-(t\*(s + t)\*(2\*(-4 + d)\*s + 3\*(-2 + d)\*t)) + 
        m_{\text{W}}^2\*(2\*(-2 + d)\*s^2 + (-14 + 3\*d)\*s\*t + 3\*(-2 + d)\*t^2)) + 
      m_{\text{t}}^2\*(t\*(s + t)\*(2\*(-4 + d)\*s + (2 + d)\*t) - 
        m_{\text{W}}^2\*(2\*(-2 + d)\*s^2 + (-22 + 5\*d)\*s\*t + (2 + d)\*t^2))))/
    (2\*(-3 + d)\*s\*t^2\*(-m_{\text{t}}^2 + s + t)) + 
   (S_2\*(-((s + t)\*(t\*(s + t)\*(4\*(-4 + d)\*s^2 + 2\*(-14 + 5\*d)\*s\*t + 
           (-4 + d^2)\*t^2) + m_{\text{W}}^2\*(-4\*(-2 + d)\*s^3 + 2\*(18 - 5\*d)\*s^2\*t + 
           (24 + 2\*d - 3\*d^2)\*s\*t^2 - (-4 + d^2)\*t^3))) + 
      m_{\text{t}}^2\*(t\*(s + t)^2\*(4\*(-4 + d)\*s + (-4 + d^2)\*t) + 
        m_{\text{W}}^2\*(-4\*(-2 + d)\*s^3 + (44 - 16\*d + d^2)\*s^2\*t - 
          4\*(-4 - 2\*d + d^2)\*s\*t^2 - (-4 + d^2)\*t^3))))/
    (2\*(-3 + d)\*s^2\*t^2\*(s + t)\*(-m_{\text{t}}^2 + s + t)))\*C_0(b_3,2,3,4)\,,
\end{math}
}
\\
\\
{\scriptsize
\begin{math}
\ampli^{\text{\tiny 4. Box}} =  
D_0(b_4)\*(((-2 + d)\*m_{\text{t}}\*S_5\*(m_{\text{W}}^2 - t)^2\*(s + t)^2)/(4\*(-3 + d)\*s\*t\*
     (-m_{\text{t}}^2 + s + t)^2) + (S_7\*(m_{\text{W}}^2 - t)^2\*(s + t)^2)/
    (4\*(-3 + d)\*s\*t\*(-m_{\text{t}}^2 + s + t)) - 
   (m_{\text{t}}\*S_1\*(m_{\text{W}}^2 - t)\*(s + t)^2\*(4\*(-3 + d)\*m_{\text{t}}^2 + d^2\*(m_{\text{W}}^2 - t) + 
      12\*(s + t) - 2\*d\*(m_{\text{W}}^2 + 2\*s + t)))/(2\*(-3 + d)\*s\*t\*
     (-m_{\text{t}}^2 + s + t)^2) - (S_6\*(m_{\text{W}}^2 - t)^2\*(s + t)\*
     (m_{\text{t}}^2\*((-4 + d)\*s - (-2 + d)\*t) + (s + t)\*(2\*s + (-2 + d)\*t)))/
    (4\*(-3 + d)\*s^2\*t\*(-m_{\text{t}}^2 + s + t)^2) + 
   (m_{\text{t}}\*S_4\*(m_{\text{W}}^2 - t)\*(s + t)\*(-(s^2\*t) + (-2 + d)\*t^3 + 
      (-2 + d)\*m_{\text{W}}^2\*(s^2 - t^2)))/((-3 + d)\*s^2\*t^2\*(-m_{\text{t}}^2 + s + t)) + 
   (S_3\*(s + t)\*(2\*(-3 + d)\*m_{\text{t}}^4\*s\*t\*(m_{\text{W}}^2 + t) + 
      (s + t)^2\*((-2 + d)\*m_{\text{W}}^4\*(2\*s - 3\*t) + (-2 + d)\*(2\*s - 3\*t)\*t^2 + 
        2\*m_{\text{W}}^2\*t\*(-2\*s + 3\*(-2 + d)\*t)) - m_{\text{t}}^2\*(s + t)\*
       ((-2 + d)\*m_{\text{W}}^4\*(2\*s - 5\*t) + t^2\*(2\*(-5 + 2\*d)\*s - 5\*(-2 + d)\*t) + 
        2\*m_{\text{W}}^2\*t\*((-5 + d)\*s + 5\*(-2 + d)\*t))))/(2\*(-3 + d)\*s\*t^2\*
     (-m_{\text{t}}^2 + s + t)^2) + (S_2\*(m_{\text{W}}^2 - t)\*(s + t)\*
     (-((s + t)\*(-((-2 + d)\*m_{\text{W}}^2\*(-4\*s^2 + 2\*s\*t + (2 + d)\*t^2)) + 
         t\*(4\*(-4 + d)\*s^2 + 2\*(-14 + 5\*d)\*s\*t + (-4 + d^2)\*t^2))) + 
      m_{\text{t}}^2\*((-2 + d)\*m_{\text{W}}^2\*(4\*s^2 + (-4 + d)\*s\*t - (2 + d)\*t^2) + 
        t\*(4\*(-4 + d)\*s^2 - (32 - 14\*d + d^2)\*s\*t + (-4 + d^2)\*t^2))))/
    (2\*(-3 + d)\*s^2\*t^2\*(-m_{\text{t}}^2 + s + t)^2)) + 
 ((-2\*d\*m_{\text{t}}\*S_1)/((m_{\text{t}}^2 - t)\*(m_{\text{t}}^2 - s - t)) + 
   (m_{\text{t}}\*S_5)/((m_{\text{t}}^2 - t)\*(m_{\text{t}}^2 - s - t)) + (4\*S_3\*(s + t))/
    ((m_{\text{t}}^2 - t)\*(m_{\text{t}}^2 - s - t)) + S_6/(s\*(-m_{\text{t}}^2 + s + t)) - 
   (4\*m_{\text{t}}\*S_4)/(m_{\text{t}}^2\*s - s\*t) - (S_2\*(2\*(2 + d)\*m_{\text{t}}^2 - 2\*(4\*s + (2 + d)\*t)))/
    (s\*(m_{\text{t}}^2 - t)\*(-m_{\text{t}}^2 + s + t)))\*B_0(b_4,1,3) + 
 ((2\*d\*m_{\text{t}}\*S_1)/((m_{\text{t}}^2 - t)\*(m_{\text{t}}^2 - s - t)) - 
   (m_{\text{t}}\*S_5)/((m_{\text{t}}^2 - t)\*(m_{\text{t}}^2 - s - t)) + (4\*m_{\text{t}}^2\*S_3\*(m_{\text{t}}^2 - s - 2\*t))/
    (t\*(-m_{\text{t}}^2 + t)\*(-m_{\text{t}}^2 + s + t)) - (m_{\text{t}}^2\*S_6)/(s\*(s + t)\*(-m_{\text{t}}^2 + s + t)) + 
   (4\*m_{\text{t}}^3\*S_4)/(m_{\text{t}}^2\*s\*t - s\*t^2) - 
   (2\*m_{\text{t}}^2\*S_2\*(4\*s^2 + 8\*s\*t + (2 + d)\*t^2 - m_{\text{t}}^2\*(4\*s + (2 + d)\*t)))/
    (s\*(m_{\text{t}}^2 - t)\*t\*(s + t)\*(-m_{\text{t}}^2 + s + t)))\*B_0(b_4,1,4) + 
 ((-4\*S_3)/t - (4\*m_{\text{t}}\*S_4)/(s\*t) + (S_2\*(8\*s + 2\*(2 + d)\*t))/(s\*t\*(s + t)) - 
   S_6/(s^2 + s\*t))\*B_0(b_4,2,4) + 
 (-((-2 + d)\*m_{\text{t}}\*S_5\*(m_{\text{W}}^2 - t)\*(s + t))/(4\*(-3 + d)\*s\*(-m_{\text{t}}^2 + s + t)^2) - 
   (S_7\*(m_{\text{W}}^2 - t)\*(s + t))/(4\*(-3 + d)\*s\*(-m_{\text{t}}^2 + s + t)) + 
   (m_{\text{t}}\*S_1\*(s + t)\*(4\*(-3 + d)\*m_{\text{t}}^2 + d^2\*(m_{\text{W}}^2 - t) + 12\*(s + t) - 
      2\*d\*(m_{\text{W}}^2 + 2\*s + t)))/(2\*(-3 + d)\*s\*(-m_{\text{t}}^2 + s + t)^2) + 
   (S_6\*(m_{\text{W}}^2 - t)\*(m_{\text{t}}^2\*((-4 + d)\*s - (-2 + d)\*t) + 
      (s + t)\*(2\*s + (-2 + d)\*t)))/(4\*(-3 + d)\*s^2\*(-m_{\text{t}}^2 + s + t)^2) - 
   (m_{\text{t}}\*S_4\*(-(s^2\*t) + (-2 + d)\*t^3 + (-2 + d)\*m_{\text{W}}^2\*(s^2 - t^2)))/
    ((-3 + d)\*s^2\*t\*(-m_{\text{t}}^2 + s + t)) + 
   (S_3\*(2\*(-3 + d)\*m_{\text{t}}^4\*s\*t + (-2 + d)\*(2\*s - 3\*t)\*(-m_{\text{W}}^2 + t)\*(s + t)^2 + 
      m_{\text{t}}^2\*(s + t)\*((-2 + d)\*m_{\text{W}}^2\*(2\*s - 5\*t) + 
        t\*((10 - 4\*d)\*s + 5\*(-2 + d)\*t))))/(2\*(-3 + d)\*s\*t\*
     (-m_{\text{t}}^2 + s + t)^2) + 
   (S_2\*(m_{\text{t}}^2\*(-((-2 + d)\*m_{\text{W}}^2\*(4\*s^2 + (-4 + d)\*s\*t - (2 + d)\*t^2)) + 
        t\*(-4\*(-4 + d)\*s^2 + (32 - 14\*d + d^2)\*s\*t - (-4 + d^2)\*t^2)) + 
      (s + t)\*(-((-2 + d)\*m_{\text{W}}^2\*(-4\*s^2 + 2\*s\*t + (2 + d)\*t^2)) + 
        t\*(4\*(-4 + d)\*s^2 + 2\*(-14 + 5\*d)\*s\*t + (-4 + d^2)\*t^2))))/
    (2\*(-3 + d)\*s^2\*t\*(-m_{\text{t}}^2 + s + t)^2))\*C_0(b_4,1,2,3) + 
 (((-2 + d)\*m_{\text{t}}\*S_5\*(m_{\text{W}}^2 - t)\*(s + t)^2)/(4\*(-3 + d)\*s\*t\*(-m_{\text{t}}^2 + s + t)^2) + 
   (S_7\*(m_{\text{W}}^2 - t)\*(s + t)^2)/(4\*(-3 + d)\*s\*t\*(-m_{\text{t}}^2 + s + t)) + 
   (m_{\text{t}}\*S_1\*(s + t)^2\*(-4\*(-3 + d)\*m_{\text{t}}^2 + d^2\*(-m_{\text{W}}^2 + t) - 12\*(s + t) + 
      2\*d\*(m_{\text{W}}^2 + 2\*s + t)))/(2\*(-3 + d)\*s\*t\*(-m_{\text{t}}^2 + s + t)^2) - 
   (S_6\*(m_{\text{W}}^2 - t)\*(s + t)\*(m_{\text{t}}^2\*((-4 + d)\*s - (-2 + d)\*t) + 
      (s + t)\*(2\*s + (-2 + d)\*t)))/(4\*(-3 + d)\*s^2\*t\*(-m_{\text{t}}^2 + s + t)^2) - 
   (m_{\text{t}}\*S_4\*(s + t)\*(-((-2 + d)\*m_{\text{W}}^2\*(s^2 - t^2)) + t\*(s^2 - (-2 + d)\*t^2)))/
    ((-3 + d)\*s^2\*t^2\*(-m_{\text{t}}^2 + s + t)) + 
   (S_3\*(s + t)\*(2\*(-3 + d)\*m_{\text{t}}^4\*s\*t + (s + t)^2\*((-2 + d)\*m_{\text{W}}^2\*(2\*s - 3\*t) + 
        t\*(2\*(-4 + d)\*s + 3\*(-2 + d)\*t)) - m_{\text{t}}^2\*(s + t)\*
       ((-2 + d)\*m_{\text{W}}^2\*(2\*s - 5\*t) + t\*(2\*(-7 + 2\*d)\*s + 5\*(-2 + d)\*t))))/
    (2\*(-3 + d)\*s\*t^2\*(-m_{\text{t}}^2 + s + t)^2) + 
   (S_2\*(s + t)\*(-((s + t)\*(-((-2 + d)\*m_{\text{W}}^2\*(-4\*s^2 + 2\*s\*t + (2 + d)\*t^2)) + 
         t\*(4\*(-4 + d)\*s^2 + 2\*(-14 + 5\*d)\*s\*t + (-4 + d^2)\*t^2))) + 
      m_{\text{t}}^2\*((-2 + d)\*m_{\text{W}}^2\*(4\*s^2 + (-4 + d)\*s\*t - (2 + d)\*t^2) + 
        t\*(4\*(-4 + d)\*s^2 - (32 - 14\*d + d^2)\*s\*t + (-4 + d^2)\*t^2))))/
    (2\*(-3 + d)\*s^2\*t^2\*(-m_{\text{t}}^2 + s + t)^2))\*C_0(b_4,1,2,4) + 
 (-(S_7\*(m_{\text{W}}^2 - t)\*(m_{\text{t}}^2\*(s - t) + t\*(s + t)))/(4\*(-3 + d)\*s\*t\*
     (-m_{\text{t}}^2 + s + t)) + (m_{\text{t}}\*S_5\*(m_{\text{W}}^2 - t)\*(-((-2 + d)\*m_{\text{t}}^4\*(s - t)) - 
      2\*m_{\text{t}}^2\*t\*(2\*(-3 + d)\*s + (-2 + d)\*t) + 
      t\*(s + t)\*(4\*(-3 + d)\*s + (-2 + d)\*t)))/(4\*(-3 + d)\*s\*(m_{\text{t}}^2 - t)\*t\*
     (-m_{\text{t}}^2 + s + t)^2) + (S_6\*(m_{\text{W}}^2 - t)\*(t\*(s + t)\*(2\*s + (-2 + d)\*t) + 
      m_{\text{t}}^4\*((-4 + d)\*s + (-2 + d)\*t) + 2\*m_{\text{t}}^2\*(s^2 - (-2 + d)\*s\*t - 
        (-2 + d)\*t^2)))/(4\*(-3 + d)\*s^2\*t\*(-m_{\text{t}}^2 + s + t)^2) - 
   (m_{\text{t}}\*S_4\*(t^2\*(-((-2 + d)\*m_{\text{W}}^2\*(s^2 - t^2)) + 
        t\*((-5 + 2\*d)\*s^2 - (-2 + d)\*t^2)) - 
      2\*m_{\text{t}}^2\*t\*(-(t\*((-3 + d)\*s^2 + (-4 + d)\*s\*t + (-2 + d)\*t^2)) + 
        m_{\text{W}}^2\*(2\*(-3 + d)\*s^2 + (-4 + d)\*s\*t + (-2 + d)\*t^2)) + 
      m_{\text{t}}^4\*(-(t\*(s^2 + 2\*(-4 + d)\*s\*t + (-2 + d)\*t^2)) + 
        m_{\text{W}}^2\*((-2 + d)\*s^2 + 2\*(-4 + d)\*s\*t + (-2 + d)\*t^2))))/
    ((-3 + d)\*s^2\*(m_{\text{t}}^2 - t)\*t^2\*(-m_{\text{t}}^2 + s + t)) + 
   (S_3\*(-2\*(-3 + d)\*m_{\text{t}}^8\*s\*t + (-2 + d)\*(2\*s - 3\*t)\*(m_{\text{W}}^2 - t)\*t^2\*
       (s + t)^2 + m_{\text{t}}^6\*(t\*(2\*(-7 + 2\*d)\*s^2 + (-8 + 5\*d)\*s\*t - 
          5\*(-2 + d)\*t^2) + m_{\text{W}}^2\*(2\*(-2 + d)\*s^2 + (-10 + d)\*s\*t + 
          5\*(-2 + d)\*t^2)) + m_{\text{t}}^2\*t\*(s + t)\*
       (t\*(-4\*(-3 + d)\*s^2 + (30 - 8\*d)\*s\*t - 11\*(-2 + d)\*t^2) + 
        m_{\text{W}}^2\*(8\*(-3 + d)\*s^2 + 2\*(-18 + 5\*d)\*s\*t + 11\*(-2 + d)\*t^2)) + 
      m_{\text{t}}^4\*(m_{\text{W}}^2\*(-2\*(-2 + d)\*s^3 + (42 - 13\*d)\*s^2\*t + 6\*(10 - 3\*d)\*s\*t^2 - 
          13\*(-2 + d)\*t^3) + t\*(-2\*(-4 + d)\*s^3 + (-6 + d)\*s^2\*t + 
          6\*(-7 + 2\*d)\*s\*t^2 + 13\*(-2 + d)\*t^3))))/
    (2\*(-3 + d)\*s\*(m_{\text{t}}^2 - t)\*t^2\*(-m_{\text{t}}^2 + s + t)^2) + 
   (S_2\*(-(t^2\*(s + t)\*(-((-2 + d)\*m_{\text{W}}^2\*(-4\*s^2 + 2\*s\*t + (2 + d)\*t^2)) + 
         t\*(4\*(-4 + d)\*s^2 + 2\*(-14 + 5\*d)\*s\*t + (-4 + d^2)\*t^2))) + 
      m_{\text{t}}^6\*(-(m_{\text{W}}^2\*(4\*(-2 + d)\*s^2 + (-24 + 2\*d + d^2)\*s\*t + 
           (-4 + d^2)\*t^2)) + t\*(-4\*(-4 + d)\*s^2 + (-48 + 10\*d + d^2)\*s\*t + 
          (-4 + d^2)\*t^2)) + m_{\text{t}}^4\*(t\*(4\*(-4 + d)\*s^3 + (84 - 26\*d)\*s^2\*t + 
          (128 - 30\*d - 3\*d^2)\*s\*t^2 - 3\*(-4 + d^2)\*t^3) + 
        m_{\text{W}}^2\*(4\*(-2 + d)\*s^3 + 2\*(-42 + 13\*d)\*s^2\*t + (-56 + 6\*d + 3\*d^2)\*s\*
           t^2 + 3\*(-4 + d^2)\*t^3)) + 
      m_{\text{t}}^2\*t\*(m_{\text{W}}^2\*(-16\*(-3 + d)\*s^3 + 4\*(18 - 5\*d)\*s^2\*t + 
          (40 - 6\*d - 3\*d^2)\*s\*t^2 - 3\*(-4 + d^2)\*t^3) + 
        t\*(16\*(-3 + d)\*s^3 + 4\*(-36 + 11\*d)\*s^2\*t + (-112 + 30\*d + 3\*d^2)\*s\*
           t^2 + 3\*(-4 + d^2)\*t^3))))/(2\*(-3 + d)\*s^2\*(m_{\text{t}}^2 - t)\*t^2\*
     (-m_{\text{t}}^2 + s + t)^2) + (m_{\text{t}}\*S_1\*(4\*(-3 + d)\*m_{\text{t}}^6\*(s - t) + 
      m_{\text{t}}^4\*(d^2\*(m_{\text{W}}^2 - t)\*(s - t) + 12\*(s^2 - 3\*t^2) + 
        2\*d\*(-2\*s^2 + s\*t + 5\*t^2 + m_{\text{W}}^2\*(-s + t))) + 
      2\*m_{\text{t}}^2\*t\*(18\*t\*(s + t) + d^2\*(m_{\text{W}}^2 - t)\*(2\*s + t) - 
        2\*d\*(2\*t^2 + m_{\text{W}}^2\*(3\*s + t))) + t\*(s + t)\*(-12\*t\*(s + t) - 
        d^2\*(m_{\text{W}}^2 - t)\*(4\*s + t) + 2\*d\*(t\*(-4\*s + t) + m_{\text{W}}^2\*(6\*s + t)))))/
    (2\*(-3 + d)\*s\*(m_{\text{t}}^2 - t)\*t\*(-m_{\text{t}}^2 + s + t)^2))\*C_0(b_4,1,3,4) + 
 ((S_7\*(m_{\text{W}}^2\*s - m_{\text{W}}^2\*t + s\*t + t^2))/(12\*s\*t - 4\*d\*s\*t) - 
   ((-2 + d)\*m_{\text{t}}\*S_5\*(m_{\text{W}}^2\*(s - t) + t\*(s + t)))/
    (4\*(-3 + d)\*s\*t\*(-m_{\text{t}}^2 + s + t)) + 
   (m_{\text{t}}\*S_4\*(t\*((-7 + 2\*d)\*s^2 + (-2 + d)\*t^2) - 
      m_{\text{W}}^2\*((-2 + d)\*s^2 + 2\*(-4 + d)\*s\*t + (-2 + d)\*t^2)))/
    ((-3 + d)\*s^2\*t^2) + (m_{\text{t}}\*S_1\*(4\*(-3 + d)\*m_{\text{t}}^2\*(s - t) + 12\*(s^2 - t^2) - 
      2\*d\*(2\*s^2 + m_{\text{W}}^2\*(s - t) + s\*t - t^2) + 
      d^2\*(m_{\text{W}}^2\*(s - t) + t\*(s + t))))/(2\*(-3 + d)\*s\*t\*(-m_{\text{t}}^2 + s + t)) - 
   (S_6\*(-(t\*(s + t)\*(2\*s + (-2 + d)\*t)) - m_{\text{W}}^2\*(2\*s^2 + (8 - 3\*d)\*s\*t - 
        (-2 + d)\*t^2) - m_{\text{t}}^2\*(t\*((-4 + d)\*s - (-2 + d)\*t) + 
        m_{\text{W}}^2\*((-4 + d)\*s + (-2 + d)\*t))))/(4\*(-3 + d)\*s^2\*t\*
     (-m_{\text{t}}^2 + s + t)) - 
   (S_3\*(2\*(-3 + d)\*m_{\text{t}}^4\*s\*t + (s + t)\*
       (-(t\*(s + t)\*(2\*(-4 + d)\*s + 3\*(-2 + d)\*t)) + 
        m_{\text{W}}^2\*(2\*(-2 + d)\*s^2 + (-14 + 3\*d)\*s\*t + 3\*(-2 + d)\*t^2)) - 
      m_{\text{t}}^2\*(t\*(s + t)\*(2\*s - 5\*(-2 + d)\*t) + m_{\text{W}}^2\*(2\*(-2 + d)\*s^2 + 
          (-10 + d)\*s\*t + 5\*(-2 + d)\*t^2))))/(2\*(-3 + d)\*s\*t^2\*
     (-m_{\text{t}}^2 + s + t)) - 
   (S_2\*(t\*(s + t)\*(4\*(-4 + d)\*s^2 + 2\*(-14 + 5\*d)\*s\*t + (-4 + d^2)\*t^2) + 
      m_{\text{W}}^2\*(-4\*(-2 + d)\*s^3 + 2\*(18 - 5\*d)\*s^2\*t + (24 + 2\*d - 3\*d^2)\*s\*t^2 - 
        (-4 + d^2)\*t^3) + m_{\text{t}}^2\*(t\*(-4\*(-4 + d)\*s^2 + (32 - 14\*d + d^2)\*s\*t - 
          (-4 + d^2)\*t^2) + m_{\text{W}}^2\*(4\*(-2 + d)\*s^2 + (-24 + 2\*d + d^2)\*s\*t + 
          (-4 + d^2)\*t^2))))/(2\*(-3 + d)\*s^2\*t^2\*(-m_{\text{t}}^2 + s + t)))\*C_0(b_4,2,3,4)\,.
\end{math}
}
	\bibliographystyle{utphysmod}
	\bibliography{bibliography}
	\section*{Danksagung}
An dieser Stelle möchte ich mich gern aufrichtig und persönlich bei bestimmten Menschen bedanken:

Herrn Prof. Dr. Uwer danke ich für die Möglichkeit der Bearbeitung dieses Themas sowie 
für die fordernde Betreuung und Unterstützung während des letzten Jahres.
So konnte ich in dem vergangenen Jahr sehr viel dazu lernen, diese Erfahrungen möchte
ich nicht missen müssen.

Herrn Prof. Dr. Plefka möchte ich ebenfalls für jedes Entgegenkommen und 
jede Unterstützung seinerseits meinen Dank aussprechen.

Auch möchte ich an dieser Stelle die Rosa-Luxemburg-Stiftung wissen lassen,
dass ich ohne die Möglichkeit des
Stipendiums dem Studium nicht so hätte nachgehen können, wie ich es getan habe: 
Mit der Freiheit und Sicherheit konnte ich mich auf die Schönheit der Physik konzentrieren.
Vielen Dank dafür.

Ein besonderer Dank geht an meine Arbeitsgruppe, an jeden, der bereit war mir bei
meinen Fragen weiterzuhelfen und dies in den vielen Gesprächen auch oft getan hat.
Besonders Mohammad Assadsolimani möchte ich danken für seine Geduld mit meiner ungeduldigen Art.
Auch Bas Tausk stand mir bei Fragen helfend zur Seite, dafür möchte ich ihm herzlich danken.

Thomas Kintscher möchte ich auch meinen Dank für fast sechs Jahre Zusammenarbeit aussprechen.
Vom Mathebrückenkurs angefangen bis zur Abgabe der Masterarbeit, bei schwierigen, 
physikalischen, technischen Fragen war er immer mein Ansprechpartner erster Wahl.

Thomas Hähnert, der mir meine Leidenschaft für die Physik als erster offenbart hat, 
gilt mein herzlicher Dank.

Den Menschen, die mich auch außerhalb der Physik unterstützt haben, gilt mein
tiefer Dank:
Meine Freunde, die mich trotz der vielen Arbeit nicht vergessen haben.
Meinem Vater und meinen Brüdern, die nie müde werden mich mit kopfzerbrechenden Fragen
herauszufordern.

Meinem Freund, der mich in den letzten Monaten viel zu wenig zu Gesicht bekommen hat und
doch in den wichtigsten Momenten immer da war.

Zum Schluss möchte ich meiner Mutter danken, dafür dass sie mich auf diese Welt gebracht hat
und mich immer ermutigt hat, dass zu tun, was ich tun möchte.
	
	
	\chapter*{Selbständigkeitserklärung}

Hiermit versichere ich,
dass ich die vorliegende Arbeit selbständig verfasst und keine
anderen als die angegebenen Quellen und Hilfsmittel verwendet habe.

Ich erkläre die Beachtung der Prüfungsordnung und das erstmalige 
Einreichen einer Masterarbeit in diesem Studiengebiet.

\bigskip
Berlin, den \abgabedatum

\end{document}